\documentclass[]{interact}

\usepackage[caption=false]{subfig}

\usepackage[numbers,sort&compress]{natbib}
\bibpunct[, ]{[}{]}{,}{n}{,}{,}
\makeatletter
\def\NAT@def@citea{\def\@citea{\NAT@separator}}
\makeatother

\usepackage{color}

\definecolor{copycolor} 		  	{rgb} {0.5, 0.25, 0.25}

\usepackage[
  linktocpage=false,        
  colorlinks=true,          
  linkcolor=blue,
  citecolor=blue,
  urlcolor=blue,
  breaklinks=true,          
  bookmarks,                
  hyperfootnotes=true      
]{hyperref}

\usepackage{siunitx}[=v2]
\DeclareSIUnit{\au}{a.u.}

\usepackage{etoolbox}
\usepackage{xspace}

\newcommand{\Up}{\ensuremath{U_\text{p}}}
\newcommand{\Uploc}{\ensuremath{U_\text{p}^\text{loc}}}
\newcommand{\Ip}{\ensuremath{I_\text{p}}}
\newcommand{\cep}{\ensuremath{\varphi_\text{ce}\xspace}}

\newcommand{\per}{\ensuremath{\varepsilon_\text{r}\xspace}}

\newcommand{\panel}[1]{\textbf{(#1)}~\ignorespaces}

\renewcommand{\vec}     [1]	{\boldsymbol{#1}}				                            
\newcommand{\vecfield}  [1] {\ifstrequal{#1}{E}{\boldsymbol{\mathcal{#1}}}{\vec{#1}}}
\newcommand{\field}     [1] {\ifstrequal{#1}{E}{\mathcal{#1}}{#1}}

\newcommand{\silica}{SiO\textsubscript{2}\xspace}
\newcommand{\MCCC}{M\textsuperscript{3}C\xspace}

\newcommand{\etal}{\textit{~et~al.}}

\NewDocumentCommand{\Figure}{mg}{\hyperref[{#1}]{Figure~\ref*{#1}\IfValueT{#2}{(#2)}}}
\NewDocumentCommand{\Fig}   {mg}{\hyperref[{#1}]{Fig.~\ref*{#1}\IfValueT{#2}{(#2)}}}

\NewDocumentCommand{\Section}{mg}{\hyperref[{#1}]{Section~\ref*{#1}\IfValueT{#2}{(#2)}}}
\NewDocumentCommand{\Sec}   {mg}{\hyperref[{#1}]{Sec.~\ref*{#1}\IfValueT{#2}{(#2)}}}

\NewDocumentCommand{\Equation}{mg}{\hyperref[{#1}]{Equation~\ref*{#1}\IfValueT{#2}{(#2)}}}
\NewDocumentCommand{\Eq}   {mg}{\hyperref[{#1}]{Eq.~\ref*{#1}\IfValueT{#2}{(#2)}}}

\begin{document}
\articletype{ARTICLE TEMPLATE}

\title{Strong-field physics with nanospheres}

\author{
	\name{
		Lennart Seiffert\textsuperscript{1},
		Sergey Zherebtsov\textsuperscript{2,3},
    	Matthias F. Kling\textsuperscript{2,3},
		Thomas Fennel\textsuperscript{1,4,5}\thanks{CONTACT Thomas Fennel: thomas.fennel@uni-rostock.de}
	}
	\affil{
		\textsuperscript{1}Institute of Physics, University of Rostock, 18059 Rostock, Germany\\
		\textsuperscript{2}Max-Planck-Institute of Quantum Optics, 85748 Garching, Germany\\
		\textsuperscript{3}Physics Department, Ludwig-Maximilians-Universität Munich, 85748 Garching, Germany\\
		\textsuperscript{4}Department of Life, Light and Matter, University of Rostock, 18059 Rostock, Germany\\
        \textsuperscript{5}Max Born Institute, 12489 Berlin, Germany
	}
}

\maketitle

\begin{abstract}
When intense laser fields interact with nanoscale targets, strong-field physics meets plasmonic near-field enhancement and sub-wavelength localization of light. Photoemission spectra reflect the associated attosecond optical and electronic response and encode the collisional and collective dynamics of the solid. Nanospheres represent an ideal platform to explore the underlying attosecond nanophysics because of their particularly simple geometry. This review summarizes key results from the last decade and aims to provide the essential stepping stones for students and researchers to enter this field.
\end{abstract}



\section{Introduction}
\begin{figure}[h!]
\centering
\includegraphics[width=1.0\textwidth]{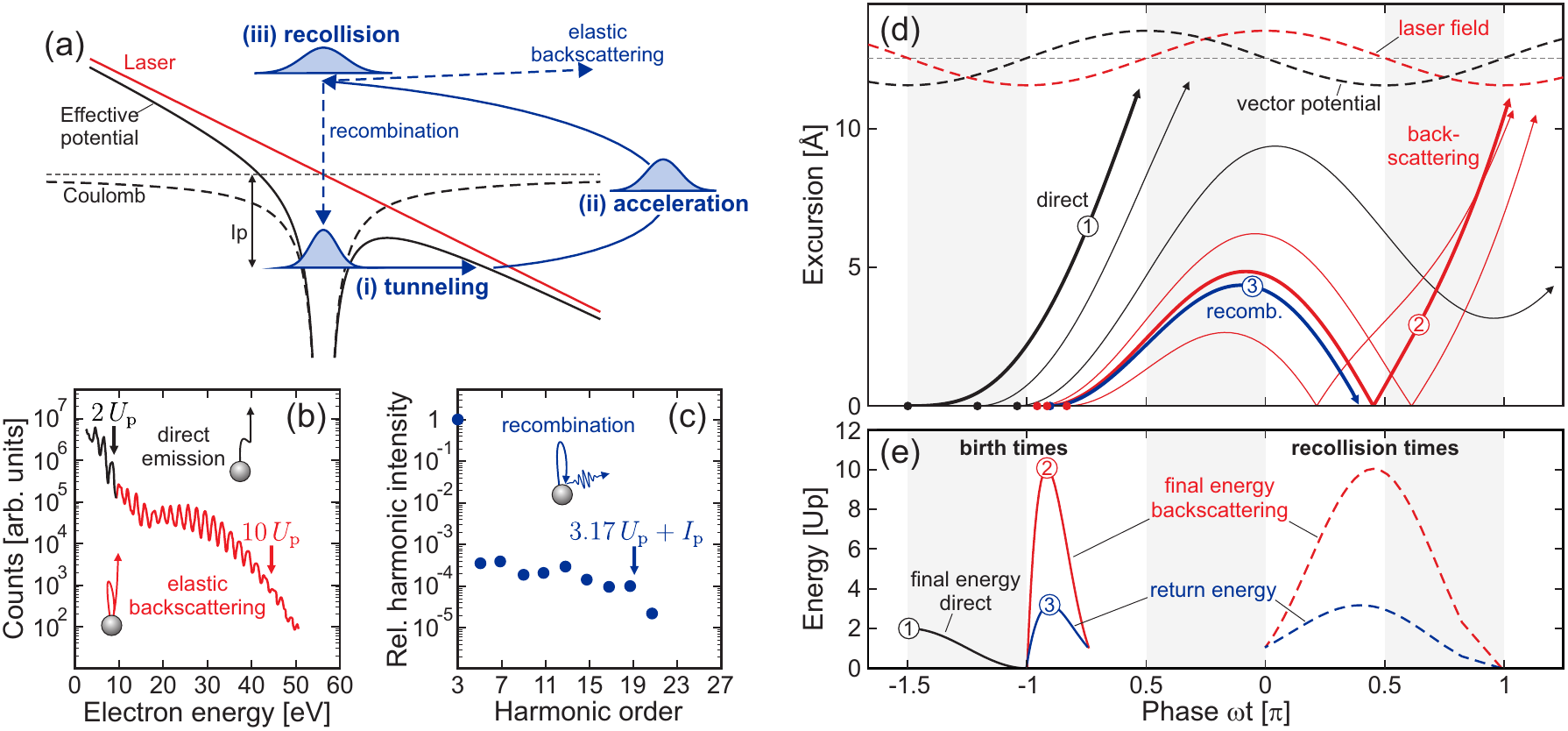}
\caption{Strong-field physics with atomic targets. \panel{a} Three-step model. An electron can (i) tunnel through the finite barrier of the effective potential (solid black curve), (ii) be accelerated by the laser field (red curve), and (iii) recollide with the ion resulting in elastic backscattering or recombination. \panel{b} HATI spectrum from Ar under $\SI{40}{fs}$ $\SI{630}{nm}$ pulses at $I = \SI{1.2e14}{W/cm^2}$. Adapted from~\cite{Paulus_PRL72_1994}. \panel{c} Intensities of harmonics of $\SI{1064}{nm}$ pulses generated in Xe at $I\approx \SI{3e13}{W/cm^2}$. Adapted from~\cite{Ferrey_JPB21_1988}. Arrows indicate the spectral cut-offs as predicted by the three-step model. \panel{d}~Trajectories for direct emission (black) and with one recollision (red) for different birth times (colored dots). Bold curves reflect optimal trajectories for direct emission '1', elastic backscattering '2' and recombination '3'. Dashed red and black curves visualize the evolution of laser electric field and its vector potential. White and gray areas indicate quarter-cycles of the field. \panel{e}~Birth time-dependent final energies for direct emission (black curve) and backscattering (solid red), and recollision energies (solid blue). Recollision timings are indicated by respective dashed curves. From~\cite{Seiffert_PhD_2018}.}
\label{fig:introduction_threestep}
\end{figure}

The availability of intense and well-controlled laser fields with durations near to or even shorter than one optical cycle and with magnitudes approaching those in atoms has fueled a remarkable development of strong-field and attosecond science. Nowadays, strong-field physics with atomic and molecular targets is well established and discussed in great detail by various comprehensive reviews (see e.g.~\cite{Brabec_RMP72_2000, Becker_AAMOP48_2002, Scrinzi_JPB39_2005, Milosevic_JPB39_2006, Krausz_RMP81_2009}). In remarkably many cases much of the basic physics can be captured in terms of classical interpretations via trajectories or respective quantum trajectories~\cite{Lewenstein_PRA49_1994}. The most prominent and successful example is the famous ’three-step model’ or ’simple man’s model’ (SMM) of strong-field science suggested by Paul Corkum in 1993~\cite{Corkum_PRL71_1993}. It provides the intuitive picture of an electron being (i) liberated from a parent ion via tunnelling, (ii) the laser field accelerating the electron away and then back towards the residual ion where it can (iii) recollide. In the latter step the electron may recombine or rescatter, e.g. via elastic backscattering, see \Fig{fig:introduction_threestep}{a}. The impact of the three-step model on the field of attosecond physics is highlighted by a recent special issue ’celebrating 25 years of recollision physics’~\cite{Schafer_JPB50_2018}. Its most important aspects were the clarification of the cutoffs in high harmonic generation (HHG) spectra~\cite{McPherson_JOSAB4_1987, Ferrey_JPB21_1988, Li_PRA39_1989, Krause_PRL68_1992, Huilier_PRA48_1993, Corkum_PRL71_1993, Lewenstein_PRA49_1994} and the energy spectra of emitted electrons via high-order above-threshold ionization (HATI)~\cite{Paulus_PRL72_1994, Paulus_JPB27_1994, Becker_JPB51_2018}. For example, the agreement of the two pronounced step-like features in the photoelectron spectrum in \Fig{fig:introduction_threestep}{b} with the predicted cut-offs at $\SI{2}{\Up}$ and $\SI{10}{\Up}$ from the three-step model enables an unambiguous assignment of the low and high energy signals to direct electron emission and elastic backscattering~\cite{Paulus_JPB27_1994}. Here $\Up = e^2\field{E}_0^2 / (4m_\text{e}\omega^2)$ is the ponderomotive potential which reflects the mean kinetic energy of the quiver motion of an electron (with elementary charge $e$ and electron mass $m_\text{e}$) in a continuous field with field strength $\field{E}_0$ and frequency $\omega$. The waveform-sensitivity of the backscattering process for example enabled to characterize the carrier-envelope phase (CEP) of few-cycle laser pulses~\cite{Paulus_PRL91_2003}. Waveform-controlled backscattering was also demonstrated using multi-color~\cite{Kitzler_PRL95_2005, Skruszewicz_PRL115_2015} and bicircular laser fields~\cite{Medisauskas_PRL115_2015, Milosevic_OL40_2015}. Further, the cut-off of the harmonic emission in \Fig{fig:introduction_threestep}{c} around $\SI{3.17}{\Up} + \Ip$ can be associated with the maximal return energy of electrons prior to recombination~\cite{Corkum_PRL71_1993} plus the ionization energy $\Ip$ of the atom. Typical trajectories for direct emission, backscattering and recombination as predicted by SMM calculations are shown in \Fig{fig:introduction_threestep}{d} (as indicated). The respective final and return energies are shown in dependence of the time of liberation (the 'birth' time) in \Fig{fig:introduction_threestep}{e}, visualizing the well-known maximum values of $2$, $10$ and \SI{3.17}{\Up} for the mentioned three characteristic trajectory classes.

\begin{figure}[h!]
\centering
\includegraphics[width=0.75\textwidth]{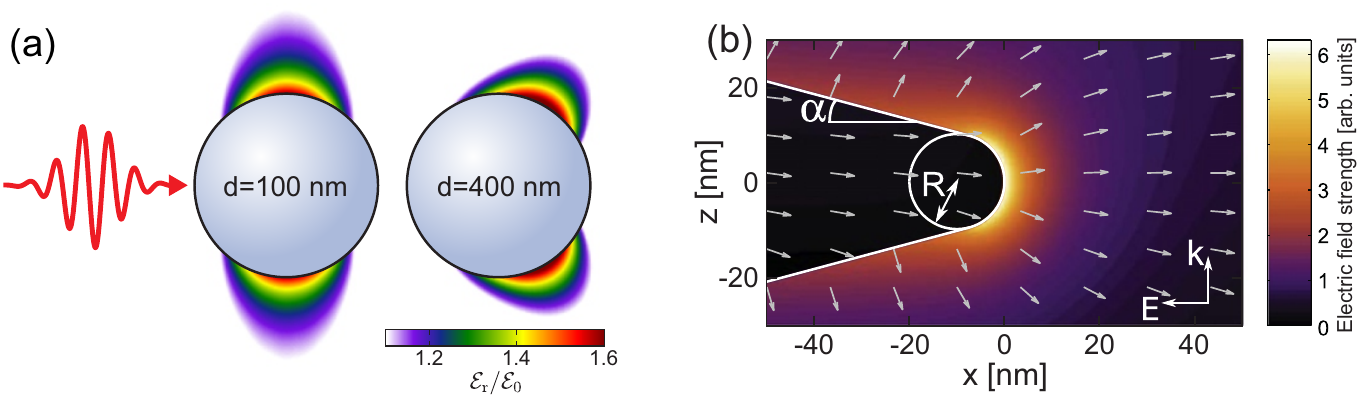}
\caption{Enhanced near-fields at nanostructures. \panel{a} Maximum enhancements of the radial near-fields at small and large  silica nanospheres under  \SI{5}{fs} \SI{720}{nm} few-cycle pulses with respect to the peak field strength of the incident field (Mie calculations). \panel{b} Near-field at a tungsten nanotip with \SI{10}{nm} apex radius and opening angle $\alpha = \ang{15}$ under a \SI{5}{fs} \SI{800}{nm} few-cycle laser pulse propagating in $z$-direction and polarized in $x$-direction. Gray arrows indicate the local orientation of the field at the instant of maximal enhancement. Adapted from~\cite{Thomas_NJP17_2015}.}
\label{fig:introduction_sphere_tip}
\end{figure}

About a decade ago, strong-field physics experienced a substantial boost of attention when researchers started to extend the established concepts from the atomic and molecular world to nanostructures, surfaces and solids. A key driver and promising future perspective for this trend is for example the realization of future protocols for ultrafast light-wave driven nanoelectronics~\cite{Krausz_NatPhot08_2014, Schoetz_ACSP6_2019}. The fundamental interest in strong-field physics of more complex (nano-)targets arises from two perspectives. First, nanostructures allow the generation of local near-fields which can be substantially enhanced with respect to the incident field, allow extreme localization of fields far beyond the wavelength, and can lead to pronounced field inhomogeneities (cf. examples for dielectric nanospheres and metal nanotips in \Fig{fig:introduction_sphere_tip}). Second, the nanostructure geometry and properties affect the electron dynamics, for example due to transport effects (collisions) within the material, effects resulting from the band structure, or the surface properties that influence collisional processes via the associated energy-dependent recollision probability and directionality (specular vs. isotropic reflections). All these additional aspects reach far beyond the common scope of (atomic) strong-field physics and form the basis for strong-field nanophysics.  

An application-wise particularly promising aspect is HHG using the enhanced fields from nanostructures. Great attention for this idea was generated by early work from Kim\etal~\cite{Kim_Nature453_2008}, which was highly debated and is today seen as a misinterpretation of the data. The authors claimed that resonantly enhanced plasmonic fields of individual nanostructures were used to drive HHG in a surrounding gas target employing unprecedentedly low input laser fluence. Via a systematic study of the properties of the emitted light Sivis\etal~could show that mainly incoherent line emission from the excited gas dominates the light emission as a result from the unfavorable scaling of the phase matched coherent signal with interaction volume~\cite{Sivis_Nature485_2012}. However, this limitation can be overcome by utilizing nanostructure arrays or by using the solid of the structure itself or the surrounding support material as the non-linear medium. For a recent example see~\cite{Sivis_NPI_2020}.

In this review we focus on the analysis of the strong-field induced electron dynamics itself, primarily by means of angle and energy-dependent photoemission spectra. One branch of existing studies has focused on the application-wise (microscopy) highly relevant use of nanometric needle tips (henceforth termed 'nanotips'). An overview over the advances in strong-field physics with nanotips has recently been published by Dombi\etal~\cite{Dombi_RMP92_2020}. Metal nanotips are relatively easy to manufacture~\cite{Klein_RSI68_1997, Eisele_RSI82_2011, Lopes_RSI84_2013}, exhibit high field enhancements~\cite{Thomas_NJP17_2015}, and allow for reproducible non-destructive measurements. Pioneering studies have revealed several important similarities between strong-field photoemission from nanotips and atoms including applicability of the coherent recollision pictures underlying high-order above-threshold-ionization~\cite{Krueger_Nature475_2011}. The early studies, however, also uncovered significant differences. One key aspect that leads to fundamental changes is the spatial field gradient that results from the finite extension of the optical near-field. If the spatial scale of the field-driven electronic quiver motion exceeds the decay length associated with the near-field, the quiver picture breaks down and electrons are no longer accelerated ponderomotively but, in the limit of very large nominal quiver amplitudes, by an impulsive boost. Using sufficiently long laser wavelength, this transition is reflected in fundamental changes in the scaling of the electron energy with intensity as well as in the acceleration mechanisms since surface recollisions are quenched~\cite{Herink_Nature483_2012}. Furthermore, nanotips enable intriguing applications such as the generation of coherent attosecond electron pulses which can be employed for near-field mediated quantum coherent manipulation and reconstruction of free-electron beams~\cite{Feist_Nature521_2015}.

Besides the various advantages of nanotips or deposited nanostructures, a complete modelling of their strong-field response is challenging due to the typically complex shape or the presence of a support material. A promising alternative results from the possibility to generate individual nanoparticles in the gas-phase. In fact, the first demonstration that elastic backscattering dominates the high energy part of the strong-field photoemission from nanostructures has been reported for dielectric nanospheres~\cite{Zherebtsov_NatPhys7_2011}. The central method enabling the experiments on isolated nanoparticles is aerosol injection~\cite{Liu_AST22_1995}. This concept allows to use the detection methods which were established for gas phase studies, as for example velocity map imaging (VMI)~\cite{Eppink_RSI68_1997}. Another unique advantage of nanoparticles in the gas phase is that a fresh target enters the interaction region for each laser shot, enabling studies in regimes where the targets are irreversibly modified or destroyed following the interaction with the strong field. This regime is especially relevant for applications in the field of laser material processing, where permanent modifications of the material are desired~\cite{Sugioka_LSA3_2014}.

While in principle arbitrarily shaped nanoparticles might be synthesized and delivered by aerosol injection methods~\cite{Liu_AST22_1995}, spherical particles are of paramount interest as they provide several important advantages. For example, missing knowledge about alignment, which is often required for the interpretation of experimental data, constitutes no problem when utilizing spherical particles. On the theory side, a central challenge is the accurate description of the combination of {\AA}ngström scale electron dynamics and light propagation on wavelength (nanometer) scale. However, the high symmetry provided by spherical systems allows to solve some of these problems analytically (e.g. via the Mie solution of Maxwell's equations~\cite{Mie_AnnPhys330_1908}) and thus enables efficient numerical schemes for simulating the strong-field physics at nanospheres~\cite{Zherebtsov_NatPhys7_2011, Suessmann_NatCommun6_2015, Seiffert_PhD_2018}. For all these reasons it is fair to say that nanospheres can be seen as the hydrogen atom (or in biological terms the drosophila melanogaster) for strong-field nanophysics~\cite{Ciappina_RPP80_2017}.

The objective of this review is to summarize the central aspects and developments connected with strong field physics with nanospheres from the last decade. Thereto, the central experimental methods as well as the theoretical approaches will be discussed. This way, we aim to provide the relevant stepping stones that students and researchers need to enter the field. Since a full quantum mechanical description is absolutely out of reach for the considered scenarios, the most promising approach for systematic studies are models based on classical trajectories. The required key elements and strategies will be discussed. They include the description of the near-fields (via the Mie solution of Maxwells equations and a multipole-expansion for additional contributions induced by free charges) and the electron dynamics, where the key ingredients are the accurate treatment of ionization (tunneling, photoionization and impact ionization), classical trajectory propagation, and electron-atom collisions (transport). In the following we will present a selection of central results, including studies of the combined impacts of enhancement and charge interaction, opportunities of waveform control offered by field propagation effects, and characterization of electron transport. For convenience, the discussion is divided into four sections. We will start the discussion in \Section{sec:small_spheres} by reviewing early surprises where rescattering from small dielectric nanospheres was first observed~\cite{Zherebtsov_NatPhys7_2011} and continue with inspecting the decisive impacts of charge interaction on the electron emission~\cite{Seiffert_JMO64_2017, Rupp_JMO64_2017}. \Section{sec:large_spheres} focuses on the effects induced by field propagation within larger dielectric nanospheres and resulting implications and applications~\cite{Suessmann_NatCommun6_2015, Powell_OE27_2019, Liu_NJP21_2019, Seiffert_APB122_2016, Rupp_NatCommun10_2019}. In \Section{sec:metal_spheres} we will review the ultrafast laser-induced metallization of initially dielectric spheres. Finally, in \Section{sec:streaking} we will discuss attosecond streaking~\cite{Constant_PRA56_1997, Drescher_Science291_2001, Hentschel_Nature414_2001, Itatani_PRL88_2002, Kienberger_Nature427_2004} on dielectric spheres, where irradiation with synchronized attosecond extreme ultraviolet (XUV) and femtosecond near infrared (NIR) enables to characterize the electron transport within in the material~\cite{Seiffert_NatPhys13_2017, Liu_JO20_2018}.


\section{Experimental Methods}
\subsection{Target delivery via aerosol injection}
\begin{figure}[h!]
\centering
\includegraphics[width=0.75\textwidth]{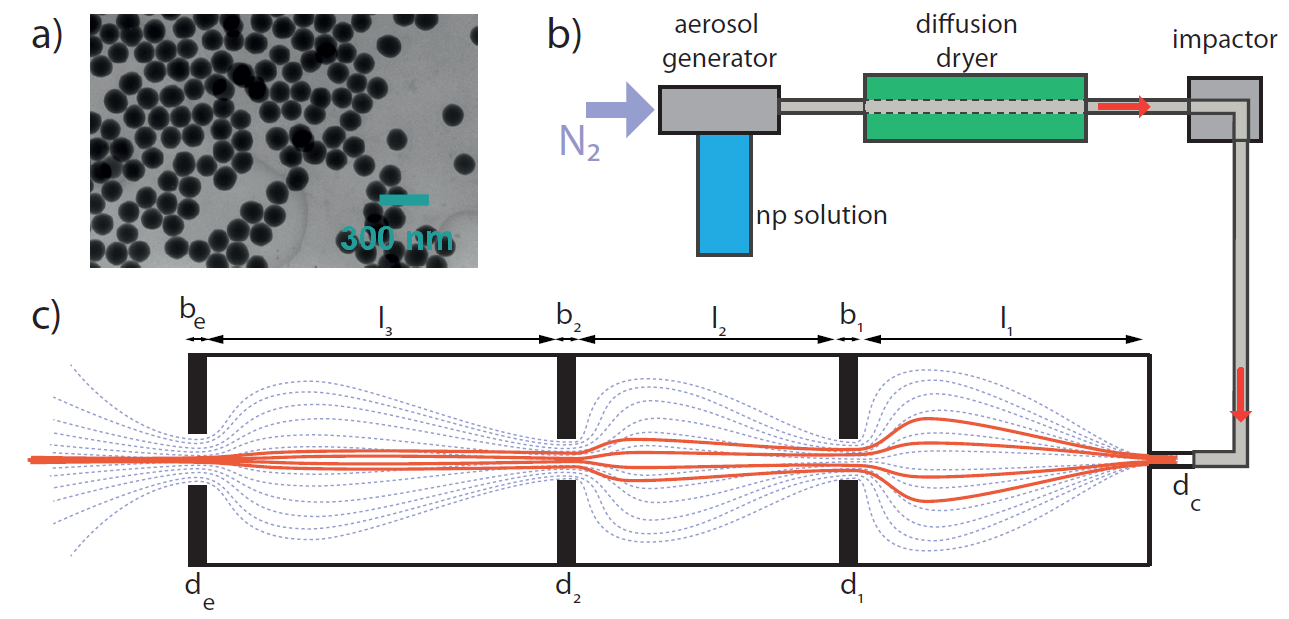}
\caption{\panel{a} TEM image of \silica nanospheres with a diameter of \SI{147}{nm}. \panel{b} Flow diagram for the generation of the nanoparticle aerosol. \panel{c} Working principle of an aerodynamic nanoparticle lens. The trajectories of gas molecules are schematically indicated by the blue dashed lines. The nanoparticle trajectories are shown in red. From~\cite{Suessmann_PhD_2013}.}
\label{fig:experiment_aerodynamic}
\end{figure}

To inspect isolated nanoparicles in the gas phase, nanoparticles can be evaporated from suspension and delivered into the interaction region by aerodynamic lensing. Nanoparticle suspensions are commercially available or can be prepared by corresponding experimental groups. Various approaches allow flexible realizations of different target types as for example dielectric, semiconductor or metal nanoparticles, core-shell particles, hollow particles or single-crystal particles with corresponding shapes. Single nanoparticles are typically preparated by chemical synthesis. For example, \silica (also referred to as silica or fused silica) nanospheres can by grown via the Stöber procedure~\cite{Stoeber_JCIS26_1968}, resulting in spheres with high surface quality with negligible roughness and size deviations that can reach below \SI{5}{\%}, see TEM image in \Fig{fig:experiment_aerodynamic}{a}.

From the nanoparticle suspension a high density aerosol (up to typically about $10^6$ particles per \si{cm^3}) can be created with a commercial aerosol generator. As the aerosol passes through a diffusion drier the solvent evaporates leaving nanoparticles suspended in the carrier gas. The stream of nanoparticles goes through an impactor where it has to pass a \ang{90} turn. This serves as a filter to remove relatively heavy nanoparticle clusters or residual contaminants. Next, the flow of nanoparticles is sent into an aerodynamic lens system (\Fig{fig:experiment_aerodynamic}{c}) in order to compress it into a narrow nanoparticle beam. The lens system consists of a set of apertures and has cylindrical symmetry. It was shown by Liu\etal~\cite{Liu_AST22_1995} that for certain particle sizes and gas flow parameters the nanoparticles follow the convergent gas motion before entering the aperture while staying confined close to the lens axis after passing the aperture. By using a set of apertures with different diameters, compression over a broad range of nanoparticle sizes can be achieved. In the works presented in this review the design proposed by Bresch~\cite{Bresch_PhD_2007} was used. With this approach isolated nanoparticle beams with diameters below one \si{mm} (full-width at half-maximum, FWHM) can be achieved for nanoparticle sizes ranging from $50$ to \SI{500}{nm} diameter.

\subsection{Single shot velocity map imaging}
\begin{figure}[h!]
  \centering
  \includegraphics[width=0.5\textwidth]{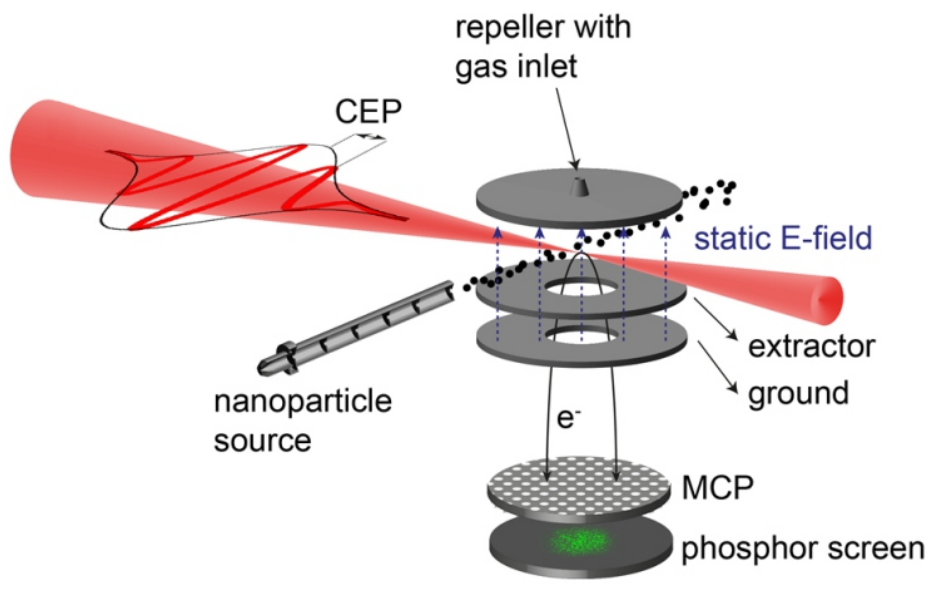}
  \caption{Schematic representation of VMI of strong-field induced ionization processes in isolated nanoparticles. Courtesy of Philipp Rupp.}
  \label{fig:experiment_VMI}
\end{figure}

Angle and energy-resolved electron momentum distributions can be measured via VMI. A typical VMI detector based on the Eppink-Parker design~\cite{Eppink_RSI68_1997} consists of an electrostatic lens system (repeller, extractor, and ground plates) and a Micro Channel Plate (MCP)/phosphor screen assembly (cf. \Fig{fig:experiment_VMI}). The interaction region is located inbetween the repeller and extractor electrodes and the momentum distribution of the emitted electrons is projected onto the MCP. The images on the phosphor screen are recorded for each laser shot by a high speed digital complementary metal-oxide-semiconductor (CMOS) camera (not shown). In order to enable storage of single shot images on a computer, flat field correction is applied to each frame and only pixels with sufficient brightness are transferred~\cite{Suessmann_RSI82_2011}. In parallel to the VMI measurements, information on the carrier-envelope phase (CEP) of each laser pulse may be recorded with a phase meter~\cite{Rathje_JPB45_2012} (not shown).

\begin{figure}[h!]
  \centering
  \includegraphics[width=0.9\textwidth]{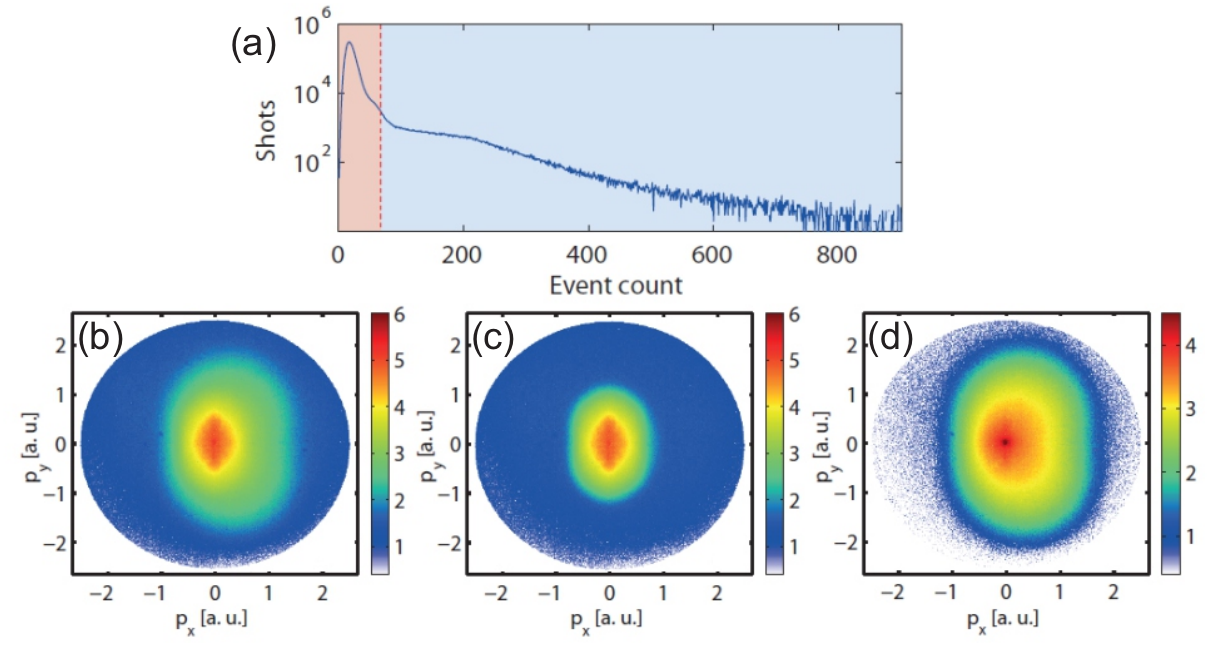}
  \caption{Illustration of efficient background suppression by histogram selection for photoemission from \SI{313}{nm} \silica nanopspheres under few-cycle laser pulses at \SI{2.7e13}{W/cm^2}. \panel{a} Histogram of the number of events per frame. \panel{b} Momentum map corresponding to the full histogram selection. \panel{c,d} Momentum maps corresponding to selection of frames with the number of events ranging from 0 to 60 (c) and from 60 to 1000 (d), as indicated by the red and blue areas in (a). Adapted from \cite{Suessmann_PhD_2013}.}
  \label{fig:experiment_singleshot}
\end{figure}

One of the major advantages of single-shot detection is the possibility of efficient data post-processing. As the nanoparticle density in the interaction volume is limited, only a small fraction of single shot images contain signal from nanoparticle targets. This is illustrated in \Fig{fig:experiment_singleshot}{a} where a histogram of the number of events per frame obtained from measurements of \SI{313}{nm} \silica nanospheres is presented. The VMI image corresponding to an average over all events is shown in \Fig{fig:experiment_singleshot}{b}. The majority of the frames contains only a low number of events with a peak centered at around 20 event per shot and can be attributed to the ATI of the background gas. This is confirmed by the analysis where the frames with the number of event per laser shot below 60 were selected. Both the momentum projection and the asymmetry map show typical features attributed to the ATI of gas (\Fig{fig:experiment_singleshot}{c}). Frame selection with the number of event per laser shot larger than 60 results in obvious reduction of the background gas contribution (\Fig{fig:experiment_singleshot}{d}) and substantially improves the analysis especially in the low energy region. 

\section{Theoretical Tools}
In general, the ultrafast electron dynamics of nanospheres under intense NIR and XUV laser pulses is fully described by highly correlated many-particle wave functions and their evolution is determined by the time-dependent Schrödinger equation (TDSE). However, for typical experimental scenarios the numerical solution of the TDSE for the full problem is neither possible (even on modern supercomputers) nor insightful as it is unclear if the relevant physics can be extracted from the high-dimensional wave functions. Hence, as in many other cases, instructive theoretical models require reasonable approximations. Based on the idea of Corkum’s three-step model, a common approximation in strong-field physics is to treat the dynamics of liberated electrons classically and include quantum effects (e.g. tunneling or collisions) via appropriate rates that can be extracted from simplified quantum models or experiments. Following this approach, the strong-field driven electron dynamics at a nanosphere may be modeled semi-classically, where an classical electron trajectory reflects the motion of the respective quantum wave packet’s center of mass. For the relevant scenarios with many active electrons, the neglect of quantum interferences is justified, as interaction will lead to rapid dephasing of individual quantum trajectories. This assumption is supported by the lack of features from individual photon orders in the HATI spectra measured for dielectric nanospheres~\cite{Zherebtsov_NatPhys7_2011, Zherebtsov_NJP14_2012}. The comprehensive semi-classical description requires addressing three key challenges: First, the evaluation of the local near-fields including field propagation and charge interaction. Second, a proper treatment of near-field driven ionization as well as electron impact ionization. Third, the description of charge transport within the dielectric material. The individual approaches that make it possible to meet these three requirements in a consisted way are outlined in more detail in the following.

In the context of strong-field laser-nanoparticle interactions, simulation models with various levels of approximations have been developed and utilized within the last decade. The probably most simple and straightforward approach is to extend the well known SMM for the description of electron emission from nanospheres. Thereto, electron trajectories are launched at rest at the nanosphere surface, are driven by the near-field outside of the sphere and can be elastically reflected when returning to the surface. Photoelectron energy spectra can by extracted by weighting each trajectory with a suitable ionization rate evaluated at the instant of the generation. This description is already suitable to inspect several key features of the electron emission as for example the typical signatures associated with direct emission and backscattering in the enhanced and inhomogeneous near-fields and even the emission directionality~\cite{Suessmann_NatCommun6_2015}. However, it fails to describe important aspects that arise from charge interaction due to the generation of free charges upon ionization as well as electron transport within the material. These effects can be accounted for in higher level descriptions such as the semi-classical Mie Mean-field Monte-Carlo (\MCCC) model~\cite{Suessmann_NatCommun6_2015, Seiffert_APB122_2016, Seiffert_PhD_2018} or via microscopic particle-in-cell (MicPic) models~\cite{Varin_PRL108_2012, Peltz_PRL113_2014}. In the following, the details of \MCCC are discussed, as this method has been utilized in most of the scenarios presented in this review~\cite{Suessmann_NatCommun6_2015, Seiffert_APB122_2016, Seiffert_JMO64_2017, Rupp_JMO64_2017, Seiffert_NatPhys13_2017, Liu_JO20_2018, Rupp_NatCommun10_2019, Liu_NJP21_2019, Liu_ACS7_2020} and was recently also extended for the description of strong-field ionization from metal nanotips~\cite{Schoetz_NP_2021}.

\subsection{Near-field description for nanospheres}
In general, the spatiotemporal electromagnetic field (i.e. $\vecfield{E}(\vec{r}, t)$ and $\vecfield{B}(\vec{r},t)$) at a nanosphere is fully described by the microscopic Maxwell equations

\noindent\hspace{5mm}
$\nabla \cdot \vecfield{E} = \frac{\varrho}{\varepsilon_0}$ \hfill
$\nabla \cdot \vecfield{B} = 0$ \hfill
$\nabla \times \vecfield{E} = -\dot{\vecfield{B}}$\hfill
$\nabla \times \vecfield{B} = \mu_0 \left[\vec{j} + \varepsilon_0\dot{\vecfield{E}}\right]$
\hspace{5mm}

\noindent with vacuum permittivity $\varepsilon_0$ and vacuum permeability $\mu_0$ and the charge and current densities $\varrho(\vec{r},t)$ and $\vec{j}(\vec{r},t)$. Key to an efficient description of the strong-field induced near-field is the convenient separation of the densities $\varrho = \varrho_\text{b}^\text{ext} + \varrho_\text{b}^\text{f} + \varrho_\text{f}$ and $\vec{j} = \vec{j}_\text{b}^\text{ext} + \vec{j}_\text{b}^\text{f} + \vec{j}_\text{f}$ into contributions from bound (subscript 'b') and free (subscript 'f') charges and further separating the bound densities into contributions which reflect the bound state polarization response to the external field (superscript 'ext') and the response to the fields from free charges (superscript 'f'). For reasons that become clear in a moment, also the fields may be separated into two contributions $\vecfield{E} = \vecfield{E}^\text{lin} + \vecfield{E}^\text{CI}$ and $\vecfield{B} = \vecfield{B}^\text{lin} + \vecfield{B}^\text{CI}$. Exploiting the linearity of the  divergence and curl operators allows to separate the Maxwell equations into two sets. The first set

\vspace{-12pt}
\begin{minipage}{0.4\textwidth}
\begin{align}
\nabla \cdot \vecfield{E}^\text{lin} &= \frac{\varrho_\text{b}^\text{ext}}{\varepsilon_0} \nonumber\\
\nabla \cdot \vecfield{B}^\text{lin} &= 0 \nonumber
\end{align}
\end{minipage}\hfill%
\begin{minipage}{0.55\textwidth}
\vspace{7.5pt}
\begin{align}
\begin{split}
\nabla \times \vecfield{E}^\text{lin} &= -\dot{\vecfield{B}}^\text{las}\\[3pt]
\nabla \times \vecfield{B}^\text{lin} &= \mu_0 \left[\vec{j_\text{b}^\text{ext}} + \varepsilon_0\dot{\vecfield{E}}^\text{lin}\right] \label{eq:Maxwell_lin}
\end{split}
\end{align}
\end{minipage}
\vspace{4pt}

\noindent describes the evolution of an incident laser field and the corresponding strictly linear polarization response of the sphere. Hence the respective field is from here on termed the \emph{linear near-field}. The remaining set

\vspace{-12pt}
\begin{minipage}{0.4\textwidth}
\begin{align*}
\nabla \cdot \vecfield{E}^\text{CI} &= \frac{\varrho_\text{b}^\text{f} + \varrho_\text{f}}{\varepsilon_0}\\
\nabla \cdot \vecfield{B}^\text{CI} &= 0
\end{align*}
\end{minipage}\hfill%
\begin{minipage}{0.55\textwidth}
\vspace{7.5pt}
\begin{align}
\begin{split}
\nabla \times \vecfield{E}^\text{CI} &= -\dot{\vecfield{B}}^\text{CI}\\[3pt]
\nabla \times \vecfield{B}^\text{CI} &= \mu_0 \left[\vec{j}_\text{b}^\text{f} + \vec{j}_\text{f}+ \varepsilon_0\dot{\vecfield{E}}^\text{CI}\right] \label{eq:Maxwell_CI}
\end{split}
\end{align}
\end{minipage}
\vspace{4pt}

\noindent reflects everything that goes beyond the strictly linear response of the initial system to the external field. In our case this set covers the Coulomb fields resulting from free charges emerging in the strong-field excitation process and the associated (linear) polarization of the sphere induced by these free charges. These fields modify the full near-field only if free charges are present following ionization of the initially neutral spheres and hence provide an additional contribution due to \emph{nonlinear charge interaction} (CI). The above splitting of the fields into the strictly linear and the additional contribution from charge interaction will be key to the efficient approximate numerical solution of the field equations.

\subsubsection{Linear response near-field contribution}
Considering a sphere medium with homogeneous, linear, and isotropic local polarization response  (with relative permittivity \per~and relative permeability $\mu_\text{r}$) in an external plane wave ($\vecfield{E}(\vec{r},t) = \vecfield{E}_0 e^{i(\vec{k}\vec{r}\pm\omega t)}$ and $\vecfield{H}(\vec{r},t) = \vecfield{H}_0 e^{i(\vec{k}\vec{r}\pm\omega t)}$) the first set of Maxwell equations (\Eq{eq:Maxwell_lin}) can be converted to the wave equations $\left[\nabla^2 + k^2\right]\vecfield{E} = 0$ and $\left[\nabla^2 + k^2\right]\vecfield{H} = 0$. Thus, the remaining problem is to find solutions of these equations which fulfill the boundary conditions imposed by the sphere. Thereto, it is convenient to express the linear response near-field at a nanosphere with radius $R$ in the form
\begin{equation*}
\vecfield{E}^\text{lin}(\vec{r},t) = \left\{\begin{array}{c c c} \vecfield{E}_\text{i}(\vec{r},t) + \vecfield{E}_\text{r}(\vec{r},t) &\text{for} &r > R  \\ 
\vecfield{E}_\text{t}(\vec{r},t) &\text{for} &r\leq R \end{array}\right.
\end{equation*}
with the incident field $\vecfield{E}_\text{i}$, the reflected field $\vecfield{E}_\text{r}$ and the transmitted field $\vecfield{E}_\text{t}$ and an analog separation also for the magnetic field. Two approaches for determining the fields are discussed below.

\begin{figure}[h!]
\centering
\includegraphics[width=0.9\textwidth]{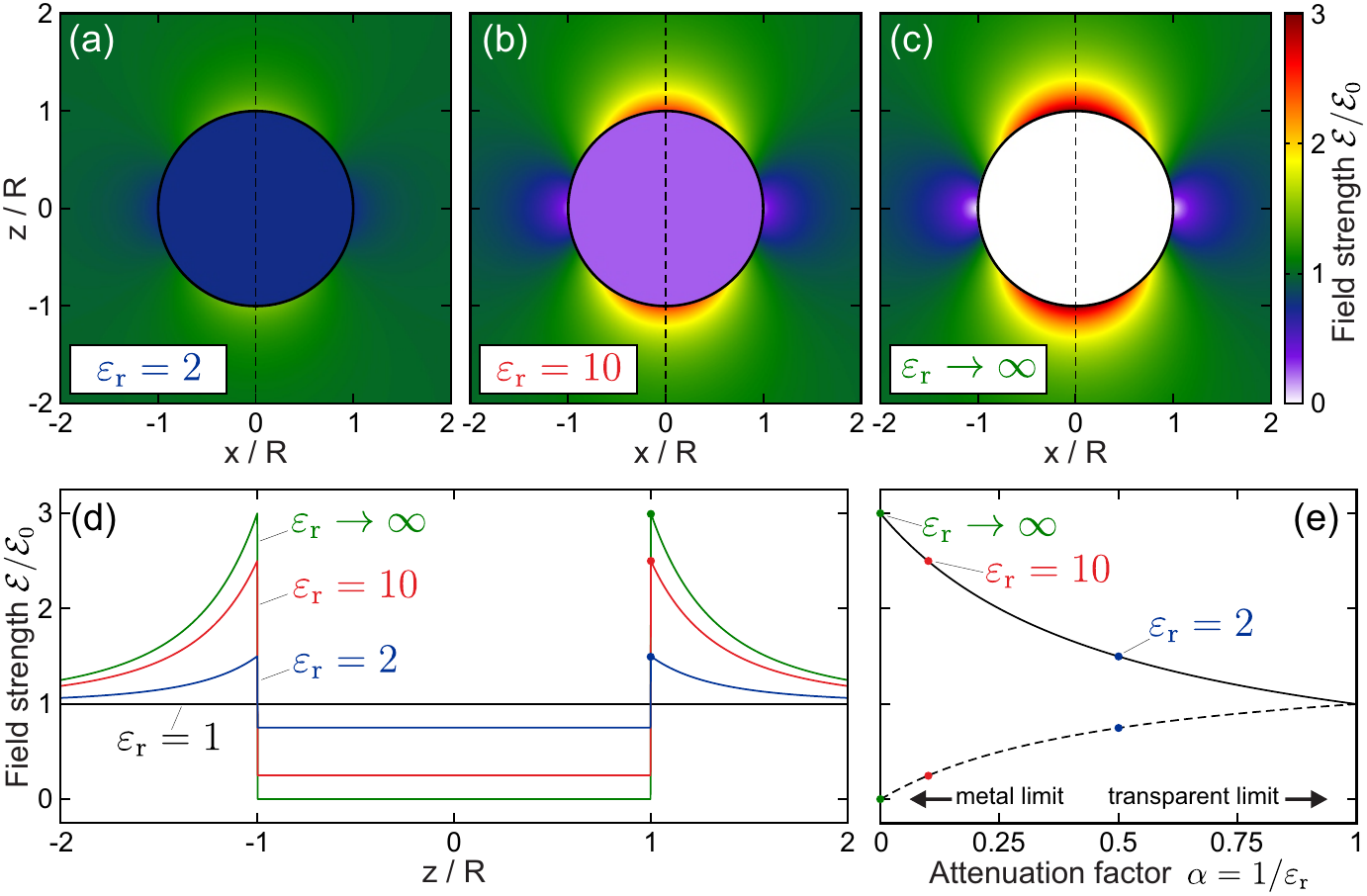}
\caption{Linear response near-fields of nanospheres in quasistatic dipole approximation. \panel{a-c} Absolute value of the near-field in the $y=0$ plane with respect to the field strength of the driving static field $\field{E}_0 \vec{e}_z$ for three different relative permittivities of the sphere with radius $R$ (as indicated). \panel{d} Field strength along the z axis, i.e. along the dashed vertical lines in (a-c). \panel{e} Maximum enhancement (solid curve) and inside field (dashed curve) in dependence of the materials attenuation factor $\alpha = 1/\varepsilon_\text{r}$. Colored dots mark the respective permittivity values of the profiles shown in (d).}
\label{fig:theory_dipolefields}
\end{figure}

\paragraph{Rayleigh limit \& quasi-static dipole approximation for small spheres}
If the dimension of the sphere is small with respect to the wavelength of the incident field, the linear response near-field may be described reasonable well within Rayleights quasistatic dipole approximation~\cite{Strutt_PhilMag41_1871}. In that case the incident field can be considered as a static electric field $\vecfield{E}_\text{i} = \field{E}_0 \vec{e}_z$, resulting in the transmitted and reflected fields 
\begin{equation*}
    \vecfield{E}_\text{t} = \field{E}_0 \left( \frac{3}{\per+2}\right) \vec{e}_z
    \hspace{1.5cm}\text{and}\hspace{1.5cm}
    \vecfield{E}_\text{r} = \field{E}_0 \left( \frac{\per-1}{\per+2}\right)  \frac{R^3}{r^5} \begin{pmatrix}3xz\\3yz\\3z^2-r^2\end{pmatrix}
\end{equation*}
as derived in standard classical electrodynamics textbooks, see eg.~\cite{Jackson_1999}. \Figure{fig:theory_dipolefields}{a-c} depicts the quasi-static linear response near-fields in the $y=0$ cut through a sphere with radius $R$ placed at the origin for three different relative permittivities. Specifically, the spatial profile of the field enhancement factor $\gamma(\vec{r}) = \field{E}(\vec{r}) / \field{E}_0$ follows from normalization of the near-fields magnitude to the amplitude of the external static field. It is obvious that higher magnitudes of the permittivity results in reduction of the internal field (which is constant within the whole sphere volume) and formation of increasingly strong enhancements located at the upper and lower poles of the nanosphere (i.e. along the polarization axis of the external field). Note that the local fields at the points of maximum enhancement (the hot spots), which are located directly at the poles, is polarized along the z-direction and is thus oriented completely radial with respect to the surface. The respective enhancement profiles along the polarization axis (cf. vertical dashed lines in \Fig{fig:theory_dipolefields}{a-c}) in \Fig{fig:theory_dipolefields}{d} illustrate the permittivity dependence, with asymptotically vanishing internal field and a maximal field enhancement value of three for the outside field. For convenience the evolution of the maximum enhancement $\gamma_0$ and the inside field in dependence of the materials field attenuation factor $\alpha = 1/\varepsilon_\text{r}$ that relates the ratio of internal to external fields at the poles is visualized in \Fig{fig:theory_dipolefields}{e}. 

\paragraph{Mie solution and field propagation effects for large spheres}
While the dipole approximation is valid for small spheres, it breaks down once the dimension of the sphere becomes comparable to the incident fields wavelength. In that case higher order multipole terms and the scattering of the incident plane wave has to be resolved by explicitly solving Maxwell’s equations. Here the spherical symmetry plays a major role as the solution can then be expressed analytically as introduced in the famous work by Gustav Mie~\cite{Mie_AnnPhys330_1908} and thoroughly discussed in many electromagnetic theory books (see e.g. Stratton~\cite{Stratton} or Bohren \& Huffman~\cite{Bohren}). The Mie solution will thus not be derived or discussed in detail here, but only the main steps of the final field evaluation are briefly outlined. We focus on the particularly relevant scenarios for this review where spheres are considered nonmagnetic ($\mu_\text{r} = 1$) and placed in free space (vacuum). The key idea is to expand the incident, reflected and transmitted fields into vector spherical harmonics. For an incident plane wave with electric field $\vecfield{E}_\text{i} = e^{ikz}\vec{e}_x$ and magnetic field $\vecfield{H}_\text{i} = \frac{k}{\mu\omega}e^{ikz}\vec{e}_y$ the respective expansions can be expressed as

\begin{figure}[h!]
\centering
\includegraphics[width=1.0\textwidth]{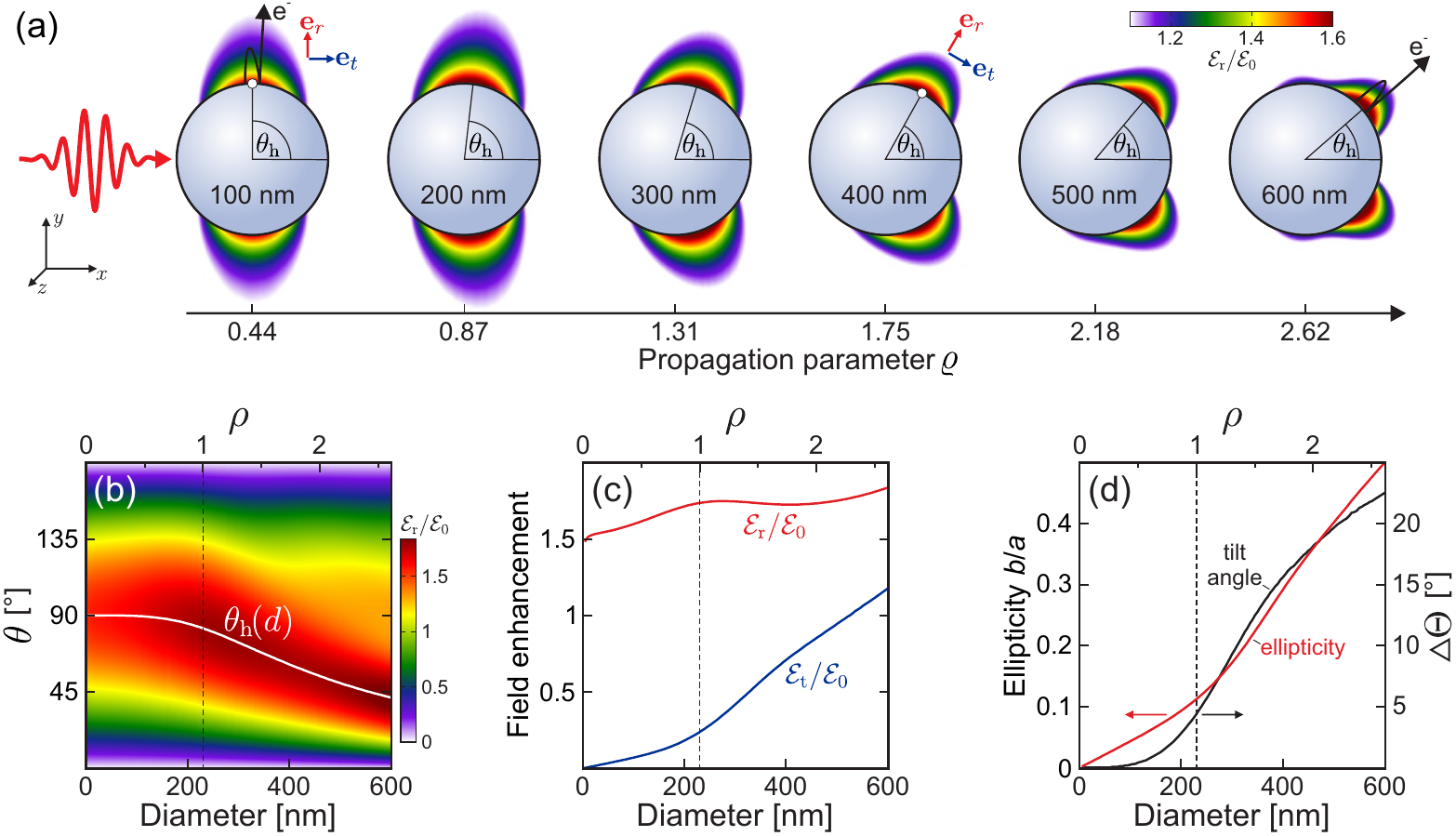}
\caption{Linear near-field at silica nanospheres. \panel{a} Enhancement of the radial linear near-field at silica spheres (false color plots) with respect to an incident \SI{4}{fs} few-cycle pulse at \SI{720}{nm} central wavelength (red curve) in dependence of sphere diameter (respective field propagation parameters $\varrho$ as indicated). The characteristic angles $\theta_\text{h}$ indicate the hot spots (defined via the maximal enhancement at the surface). Red and blue arrows indicate radial and tangential unit vectors, respectively. Adapted from\cite{Seiffert_PhD_2018}. \panel{b} Map of the radial surface fields relative enhancement $\field{E}_\text{r}(d,\theta)/\field{E}_0$ with respect to the peak field strength $\field{E}_0$ of the incident few-cycle laser pulse, in dependence of the angle $\theta$ sampled at the surface of the upper sphere half in the $z = 0$ plane. The white curve indicates the characteristic (hot-spot) angle $\theta_\text{h}(d)$ of maximum enhancement. \panel{c} Relative enhancements $\field{E}_\text{r/t}(d,\theta)/\field{E}_0$ of the radial (red) and tangential (blue) surface fields, sampled at the characteristic angle. \panel{d} Ellipticity parameter (red) and tilt angle of the local polarization ellipses. Adapted from~\cite{Suessmann_NatCommun6_2015}.}
\label{fig:Seiffert_Dissertation_2018_Fig5.1}
\end{figure}

{\vspace{-6pt}
\begin{minipage}{0.35\textwidth}
\begin{align*}
\vecfield{E}_\text{i} &= \sum\limits_{l=1}^{\infty} i^l  \frac{2l+1}{l(l+1)} (\vecfield{M}_{\text{o}l} - i\vecfield{N}_{\text{e}l})\\
\vecfield{E}_\text{r} &= \sum\limits_{l=1}^{\infty} i^l  \frac{2l+1}{l(l+1)} (ia_l\vecfield{N}^{(3)}_{\text{e}l} - b_l\vecfield{M}^{(3)}_{\text{o}l})\\
\vecfield{E}_\text{t} &= \sum\limits_{l=1}^{\infty} i^l  \frac{2l+1}{l(l+1)} (c_l\vecfield{M}_{\text{o}l} - id_l\vecfield{N}_{\text{e}l})
\end{align*}
\end{minipage}
\begin{minipage}{0.55\textwidth}
\begin{align*}
\vecfield{H}_\text{i} &= \frac{-k}{\mu\omega} \sum\limits_{l=1}^{\infty} i^l  \frac{2l+1}{l(l+1)} (\vecfield{M}_{\text{e}l} + i\vecfield{N}_{\text{o}l})\\
\vecfield{H}_\text{r} &= \frac{k}{\mu\omega} \sum\limits_{l=1}^{\infty} i^l  \frac{2l+1}{l(l+1)} (ib_l\vecfield{N}^{(3)}_{\text{o}l} + a_l\vecfield{M}^{(3)}_{\text{e}l})\\
\vecfield{H}_\text{t} &= \frac{-\sqrt{\varepsilon_\text{r}}k}{\mu\omega} \sum\limits_{l=1}^{\infty} i^l  \frac{2l+1}{l(l+1)} (d_l\vecfield{M}_{\text{e}l} + ic_l\vecfield{N}_{\text{o}l})
\end{align*}
\end{minipage}
\vspace{6pt}}

\noindent with expansion coefficients $a_l$, $b_l$, $c_l$ and $d_l$ and the vector spherical harmonics
\begin{align*}
\vecfield{M}_{^\text{o}_\text{e}l} &= \pm\frac{1}{\sin\theta} J_l P_l \hspace{-3.5pt} \def\arraystretch{0.1}\begin{array}{c}
\cos\\
\sin
\end{array} \hspace{-3.5pt}\phi \, \vec{e}_r - J_l \frac{\partial P_l}{\partial\theta} \hspace{-3.5pt} \def\arraystretch{0.5}\begin{array}{c}
\sin\\
\cos
\end{array} \hspace{-3.5pt}\phi \, \vec{e}_\theta\\
\vecfield{N}_{^\text{o}_\text{e}l} &= \frac{l(l+1)}{kr}J_lP_l \hspace{-3.5pt} \def\arraystretch{0.5}\begin{array}{c}
\sin\\
\cos
\end{array} \hspace{-3.5pt}\phi \, \vec{e}_r + \frac{1}{kr}\left[krJ_l\right]' \frac{\partial P_l}{\partial\theta} \hspace{-3.5pt} \def\arraystretch{0.5}\begin{array}{c}
\sin\\
\cos
\end{array} \hspace{-3.5pt}\phi \, \vec{e}_\theta \pm \frac{1}{kr \sin\theta} \left[krJ_l\right]'P_l \hspace{-3.5pt} \def\arraystretch{0.1}\begin{array}{c}
\cos\\
\sin
\end{array} \hspace{-3.5pt}\phi \, \vec{e}_\phi.
\end{align*}
Here, $P_l = P_l(\cos \theta)$ are the associated Legendre functions of first kind with degree l and order $m=1$ and $J_l = J_l(kr)$ are spherical Bessel functions. For the incident and transmitted fields, Bessel function of first type $J_l$ are required to assure finite solutions at $r \rightarrow 0$. The reflected fields must vanish at infinity ($r \rightarrow \infty$) which is a property of the Bessel functions of third type $J_l^{(3)}$ (often also referred to as Hankel functions). Their use is denoted by the superscript $(3)$ at the respective vector spherical harmonics. Further, note that for the transmitted field the incident field wavenumber $k$ has to be replaced by the wavenumber in the medium $\sqrt{\varepsilon_\text{r}}k$. The expansion coefficients are calculated from the interface conditions $\left[\vecfield{E}_\text{i} + \vecfield{E}_\text{r}\right]\times\vec{e}_r = \vecfield{E}_\text{t}\times\vec{e}_r$ and $\left[\vecfield{H}_\text{i} + \vecfield{H}_\text{r}\right]\times\vec{e}_r = \vecfield{H}_\text{t}\times\vec{e}_r$, which reflect the continuity of the tangential fields at the surface of the sphere, and follow as
\begin{align*}
a_l &= \frac{\varepsilon_\text{r} J_l(\sqrt{\varepsilon_\text{r}}\varrho)\left[\varrho J_l(\varrho) \right]' - J_l(\varrho)\left[\sqrt{\varepsilon_\text{r}}\varrho J_l(\sqrt{\varepsilon_\text{r}}\varrho) \right]'}
{\varepsilon_\text{r} J_l(\sqrt{\varepsilon_\text{r}}\varrho)\left[\varrho H_l(\varrho) \right]' - H_l(\varrho)\left[\sqrt{\varepsilon_\text{r}}\varrho J_l(\sqrt{\varepsilon_\text{r}}\varrho) \right]'}\\
b_l &= \frac{J_l(\sqrt{\varepsilon_\text{r}}\varrho)\left[\varrho J_l(\varrho) \right]' - J_l(\varrho)\left[\sqrt{\varepsilon_\text{r}}\varrho J_l(\sqrt{\varepsilon_\text{r}}\varrho) \right]'}
{J_l(\sqrt{\varepsilon_\text{r}}\varrho)\left[\varrho H_l(\varrho) \right]' - H_l(\varrho)\left[\sqrt{\varepsilon_\text{r}}\varrho J_l(\sqrt{\varepsilon_\text{r}}\varrho) \right]'}\\
c_l &= \frac{J_l(\varrho)\left[\varrho H_l(\varrho) \right]' - H_l(\varrho)\left[\varrho J_l(\varrho) \right]'}
{J_l(\sqrt{\varepsilon_\text{r}}\varrho)\left[\varrho H_l(\varrho) \right]' - H_l(\varrho)\left[\sqrt{\varepsilon_\text{r}}\varrho J_l(\sqrt{\varepsilon_\text{r}}\varrho) \right]'}\\
d_l &= \frac{\sqrt{\varepsilon_\text{r}} J_l(\varrho)\left[\varrho H_l(\varrho) \right]' - \sqrt{\varepsilon_\text{r}} H_l(\varrho)\left[\varrho J_l(\varrho) \right]'}
{\varepsilon_\text{r} J_l(\sqrt{\varepsilon_\text{r}}\varrho)\left[\varrho H_l(\varrho) \right]' - H_l(\varrho)\left[\sqrt{\varepsilon_\text{r}}\varrho J_l(\sqrt{\varepsilon_\text{r}}\varrho) \right]'}.
\end{align*}
Here, $[\ldots]'$ denotes the derivative of the expression within the brackets with respect to the argument of the respective Bessel function. The coefficients show that the linear near-field at a dielectric sphere (for a given frequency) only depends on its relative permittivity $\varepsilon_\text{r}$ and the dimensionless propagation parameter
\begin{equation}
\varrho = kR = \frac{2\pi R}{\lambda}
\label{eq:propagation_parameter}
\end{equation}
that sets the sphere radius $R$ in proportion to the incident fields wavelength $\lambda$. As an example, \Fig{fig:Seiffert_Dissertation_2018_Fig5.1}{a} shows the maximal relative enhancement of the radial near-field at silica nanoparticles ($\per=2.12$) with increasing diameters in the propagation-polarization plane of a driving few-cycle laser pulse. For small nanospheres ($\varrho << 1$) the linear near-field can be described well within Rayleigh’s quasi-static dipole approximation~\cite{Strutt_PhilMag41_1871}, i.e. neglecting the influence of field propagation within the nanospheres. It should be noted that terminating the series expansion after the first order ($l=1$) is formally equivalent to the dipole approximation and thus reproduces the fields obtained within the quasistatic description. However, with increasing size this approximation fails and the spatiotemporal near-fields become strongly deformed due to the  excitation of higher order multipole modes for $\varrho \gtrsim 1$~\cite{Mie_AnnPhys330_1908}. The most notable effect is the gradual shift of the hot spot regions towards the rear side of the spheres, as quantified by the characteristic (hot-spot) angle $\theta_\text{h}$, see \Fig{fig:Seiffert_Dissertation_2018_Fig5.1}{b}. The diameter-dependent increase of the radial and tangential field-enhancements sampled at the field hot spots are shown in \Fig{fig:Seiffert_Dissertation_2018_Fig5.1}{c}. While the impact of tangential components is small for $\varrho<1$ and the near-field is mostly linearly polarized in radial direction, the increasing tangential fields at larger spheres result in elliptically polarized near-fields. The latter can be quantified by the ellipticity parameter (the ratio of the small to long half axis of the polarization ellipse), and the tilt angle of the ellipse (the angle of the long half axis with respect to the radial unit vector), as shown in \Fig{fig:Seiffert_Dissertation_2018_Fig5.1}{d}.

\paragraph{Spectral decomposition for finite pulses and material dispersion}
\label{sec:spectral_decomposition}
The Mie solution reflects the spatial mode structure of the linear response near-field for a single frequency. To describe the linear near-field of a sphere exposed to laser pulses with finite duration and for including material dispersion a set of these spatial modes can be combined via spectral field decomposition. Therefore, the incident pulse is described via the spectral modes of free space, i.e. by plane waves. The complex electric field in the Fourier domain reads
\begin{align*}
\vecfield{E}(\vec{r},\omega) &= \vecfield{E}_0 f(\omega) e^{i\vec{k}\vec{r}} e^{-i\varphi(\omega)},
\end{align*}
where $\vecfield{E}_0$ reflects the field's peak amplitude and with spectral amplitude profile $f(\omega)$, spectral phase $\varphi(\omega)$ and the spatial plane wave mode $e^{i\vec{k}\vec{r}}$. Commonly, pulses are considered to have a Gaussian amplitude spectrum $f(\omega) = \frac{1}{\sigma_\omega} e^{-\frac{1}{2}\left(\frac{\omega-\omega_0}{\sigma_\omega}\right)^2}$ with spectral width $\sigma_\omega = \frac{2\sqrt{\text{ln}2}}{\tau}$ defined by the FWHM $\tau$ of the temporal intensity envelope. The spectral phase may be approximated as $\varphi(\omega) = \cep + \frac{\zeta}{2}(\omega-\omega_0)^2$, i.e. by including the carrier envelope phase (CEP) $\cep$ and/or a chirp quantified by the chirp parameter $\zeta$  (note that a linear temporal chirp corresponds to a parabolic spectral phase). Similar to the incident field, also the linear near-field of a sphere can be expressed via spectral decomposition when replacing the plane wave spatial modes by the previously discussed Mie-solutions (or the results obtained within the dipole approximation). Dispersion may be incorporated by evaluating the single frequency Mie solutions for the respective optical properties (i.e. $\per(\omega)$). The spatio-temporal evolution of the electric field is obtained via the Fourier transform
\begin{align*}
\vecfield{E}(\vec{r},t) &= \frac{1}{2}\frac{1}{\sqrt{2\pi}} \int\limits_{-\infty}^{\infty} \vecfield{E}(\vec{r},\omega) e^{-i\omega t}\,\text{d}\omega + c.c.
\end{align*}
and allows one to calculate the local instantaneous electric field. This instantaneous near-field is calculated for example to evaluate instantaneous tunnel ionization rates for long-wavelength fields or for the integration of the classical trajectories. For evaluating vertical ionization, e.g. in case of  XUV-induced photoionization, also the instantaneous spectral profile is needed. Therefore, the combined spectral and temporal evolution of the local intensity is characterized by the Wigner distribution~\cite{Wigner_PR40_1932, Cohen_IEEE77_1989, Hong_APB74_2002}
\begin{equation*}
W(\vec{r}, t, \omega) = \frac{\varepsilon_0 c_0}{2} \frac{1}{2\pi} \int\limits_{-\infty}^{\infty} \vecfield{E}(\vec{r},\omega+\frac{\omega'}{2}) \vecfield{E}^*(\vec{r},\omega-\frac{\omega'}{2}) e^{-i\omega' t}\,\text{d}\omega'.
\end{equation*}
The local intensity evolution $I(\vec{r}, t) = \int W(\vec{r}, t, \omega)\,\text{d}\omega$ and spectral intensity profile $I(\vec{r}, \omega)= \int W(\vec{r}, t, \omega)\,\text{d}t$ follow from frequency and time integration over the Wigner distribution. The Wigner distributions were used to evaluate spectral photoionization rates and to analyze the impact of the XUV chirp on attosecond streaking from nanospheres~\cite{Seiffert_NatPhys13_2017, Liu_JO20_2018}.

\subsubsection{Treatment of field contributions beyond linear response}
Besides the above discussed linear response near-field of a neutral dielectric sphere in an external laser field, free charges that are generated via ionization (i.e. liberated electrons and residual ions) can generate two additional contributions to the near-field. First, the Coulomb fields of the free charges themselves and second, the additional linear sphere polarization resulting from fields stemming from these free charges. As both additional contributions are not included in the strictly linear part discussed before the additional terms are termed nonlinear contributions due to the interactions resulting from generated free charges. In the scenarios considered here, this additional charge interaction terms are evaluated in mean-field approximation. 

In practice, the evaluation of both the Coulomb fields of the free charges and the associated additional sphere polarization is required to solve the second set of Maxwell equations (\Eq{eq:Maxwell_CI}). Under the assumption that field retardation effects are negligible for the  additional nonlinear field terms , this task can be treated in quasi-electrostatic approximation. In this case the problem reduces to finding solutions to the divergence equation for the electric displacement field that can be expressed as $\nabla \cdot \vecfield{D} = \varrho_\text{f}$. Considering a homogeneous dielectric sphere with radius $R$ and permittivity \per~surrounded by vacuum leads to two divergence equations $\nabla \cdot \vecfield{E}_\text{out} = \frac{\varrho_\text{f}}{\varepsilon_0}$ and $\nabla \cdot \vecfield{E}_\text{in} = \frac{\varrho_\text{f}}{\varepsilon_0 \varepsilon_\text{r}}$ for the fields outside and inside of the sphere, respectively. In electrostatic approximation, where the electric field is curl-free and can be fully characterized by the associated electrostatic potential $\Phi(\vec{r})$ via $\vecfield{E} = - \nabla\Phi$, the problem leads to two corresponding Poisson equations $\Delta\Phi_\text{out} = -\frac{\varrho_\text{f}}{\varepsilon_0}$ and $\Delta\Phi_\text{in} = -\frac{\varrho_\text{f}}{\varepsilon_0\varepsilon_\text{r}}$. The electrostatic potential is thus generated by the free charges and also includes the static polarization of the medium induced by the free charges via the relative permittivity. Further, it has to fulfill the interface conditions $\Phi_\text{in}(\vec{R}) = \Phi_\text{out}(\vec{R})$ and $\varepsilon_\text{r}\frac{\partial}{\partial r}\Phi_\text{in}(\vec{R}) = \frac{\partial}{\partial r}\Phi_\text{out}(\vec{R})$ at the surface, which follow directly from the field continuity conditions.

An efficient approximate solution of the Poisson equation is possible using a high-order multipole expansion, see Ref.~\cite{Seiffert_PhD_2018} for a detailed derivation. In brief, the solution of the Laplace equation $\Delta \Phi = 0$ is expressed via a series expansion $\Phi(\vec{r}) = \sum\limits_{l=0}^\infty \left[A_lr^l + B_lr^{-(l+1)}\right] P_l(\cos\theta)$ with Legendre polynomials $P_l(\cos\theta)$ with $\theta$ being the angle between $\vec{r}$ and the $z$-axis (w.l.o.g). The expansion coefficients $A_l$ and $B_l$ are determined by matching this solution to the electrostatic potential for a point charge following from Coulomb's law $\Phi(r) = \frac{1}{4\pi\varepsilon_0\varepsilon_\text{r}}\frac{q_i}{|r-r_i|}$. Summation over an ensemble of individual point charges yields the resulting electrostatic potentials in- and outside of the sphere
\begin{align*}
\Phi_\text{in}(\vec{r}) &= \sum_{l=0}^\infty \biggl[ \underbrace{\sum_{r_i \leq R \atop r_i \leq r} \frac{q_i}{4\pi\varepsilon_0\varepsilon_r} \frac{r_i^l}{r^{l+1}} + \sum_{r_i \leq R \atop r_i > r} \frac{q_i}{4\pi\varepsilon_0\varepsilon_r} \frac{r^l}{r_i^{l+1}} + \sum_{r_i \leq R \atop} A_{l,i}r^l}_\text{charges inside} + \underbrace{\sum_{r_i>R \atop\mathstrut} C_{l,i} r^l}_\text{charges outside} \biggr] P_l(\cos\theta_{rr_i})
\intertext{and}
\Phi_\text{out}(\vec{r}) &= \sum_{l=0}^\infty \biggl[ \underbrace{\sum_{r_i>R \atop r_i \leq r} \frac{q_i}{4\pi\varepsilon_0} \frac{r_i^l}{r^{l+1}} + \sum_{r_i>R \atop r_i > r} \frac{q_i}{4\pi\varepsilon_0} \frac{r^l}{r_i^{l+1}} + \sum_{r_i>R \atop} D_{l,i} r^{-(l+1)}}_\text{charges outside} + \underbrace{\sum_{r_i \leq R \atop\mathstrut} B_{l,i} r^{-(l+1)}}_\text{charges inside} \biggr] P_l(\cos\theta_{rr_i}).
\end{align*}
Each term includes all charges which fulfill all conditions below the respective sum. For example, the first sum contains all charges within the sphere ($r_i \leq R$) and closer to the origin than the point where the potential is sampled ($r_i \leq r$). The angle $\theta_{rr_i}$ reflects the angle between the vectors to the charge and the sample point and the four expansion coefficients

{\vspace{-6pt}
\begin{minipage}{0.5\textwidth}
\begin{align*}
A_{l,i} &= \frac{q_i}{4\pi\varepsilon_0} \frac{(l+1)(\varepsilon_\text{r}-1)}{(1+\varepsilon_r)l+1} \frac{r_i^l}{R^{2l+1}}\\
B_{l,i} &= \frac{q_i}{4\pi\varepsilon_0} \frac{2l+1}{(1+\varepsilon_r)l+1} r_i^l
\end{align*}
\end{minipage}
\begin{minipage}{0.5\textwidth}
\begin{align*}
C_{l,i} &= \frac{q_i}{4\pi\varepsilon_0} \frac{2l+1}{(1+\varepsilon_r)l+1} \frac{1}{r_i^{l+1}}\\
D_{l,i} &= \frac{q_i}{4\pi\varepsilon_0} \frac{l(1-\varepsilon_\text{r})}{(1+\varepsilon_r)l+1} \frac{R^{2l+1}}{r_i^{l+1}}
\end{align*}
\end{minipage}
\vspace{6pt}}

\noindent can again be derived from the interface conditions. Hence, for a given mean distribution of representative charges, the potential and the respective mean-field $\vecfield{E}_\text{mf}(\vec{r}) = -\nabla \Phi(\vec{r})$ can be calculated at any point $\vec{r}$. Obviously, this requires the summation over all charges and one evaluation of the Legendre polynomial $P_l(\cos\theta_{rr_i})$ per charge which is numerically particularly demanding. In a typical simulation, where the potential needs to be evaluated at the positions of all $n$ charges, the direct evaluation of the potential thus results in unfeasible numerical effort of the order $\mathcal{O}(n^2)$. However, this effort can be drastically reduced via an efficient numerical implementation based on lookup tables (LUTs) described briefly in the following. The Legendre polynomials can be written as
\begin{align*}
	P_l(\alpha_i) &= \sum_{s=0}^{l/2} \underbrace{(-1)^s \frac{(2l-2s)!}{(l-s)! \, (l-2s)!\, s! \,2^l}}_{L_{ls}} \alpha_i^{k},
\end{align*}
with $\alpha_i = \cos(\theta_{rr_i}) = \frac{xx_i+yy_i + zz_i}{rr_i}$ and $k=l-2s$. Hence, the only task is to evaluate $\alpha_i^{k}$ which can be achieved via the multinomial expansion
\begin{align*}
	\alpha_i^k	&= \left(\frac{xx_i + yy_i + zz_i}{rr_i}\right)^k\\
				&= \sum_{q=0}^k \sum_{p=0}^q
					\underbrace{
					\left(\hspace{-2mm}\begin{array}{c}
						k\\q
					\end{array}\hspace{-2mm}\right)
					\left(\hspace{-2mm}\begin{array}{c}
						q\\p
					\end{array}\hspace{-2mm}\right)
					}_{M_{kqp}}
					\frac{x^{k-q} \, y^{q-p} \, z^p}{r^k} \,\, \frac{x_i^{k-q} \, y_i^{q-p} \, z_i^p}{r_i^k},
\end{align*}
with the multinomial coefficient $M_{kqp} = \frac{k!}{(k-q)!(q-p)!p!}$. Using this, the first term of the inside potential (this term is chosen here for simplicity, all other terms follow analogously) can be written in the form
\begin{align*}
	\Phi^1_\text{in}(\vec{r}) &= \frac{1}{4\pi\varepsilon_0\varepsilon_\text{r}} \sum_{l=0}^\infty \sum_{s=0}^{l/2} \sum_{q=0}^{k=l-2s} \sum_{p=0}^q L_{ls} \, M_{kqp} \, \frac{x^{k-q} \, y^{q-p} \, z^p}{r^{l+k+1}} \underbrace{\sum_{r_i \leq R \atop r_i \leq r} q_i \frac{x_i^{k-q} \, y_i^{q-p} \, z_i^p}{r_i^{k-l}}}_{S_{lkqp}(r)}.
\end{align*}
This representation reveals why the multipole expansion is very attractive for numerical simulations. In principle, all variables related to the charges ($x_i$, $y_i$, $z_i$ and $q_i$) are 'decoupled' from the position where the potential is evaluated ($x$, $y$ and $z$). The only coupling appears in the summation, where in this case only charges with $r_i \leq r$ are taken into account. For a given distribution of charges, the sum $S_{lkqp}(r)$ for a specific set of $[l,k,q,p]$ only depends on the radial coordinate $r$. Hence, it is very convenient to precalculate $S_{lkqp}(r)$ in dependence of $r$, store the results in radial lookup tables, and extract the necessary table entry upon evaluation of the potential for a specific $r$. Three steps are required. First, all charges are injected into the lookup table, i.e. $q_i x_i^{k-q} y_i^{q-p} z_i^p / r_i^{k-l}$ is added to the respective radial bin for each charge. This operation obviously requires one loop through the full list of charges and thus scales linearly with the number of charges ($\mathcal{O}(n)$). In the second step, a cumulative sum of the LUT is calculated in order to reflect the summation over the charges with $r_i \leq r$ for each individual bin of the LUT. A second LUT containing the cumulative sum from the right needs to be calculated in order to store the summation over charges with $r_i > r$. Evaluation of the two cumulative sums requires to loop over the number of LUT entries, which is in typical simulations one or two orders of magnitude lower than the number of charges and thus negligible for the numerical effort. In the third step, upon evaluation of the potential at a given position $\vec{r}$ the respective LUT entries are extracted. The evaluation of the potential for each charge in the simulation obviously also scales linear with the number of charges, such that the overall computational cost is of the order $\mathcal{O}(n)$, though the a large number of included multipole order may generate a large pre-factor. Nevertheless, the multipole expansion is the enabling technology for a fast field evaluation needed for systematic trahjectory simulations.

\subsection{Field-driven ionization}
Within the \MCCC model, atomic ionization models are employed to treat local ionization events that result in free carrier generation starting from an initially neutral sphere. Atomic photoemission is often described following the famous work of Keldysh~\cite{Keldysh_JETP20_1965} that connects the regimes of multiphoton ionization and ionization via field-induced tunneling. The dimensionless Keldysh parameter $\kappa = \sqrt{\Ip / 2\Up}$ that sets the relevant energy scales of atom ($\Ip$) and laser field ($\Up$) into proportion, allows one to roughly estimate the relevant regime for a specific scenario. Typically, ionization can be described in the multiphoton picture for $\kappa\gg1$, while tunneling is considered for $\kappa\lesssim1$. In the simulations considered here, ionization sites are sampled randomly within the volume and the probability for a successful ionization event is determined based on the local near-field.

For the strong-field scenarios with optical fields discussed in \Section{sec:small_spheres} -- \Section{sec:metal_spheres}, ionization was considered to be driven by the combined NIR near-field and the additional nonlinear mean-field. As the intensities of the considered NIR pulses were $\gtrsim\SI{e13}{W/cm^2}$, corresponding to a Keldysh parameter near unity when taking into account a field enhancement of ($\approx2$) at the nanosphere surface, ionization was treated in the tunneling picture. The ionization probabilities were determined utilizing the atomic ADK-rate~\cite{Ammosov_JETP64_1986} calculated from the local instantaneous near-field and the atomic ionization potential. For the case of \silica nanospheres an effective ionization energy of $\Ip \approx \SI{9}{eV}$ was assumed to model the wide band gap measured via soft X-ray photoemission~\cite{Antonsson_PhD_2011}. For the attosecond streaking simulations presented in \Section{sec:streaking}, both XUV and NIR fields had intensities around $\SI{e12}{W/cm^2}$ where NIR-driven tunneling could be safely neglected due to vanishing ionization probabilities. Ionization from the XUV near-field was treated in the photon picture as vertical ionization because of large  respective Keldysh parameters ($>100$). More specifically, the XUV near-field was considered to drive only single-photon ionization as the photon energy was significantly larger than the ionization potential.

\subsubsection{Tunnel ionization}
\begin{figure}[h!]
\centering
\includegraphics[width=0.45\textwidth]{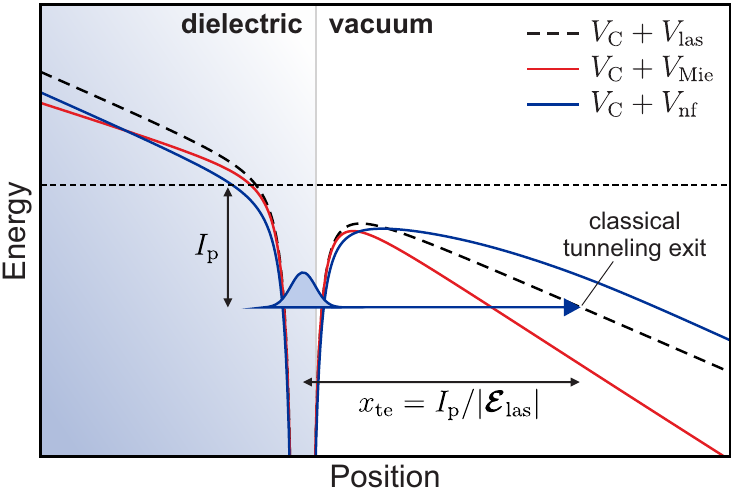}
\caption{Schematic representation of tunneling from the surface of a dielectric. The dashed black curve reflects the effective potential for the atomic case (Coulomb + Laser). Solid red and blue curves represent the effective potentials when considering the enhanced linear (Mie) near-field and the full near-field for an atom located near the surface of a dielectric, respectively. From~\cite{Seiffert_PhD_2018}.}
\label{fig:theory_tunneling}
\end{figure}

For atomic tunneling a widely utilized tunneling rate has been derived by Ammosov, Delone and Krainov~\cite{Ammosov_JETP64_1986}. The respective ADK rate (here given in atomic units) $\Gamma^\text{ADK} \sim \exp\left( -\frac{2^{5/2}\Ip^{3/2}}{3\field{E}} \right)$ scales highly nonlinearly with field-strength. For the intensities considered in most of the studies discussed in this review, ionization in the nanospheres is restricted to tunneling from the surface into the surrounding vacuum. In the atomic case, the starting point of a classical electron trajectory is typically chosen as the classical tunneling exit $x_\text{te} = \Ip / |\field{E}_\text{las}|$ (see \Fig{fig:theory_tunneling}) and its statistical weight is reflected by the tunneling probability obtained from the ADK-rate calculated for the lased field. For an atom located close to the surface within a dielectric material the effective, modified potential which determines the tunneling barrier is determined by the local near-field instead of the incident laser field. When considering only the enhanced linear near-field, the effective potential exhibits a steeper slope in the vacuum region (compare solid red to dashed black curve in \Fig{fig:theory_tunneling}). This typically results in a shorter tunneling exit and an increased tunneling probability. Hence, tunneling is generally most pronounced in the localized field hot spots at the nanosphere surface. However, considering the full near-field (linear near-field \& mean-field, see blue curve in \Fig{fig:theory_tunneling}) can lead to a reduced tunneling probability and may even result in complete quenching of tunnel ionization if the mean-field becomes comparable to the linear near-field. As the straight-forward approach to determine the classical tunneling exit and the tunneling rate from the laser field is not applicable for tunneling from a nanosphere, an alternative approach is to sample the tunneling path along the effective near-field and to determine tunneling exit and rate from the average field along this path. This approach can also be generalized for the description of volume tunneling as required for the simulations discussed in \Section{sec:metal_spheres}.

\subsubsection{Photoionization}
Photoionization can be implemented via evaluation of a suitable ionization rate at randomly sampled points inside the sphere in each timestep of the simulation. In order to account for the spectral width and a possible chirp of the driving pulse the local instantaneous total photoionization rate
$\Gamma(\vec{r},t) = \int \gamma(\omega, \vec{r},t) d\omega,$
is determined from the spectral photoionization rate
$\gamma(\omega, \vec{r},t) = \frac{\sigma(\omega) W(\vec{r},t,\omega)}{\hbar\omega}$.
The local instantaneous spectral intensity profile of XUV pulse is determined from the Wigner distribution $W(\vec{r},t,\omega)$ as described in \Section{sec:spectral_decomposition}. The rate further depends on the spectral molecular photoionization cross section
$\sigma(\omega) = \frac{2n_\text{i}\omega}{n_\text{mol}c_o}$
which can be extracted from the extinction coefficient $n_\text{i}$ (i.e. the imaginary part of the complex refractive index $n=n_\text{r} + in_\text{i}$) and includes the density $n_\text{mol}$ of potential ionization sites. For silica nanospheres, a molecular density of $n_\text{mol}=\SI{0.022}{\A^{-3}}$ was considered to reflect the density of effective \silica 'molecules'. Upon a successful ionization event the initial energy of the generated electron was sampled randomly according to the local instantaneous spectral intensity and reduced by the ionization energy. The direction of the corresponding initial momentum was initialized randomly.

\subsection{Classical trajectory propagation and transport}
In contrast to atomic systems, the strong-field approximation (SFA) is not applicable for determining electron trajectories at nanospheres. Instead, it is needed to resolve  the strongly inhomogeneous near-fields and possible additional fields due to charge interaction to capture the relevant physics. A resulting drawback is the fact that the elegant description of the final electron momenta via the vector potential often used in atomic strong-field physics is no longer applicable. Hence, the complete evolution of electron trajectories needs to be resolved, e.g. via numerical integration of the classical equation of motion $m \ddot{\vec{r}} = -e\vecfield{E}\textsubscript{nf}(\vec{r}, t)$ in the spatio-temporal near-field $\vecfield{E}_\text{nf}$ of the sphere. The effective mass $m$ can be chosen as the electron mass $m\textsubscript{e}$ for the considered materials in the relevant energy ranges~\cite{Schreiber_JESRP124_2002}. Further, the magnetic contribution in the Lorentz force is typically neglected as the impact of magnetic fields is negligible at the considered intensities. The numerical integration can for example be performed via the celebrated Velocity-Verlet algorithm~\cite{Verlet_PR159_1967, Swope_JCP76_1982}.

\subsubsection{Transport via electron-atom scattering}
The collisional transport effects associated with liberated electrons moving within the nanosphere can be accounted for via elastic and inelastic electron-atom collisions, which were treated as instantaneous scattering events sampled using Monte-Carlo methods. The probability for both scattering processes is evaluated from respective energy-dependent mean free paths (MFP) $L_\text{el/inel}(E) = \frac{1}{n_\text{mol} \sigma_\text{el/inel}(E)}$
which describe the distance an electron propagates on average between two adjacent collision
events (of the same type). They depend on the density of atomic/molecular scattering centers $n_\text{mol}$ and the corresponding scattering cross sections $\sigma_\text{el/inel}(E)$.

\paragraph{Elastic collisions}
\begin{figure}[b!]
\centering
\includegraphics[width=0.85\textwidth]{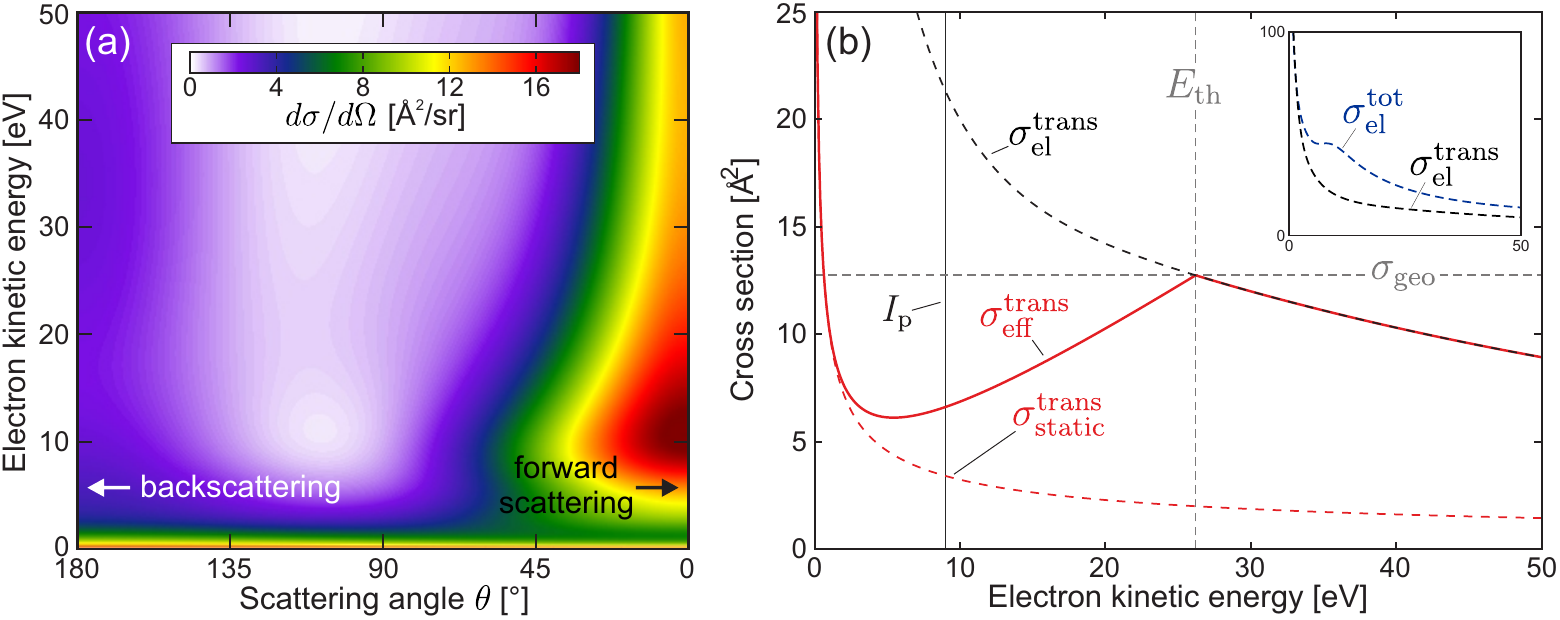}
\caption{Elastic electron-atom scattering in \silica. \panel{a} Effective differential cross section in dependence of the incoming electron’s energy $E$ and the scattering angle $\theta$. \panel{b} Energy-dependent impact ionization cross section (blue curve) and effective transport cross section for elastic scattering in \silica  (solid red curve). The dashed black and red curves represent the effective molecular transport cross section and a static cross section corresponding to a constant collision frequency, respectively. The horizontal dashed line indicates the geometric cross section of the Wigner-Seitz cell and the vertical dashed line marks the threshold energy, where molecular and geometric cross sections are equal. The solid vertical line marks the effective ionization energy of \silica. Transport and total cross section are compared in the top right inset. From~\cite{Liu_ACS7_2020}}
\label{fig:theory_elastic_scattering}
\end{figure}
A straightforward brute-force approach to account for elastic collisions within the semiclassical description offered by evaluating the full energy-dependent differential cross section (DCS) $\frac{d\sigma}{d\Omega}(E,\theta)$, which characterize the probability of an electron with kinetic energy $E$ impinging upon the differential area $d\sigma$ to be scattered into the differential solid angle element $d\Omega$. An efficient approach to model elastic scattering within different materials is to define an effective DCS via superimposing the specific DCS of the relevant atoms weighted by their stochiometric ratios. The individual atomic DCS can for example be extracted from quantum mechanical electron–atom scattering simulations for the atomic potentials, which can be obtained from density functional theory (DFT) including exchange and self-interaction correction. The combined DCS for \silica, determined from quantum mechanical partial wave scattering calculations for the individual atomic potentials and subsequent stochiometric averaging, is displayed in \Fig{fig:theory_elastic_scattering}{a}. While scattering is nearly isotropic at very low energies ($\lesssim\SI{3}{eV}$), forward scattering dominates at higher energies. The total scattering cross section $\sigma_\text{el}^\text{tot}(E) = \int\frac{d\sigma}{d\Omega}(E,\theta) d\Omega$ then follows from integration over the full solid angle $d\Omega$ (cf. blue dashed curve in the inset of \Fig{fig:theory_elastic_scattering}{b}). Implementation of anisotropic elastic scattering can be achieved by calculating the scattering probability from the total cross section at the electrons kinetic energy and subsequent random sampling of the scattering angle according to the differential cross section. It turns out, however, that for many scenarios it is sufficient to approximate the anisotropic scattering by an effective isotropic model scattering process. This description is very convenient, as the latter is fully characterized by the so-called transport cross section $\sigma_\text{el}^\text{tr}(E) = \int\frac{d\sigma}{d\Omega}(E,\theta) (1-\cos\theta) d\Omega$ which reflects the same loss of forward momentum in the isotropic scattering process when compared to the anisotropic process with the corresponding total scattering cross section. Therefore, the transport cross section equals the total cross section for isotropic scattering and is smaller/larger than the full cross section if forward/backward scattering dominates (compare black to blue dashed curve in the inset of \Fig{fig:theory_elastic_scattering}{b}). Most importantly, in order to gain physical insight into the effect of elastic collisional by means of a collision time or mean-free path, the description via the (isotropic) transport cross section is imperative. Otherwise, the physical significance of a mean free path is ambiguous unless the full DCS is known.

At low electron kinetic energies the consideration of pure atomic scattering cross sections drastically overestimates the scattering probability as solid state effects such as the finite size of the Wigner-Seitz cell are neglected. As an example, the energy-dependent effective molecular cross section for \silica is shown as  black dashed curve in \Fig{fig:theory_elastic_scattering}{b}. In this case, the latter becomes larger that the geometrical cross section $\sigma_\text{geo} = n_\text{mol}^{-2/3}$ of the molecular Wigner-Seitz cell (horizontal gray dashed line) below a threshold energy $E_\text{th}$. The resulting unphysically overestimated scattering probability prevents the buildup of collective oscillations of liberated slow electrons in the driving laser field. This limitation of the simplified description can be remedied by considering an modified energy-dependent effective transport cross section
\begin{align}
\sigma_{\rm eff}^{\rm trans} = \left\lbrace\begin{tabular}{ll}
$\sigma_{\rm stat}^{\rm trans} \left(1-\frac{E}{E_{\rm th}} \right) + \sigma_{\rm geo} \frac{E}{E_{\rm th}}$ & for $E \leq E_{\rm th}$ \\
$\sigma_{\rm el}^{\rm trans}$ & for $E > E_{\rm th}$,
\end{tabular} \right.\label{eq:el_scattering_plasmonic}
\end{align}
shown as solid blue curve in \Fig{fig:theory_elastic_scattering}{b}. For energies below $E_\text{th}$ the effective cross section is evaluated as a linear mixture of the geometric area $\sigma_\text{geo}$ and a static cross section $\sigma_\text{static} = \left(v n \tau\right)^{-1}$ (cf. red dashed curve). The latter mimics an energy independent collision frequency $\tau^{-1}$, where $v$ is the electrons velocity. In the limit of low velocities this description reflects a fixed lifetime $\tau$ of plasmonic excitations, which is typically on the order of few femtoseconds~\cite{Soennichsen_NJP4_2002, Peltz_NJP14_2012}. Above the threshold energy the effective molecular transport cross section obtained from the atomic potentials is considered.

\begin{figure}[b!]
\centering
\includegraphics[width=1.0\textwidth]{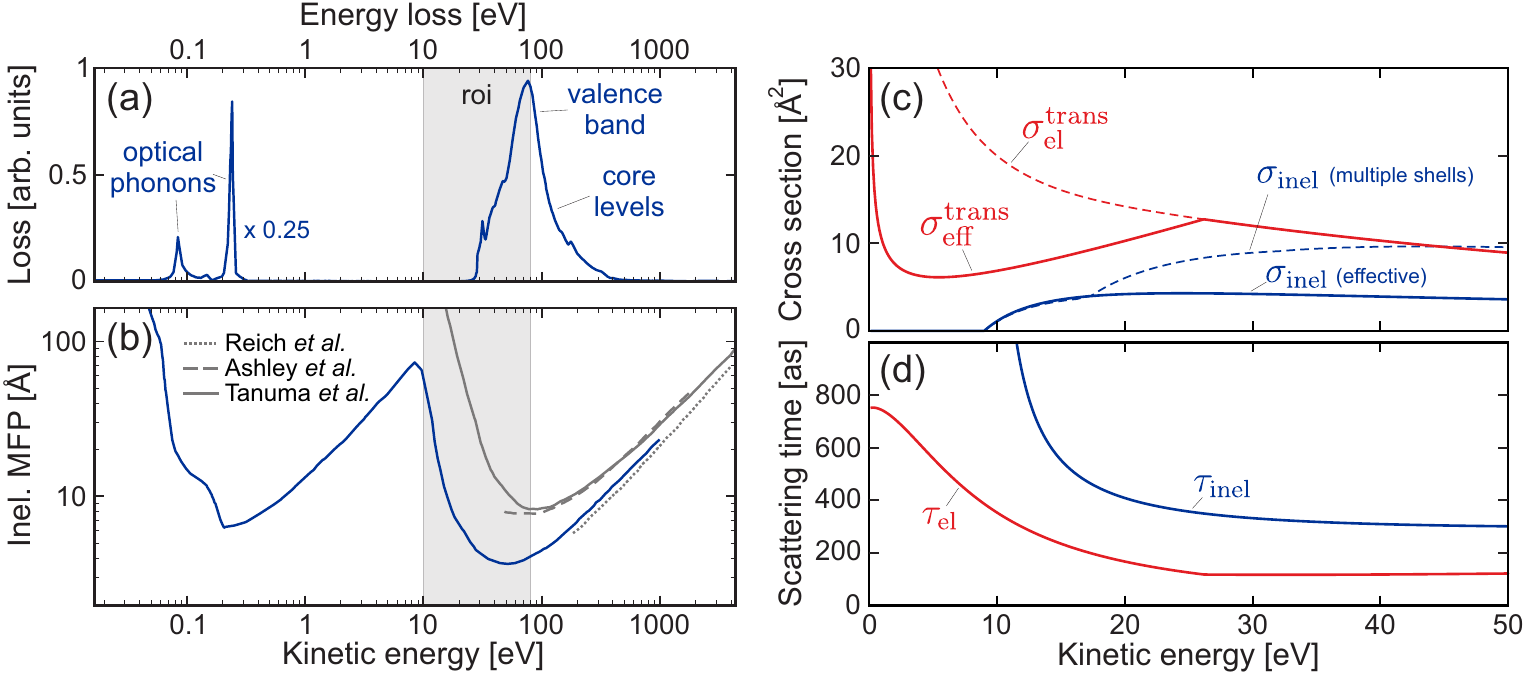}
\caption{Electron-atom scattering in \silica. \panel{a,b}~Loss function and inelastic mean free path in dependence of the electrons kinetic energy (blue curves). Adapted from~\cite{Kuhr_JESRP105_1999} (blue curves) and~\cite{Tanuma_SIA17_1991, Ashley_TNS28_1981, Reich_JESRP46_1988} (gray curves, as indicated). The gray areas indicate the energy region of interest for the scenarios outlined in this review. \panel{c,d} Energy-dependent elastic and inelastic cross sections (c) and respective scattering times (d). The dashed blue curve in (c) shows the inelastic scattering cross section including all shells of the Si and O atoms. The solid blue curve reflects an effective (scaled) cross section that only includes contributions from the shell with the lowest energy close to the band gap of the \silica nanospheres (for details see~\cite{Seiffert_NatPhys13_2017}) which was sufficient for the theoretical description of most of the considered scenarios. The inelastic scattering time in (d) corresponds to the effective cross section.}
\label{fig:theory_inelastic_scattering}
\end{figure}

\paragraph{Inelastic collisions (Impact ionization)}
For the kinetic energies relevant for typical scenarios discussed in this review, inelastic scattering of liberated electrons inside the nanospheres is mainly dominated by interband excitations~\cite{Kuhr_JESRP105_1999}, see gray areas in \Fig{fig:theory_inelastic_scattering} for the case of \silica. These interband excitations can be efficiently modeled as impact ionization of the effective \silica molecules and the respective inelastic scattering cross section can be obtained from a simplified Lotz-formula~\cite{Lotz_ZfP206_1967}
\begin{equation*}
	\sigma_\text{inel}(E) = \sum\limits_s N_\text{ele}(s)\,\SI{450}{(\AA eV)}^2\,\frac{\log(E/\Ip(s))}{E\Ip(s)},
\end{equation*}
where $E$ is the incoming electron's kinetic energy and the summation addresses all shells $s$ of the contributing atomic species. Further, $N_\text{ele}(s)$ reflects the number of electrons occupying the respective shell and $\Ip(s)$ describes the shells ionization energy. Upon a successful inelastic scattering event, the shell which resulted in the scattering is randomly sampled according to its individual contribution to the total cross section and a new pair of a liberated electron and a residual ion is generated. Further, the energy of the incoming electron is reduced by the ionization potential of the previously selected shell.

As an example, energy-dependent elastic and inelastic scattering cross sections and respective scattering times $\tau_\text{el/inel} = 1/(n_\text{mol} v(E) \sigma_\text{el/inel}(E))$, i.e. the time an electron travels on average between two scattering events of the same type, are shown in \Fig{fig:theory_inelastic_scattering}{c,d} for the case of \silica.


\section{Strong-field photoemission from small nanospheres}
\label{sec:small_spheres}
In this section we review the strong-field driven electron emission from  small dielectric nanospheres with size parameters $\varrho < 1$. In this regime, field propagation effects are negligible and the linear response near-field can be treated in dipole approximation. In particular, we discuss the spectral and angular features of the photoemission as well as the underlying mechanisms such as surface-backscattering. Further we will discuss the impact of charge interaction on the emission dynamics and the material dependence of the emission process.

\subsection{Signatures of elastic surface-backscattering from dielectric spheres}
\begin{figure}[t!]
\centering
\includegraphics[width=0.9\textwidth]{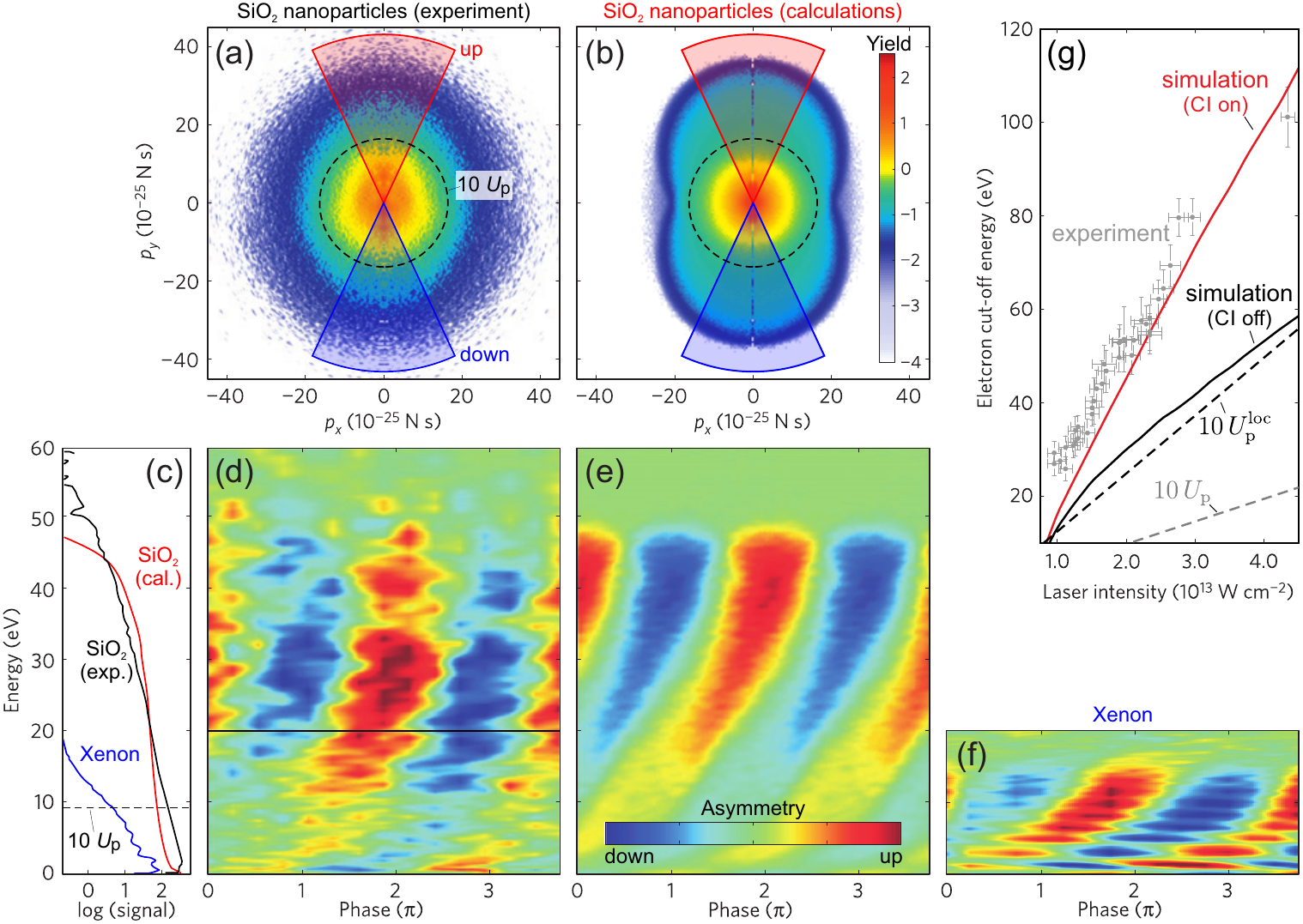}
\caption{Electron emission from silica nanoparticles under intense few-cycle pulses.
\panel{a,b} Carrier-envelope phase-averaged maps of the photoelectron momenta in the propagation-polarization (x-y) plane measured (a) from silica nanoparticles (diameter $\approx\SI{100}{nm}$) and as predicted by \MCCC for the experimental parameters (b). The dashed circles indicate the momentum corresponding to an energy of \SI{10}{\Up}. Red and blue shaded areas visualize a \ang{50} full opening angle along the laser polarization axis for upward (red) and downward (blue) emission. Electron kinetic energy spectra are extracted via integration of the data in these regions.
\panel{c} CEP-averaged photoelectron energy spectra $Y(E)$ for xenon gas and nanoparticles (as indicated).
\panel{d-f} Energy- and CEP-dependent maps of the asymmetry parameter $A = (Y_\text{up}-Y_\text{down}) / (Y_\text{up}+Y_\text{down})$, extracted from the electron yields $Y_\text{up/down}$ in up- and downward emission direction for xenon gas (d), silica nanoparticles (e) and as predicted by \MCCC (f). The limits of the asymmetry color axis in (d), (e) and (f) are set to $\pm0.4$, $\pm0.2$ and $\pm0.6$, respectively. 
\panel{g} Intensity-dependence of the measured cut-off energies of electrons emitted from $d = (100 \pm 50)\,\si{nm}$ silica spheres (gray symbols). Solid curves show respective simulation results excluding (black) and including charge interaction (red). Dashed gray and black lines illustrate the classical cut-offs of backscattered electrons for the ponderomotive energies of the incident field \Up{} and the maximally enhanced local field $\Uploc = \gamma_0^2\Up$, with peak field enhancement $\gamma_0 = 1.6$. Adapted from~\cite{Zherebtsov_NatPhys7_2011}.}
\label{fig:Zherebtsov_NatPhys7_2011_Fig1}
\end{figure}
In a pioneering study, Zherebtsov\etal~\cite{Zherebtsov_NatPhys7_2011} investigated the photoemission from small \silica nanoparticles under intense few-cycle pulses as function of pulse intensity and CEP. \Figure{fig:Zherebtsov_NatPhys7_2011_Fig1}{a} shows the momentum distribution of photoelectrons emitted from $d=\SI{109}{nm}$ \silica nanoparticles in the propagation-polarization (x-y) plane recorded via VMI. The first key observation was that the electron momenta by far exceed the value expected from the classical $\SI{10}{\Up}$ cutoff-law (cf. dashed black circle), hinting at an enhanced acceleration as compared to the cutoff prediction for conventional backscattering for the incident laser intensity. This was further corroborated by inspecting the kinetic energies of photoelectrons emitted along the laser polarization axis (black curve in \Fig{fig:Zherebtsov_NatPhys7_2011_Fig1}{c}) which extend up to around $\SI{50}{eV}$, i.e. around $5$ times larger than \SI{10}{\Up}, where $\Up \approx \SI{0.9}{eV}$ for the considered laser parameters. A systematic analysis of the respective spectral cutoff revealed that the cutoff energies scale linearly with laser intensity, following a modified cutoff law of around $\gtrsim\SI{54}{\Up}$, see symbols \Fig{fig:Zherebtsov_NatPhys7_2011_Fig1}{g}.

Besides the enhanced energies it was demonstrated that the directional yield could be controlled via the field waveform. Thereto the emission asymmetry (quantified by the asymmetry parameter as defined in the caption of \Fig{fig:Zherebtsov_NatPhys7_2011_Fig1}) was inspected as function of the CEP. The resulting asymmetry map (see \Fig{fig:Zherebtsov_NatPhys7_2011_Fig1}{d}) clearly indicated that the emission could be steered into upward (red) or downward (blue) direction by varying the CEP. The phase for optimal up- or downward emission increased with electron energy, resulting in a right tilt of the asymmetry features. Note that the low energy part of the map revealed an additional feature with an even stronger tilt, which could be associated unambiguously with electrons stemming from atomic xenon by comparison with a gas-only experiment (cf. \Fig{fig:Zherebtsov_NatPhys7_2011_Fig1}{f}).

\begin{figure}[t!]
\centering
\includegraphics[width=1.0\textwidth]{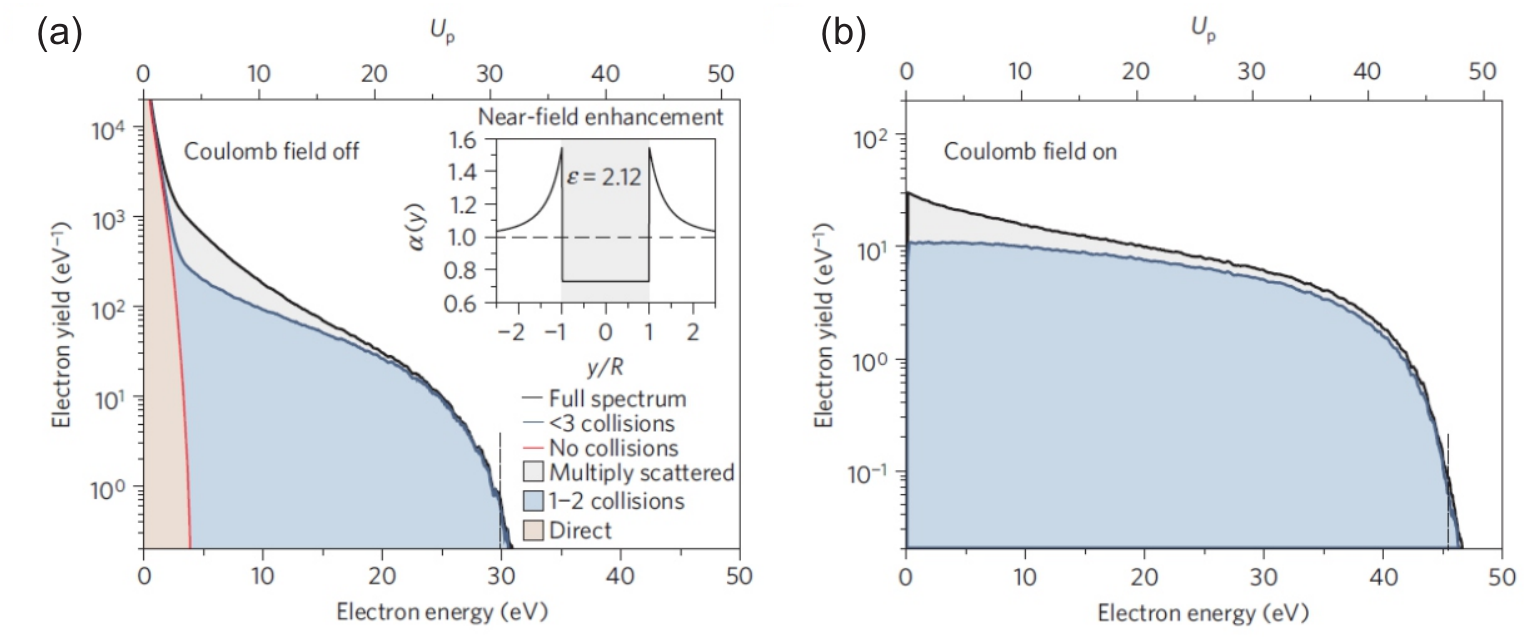}
\caption{Impact of charge interaction on the electron emission from silica nanospheres (\SI{100}{nm} diameter) under intense few-cycle fields. \panel{a,b} Simulated photoelectron energy spectra of electrons with different numbers of elastic collisions (as indicated) with charge interaction turned off (a) and on (b). The inset in (a) shows the spatial field enhancement profile along the laser fields polarization axis. From~\cite{Zherebtsov_NatPhys7_2011}.}
\label{fig:Zherebtsov_NatPhys7_2011_Recollision}
\end{figure}

To unravel the mechanism behind the enhanced energies, the experimental results were compared to \MCCC simulations for the experimental parameters. The good agreement of the momentum map (\Fig{fig:Zherebtsov_NatPhys7_2011_Fig1}{b}), as well as the correct predictions of the intensity-dependent cutoff energies (red curve in \Fig{fig:Zherebtsov_NatPhys7_2011_Fig1}{g}) and the CEP-dependent asymmetry (\Fig{fig:Zherebtsov_NatPhys7_2011_Fig1}{e}) supported that the simulations capture the relevant physics. Under this assumption, the simulations could be used to uncover the physical picture. Key for understanding the enhanced cutoff energies was the ability to selectively disable and enable charge interaction in the simulations by switching the mean-field off or on. \Figure{fig:Zherebtsov_NatPhys7_2011_Recollision}{a} shows selective energy spectra of electrons with different numbers of elastic scattering events extracted from a typical simulation with the mean-field turned off. The results showed that the low energy region of the spectrum is dominated by directly emitted electrons (red curve), similar to the direct emission from atomic or molecular systems. The high energy region is dominated by electrons that returned to the nanosphere and are emitted after one or multiple scattering events (blue and black curves). The cutoff of the latter class of trajectories extends to about $\SI{30}{\Up}$ which could be explained by the acceleration in the linearly enhanced near-field. For the considered scenario, the peak enhancement of the near-field at the nanoparticle poles was found to be $\gamma_0 \approx 1.6$ (cf. \Fig{fig:Seiffert_Dissertation_2018_Fig5.1}{c}), corresponding to a near-field intensity almost tripled with respect to the incident field. As the enhancement factor is constant in the absence of additional nonlinear charge interaction effects, the intensity-scaling of the cutoff follows as $\gamma_0^2\,\SI{10}{\Up} \approx \SI{30}{\Up}$ (cf. black dashed line in \Fig{fig:Zherebtsov_NatPhys7_2011_Fig1}{g}) or, in terms of the ponderomotive potential of the enhanced near-field $\Uploc = \gamma_0^2 \Up$, as $\SI{10}{\Uploc}$.

The respective simulation including charge interaction (\Fig{fig:Zherebtsov_NatPhys7_2011_Recollision}{b}) allowed two main conclusions. First, direct emission is suppressed due to a capacitor-like field emerging from the charge-separation at the surface (i.e. positively charged ions at the sphere surface and liberated electrons outside of the sphere) which traps low-energy electrons. Second, the cutoff of fast recollision electrons is even further enhanced. This could be explained by the combination of an enhanced acceleration during the recollision phase, later termed 'trapping field assisted backscattering' (discussed in more detail below) and the Coulomb repulsion among the electrons in the escaping bunches, which takes place after the electrons have left the vicinity of the surface.

\begin{figure}[b!]
\centering
\includegraphics[width=0.5\textwidth]{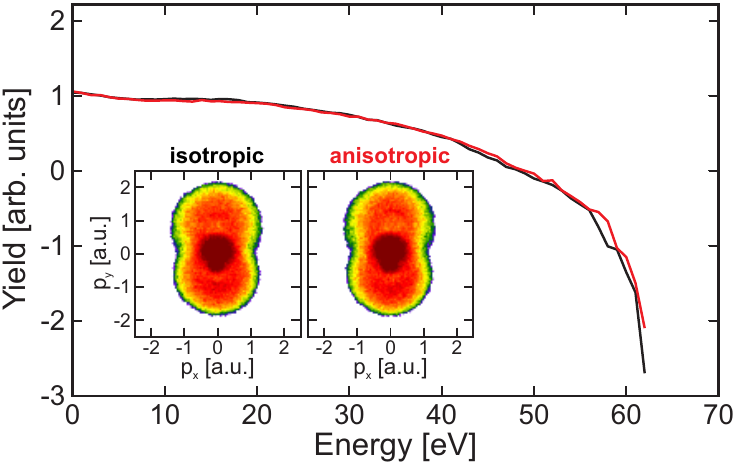}
\caption{Impact of anisotropic elastic collisions on the electron emission from $d=\SI{100}{nm}$ \silica nanoparticles under \SI{4}{fs} NIR few-cycle pulses ($\lambda=\SI{720}{nm}$, $I=\SI{3e13}{W/cm^2}$, $\cep = 0$). From~\cite{Seiffert_PhD_2018}.}
\label{fig:Seiffert_Dissertation_2018_FigC.1a}
\end{figure}

In the above discussed work, elastic collisions were approximated via an isotropic model scattering process. To inspect the impact of the angular dependence of the respective differential cross sections we compared the electron emission predicted by \MCCC simulations including isotropic and anisotropic elastic scattering for similar parameters as before. Both the electron energy spectra (compare red to black curve in \Fig{fig:Seiffert_Dissertation_2018_FigC.1a}) and the projected momentum distributions (see inset) are essentially the same, substantiating the minor importance of the collision anisotropy for the considered scenarios.

\subsection{Trapping-field assisted backscattering}
\begin{figure}[b!]
\centering
\includegraphics[width=0.9\textwidth]{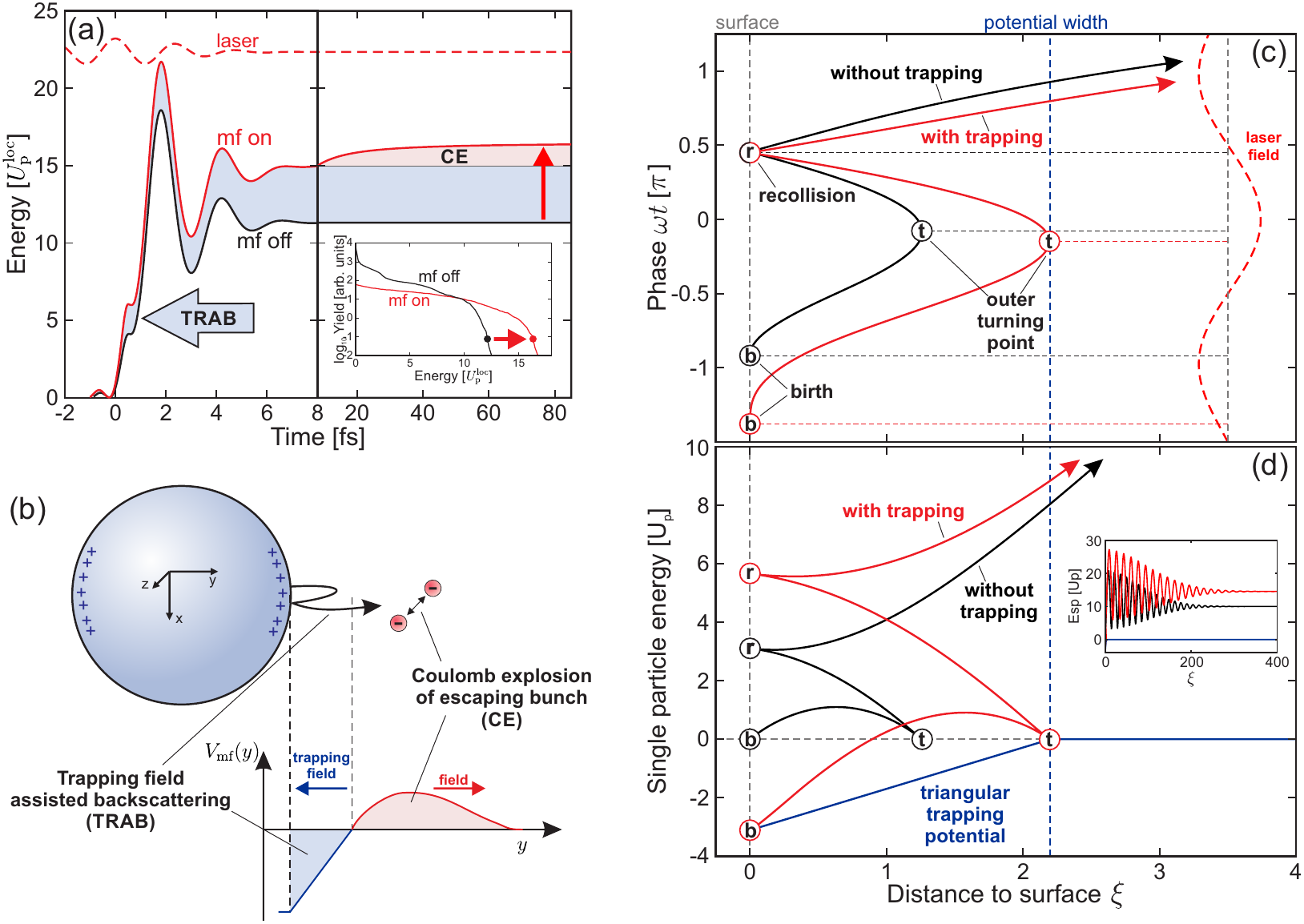}
\caption{Charge interaction effects in strong-field photoemission from dielectric nanospheres.
\panel{a} Evolution of the kinetic energies of typical recollision trajectories corresponding to the cutoff energies of spectra extracted from \MCCC simulations without (black) and with (red) charge interaction. The cut-offs are defined as the energies where the spectra of single recollision electrons drop by three orders of magnitude (symbols in the inset). Blue and red shaded areas indicate energy gains due to trapping field assisted backscattering (TRAB) and Coulomb explosion of the escaping electron bunch (CE), respectively.
\panel{b} Top: Blue plus signs represent positive surface charges from residual ions. Red spheres indicate escaping electrons under the effect of space charge repulsion. Bottom: Attractive trapping potential near the surface mediated by residual ions and emitted electrons (blue) and additional repulsive component (red) due to space-charge repulsion among the escaping electrons.
\panel{c} Trajectory analysis of TRAB. Optimal recollision trajectories calculated in the long pulse limit via the conventional SMM (solid black curve) and the SMM extended to account for a triangular trapping potential (solid red curve). The labeled circles mark birth 'b', outer turning point 't' and recollision 'r' of the respective trajectories.
\panel{d} Time evolution of the single particle energies $E\textsubscript{sp}$ corresponding to the respective trajectories in (c). The solid blue curve represents the triangular trapping potential. Vertical dashed gray and blue lines indicate the surface and the end of the trapping potential, respectively. The inset shows the evolution of the single particle energies on a longer timescale.
Adapted from~\cite{Seiffert_JMO64_2017}.}
\label{fig:Seiffert_Dissertation_2018_Fig4.7}
\end{figure}
Sparked by the initial observations of pronounced charge interaction effects on the electron acceleration process, the impact of the trapping field, which is generated by the charge separation at the surface, was inspected in more detail by Seiffert\etal~\cite{Seiffert_APB122_2016}. \Figure{fig:Seiffert_Dissertation_2018_Fig4.7}{a} shows the evolution of the kinetic energy of typical fast electrons from \MCCC simulations with charge interaction turned off (black curve) and on (red curve). While the general shape of the evolution appears similar, including charge interaction results in two additional energy gains that unfold on very different time scales. The first additional boost in energy takes place during the recollision phase (around \SI{1}{fs}) and is attributed to the enhanced acceleration in the attractive trapping field and is henceforth termed 'trapping field assisted backscattering' (TRAB, cf. blue shaded area in \Fig{fig:Seiffert_Dissertation_2018_Fig4.7}{b}). The second gain unfolds on a much longer time scale after the laser pulse ($\gtrsim\SI{10}{fs}$) and is attributed to the Coulomb explosion (CE) of the escaping electron bunch (cf. red shaded area in \Fig{fig:Seiffert_Dissertation_2018_Fig4.7}{b}). While the latter effect is intuitive, the additional acceleration by the attractive trapping potential might seem counter-intuitive at first glance.

To unravel the physical picture behind the enhanced acceleration in the trapping field assisted backscattering process, the conventional SMM was extended to account for the trapping potential. As the latter impacts the electron emission mainly during the recollision phase and only varies weakly during this short time interval, it was sufficient and most instructive to consider a static triangular model potential, which is only defined by its depth and width, see blue curve in \Fig{fig:Seiffert_Dissertation_2018_Fig4.7}{d}. \Figure{fig:Seiffert_Dissertation_2018_Fig4.7}{c} shows the optimal trajectories of recollision electrons (i.e. those reaching the highest final energies) without (black curve) and with (red curve) the trapping potential. Note that these simulations were performed in the limit of long pulses to neglect additional CEP effects. Comparison of the trajectories showed that the trapping potential results in an earlier birth time (compare circles 'b') and leads to an optimal trajectory with an outer turning point further away from the surface that is reached slight earlier (compare circles 't'). The recollision time remains the same (circle 'r') close to the zero-crossing of the driving electric field.

Key to revealing the mechanism behind the TRAB induced energy gain was to inspect the respective evolution of the single particle energies $E\textsubscript{sp}(t) = E\textsubscript{kin}(t) + E\textsubscript{pot}(t)$, where the potential energy $E\textsubscript{pot}(t) = V(x(t))$ was determined from the trapping potential (see \Fig{fig:Seiffert_Dissertation_2018_Fig4.7}{d}). Although, the trajectory starts with negative single particle energy when including the trapping potential (cf. red circle 'b'), in both cases the energies become zero at the outer turning point (which is located at the potential edge for the optimal trajectory including the trapping potential). The subsequent approach towards the surface results in the additional energy gain (compare red to black curves between the circles labeled with 't' and 'r'). This can be explained via the evolution of the single particle energy $\frac{d}{dt}E\textsubscript{sp} = -e\dot{x}\field{E}(t) + \frac{\partial}{\partial t} V(x,t)$~\cite{Passig_NC8_2017}, which scales linearly with the electrons velocity and the field strength in a static potential. The increased velocity during both approach to and departure from the surface thus leads to the observed increased energy gain. 

\begin{figure}[h!]
\centering
\includegraphics[width=1.0\textwidth]{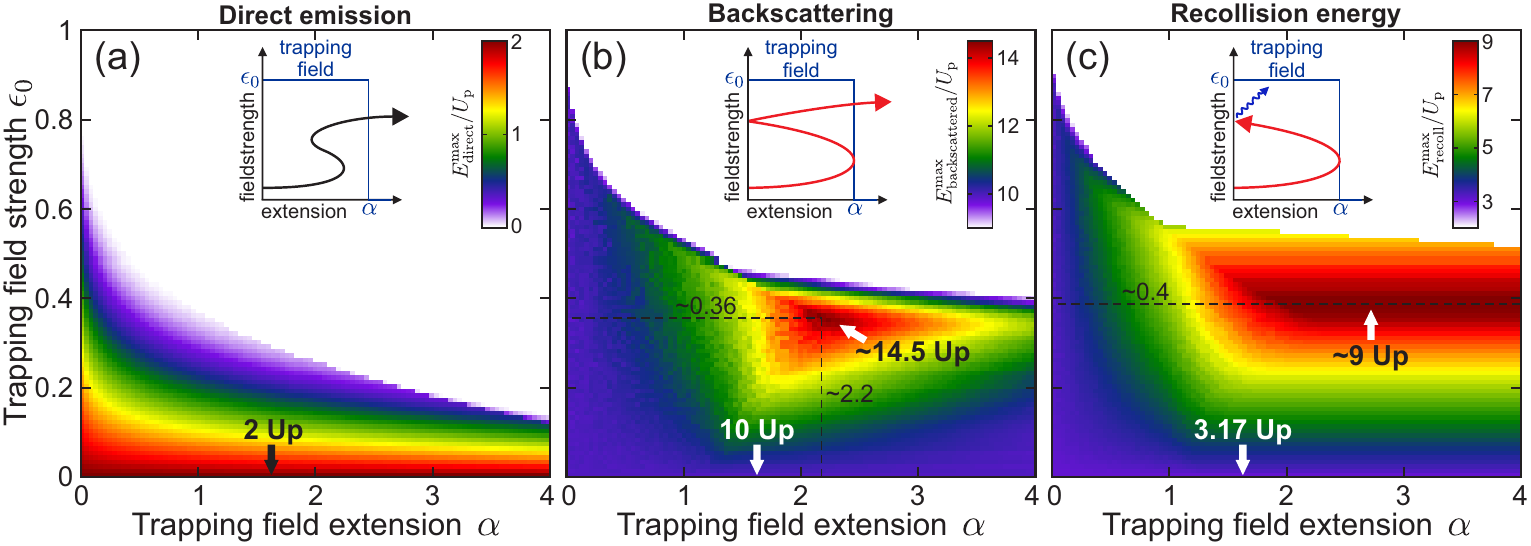}
\caption{Systematic analysis of trapping field induced quenching and enhancement of the electron emission from surfaces. \panel{a,b} Final energies of optimal trajectories of electrons emitted directly (a) and after elastic backscattering (b) in dependence of the field strength $\epsilon_0$ and extension $\alpha$ of a triangular trapping potential. \panel{c} Recollision energies of optimal backscattering trajectories in dependence of the trapping field parameters as in (a) and (b). Insets in (a-c) visualize the respective processes. Dashed black lines in (b) and (c) indicate the optimal parameters for the maximal cut-off and recollision energy, respectively. Adapted from~\cite{Seiffert_JMO64_2017}.}
\label{fig:Seiffert_Dissertation_2018_Fig4.9}
\end{figure}

A systematic analysis of the trapping fields impact on the electron dynamics is presented in \Fig{fig:Seiffert_Dissertation_2018_Fig4.9}, where the resulting final energies of directly emitted and recollision electrons as well as the return energy (i.e. the electrons energy at the moment of returning to the surface) is evaluated in dependence of the trapping potentials depth and extension. For a vanishing trapping potential, the conventional values ($\SI{2}{\Up}$ for direct emission, $\SI{10}{\Up}$ for backscattered electrons and $\SI{3.17}{\Up}$ for the return energy) are reproduced. While a non-vanishing trapping potential always reduces the final energy of directly emitted electrons and finally quenches their emission completely, the final energy of backscattered electrons can be enhanced by almost $\SI{50}{\%}$ and the return energy can almost be tripled for optimal parameters of the trapping potential. The  substantial enhancement of the latter could be relevant for future nanostructure-based HHG. Similar effects have been studied for HHG in atomic systems in static electric fields~\cite{Wang_JPB31_1998, Ciappina_RPP80_2017}.

\subsection{Transition to a universal material-independent cutoff scaling}
\label{sec:quenching_material}
Besides its impact on the recollision process, the trapping field also affects the tunneling step. In particular, the trapping field counteracts the linearly enhanced near-field, resulting in a lowered local intensity of the effective near-field. This affects the tunneling step (1) by reducing the ionization probability resulting from the highly nonlinear scaling of the rate and (2) by increasing the classical tunneling exit $x\textsubscript{te} = \Ip/|\field{E}\textsubscript{nf}|$. In general both the rate and the tunneling exit depend on the material-specific ionization energy. However, Rupp\etal~demonstrated that at sufficiently high laser intensities a charge interaction dominated emission regime is reached, effectively quenching the material dependence in the strong-field electron emission~\cite{Rupp_JMO64_2017}. Thereto, photoemission from \silica, ZnS, Fe\textsubscript{3}O\textsubscript{4} and polystyrene (PS) nanospheres was compared as function of laser intensity.

\begin{figure}[t!]
\centering
\includegraphics[width=1.0\textwidth]{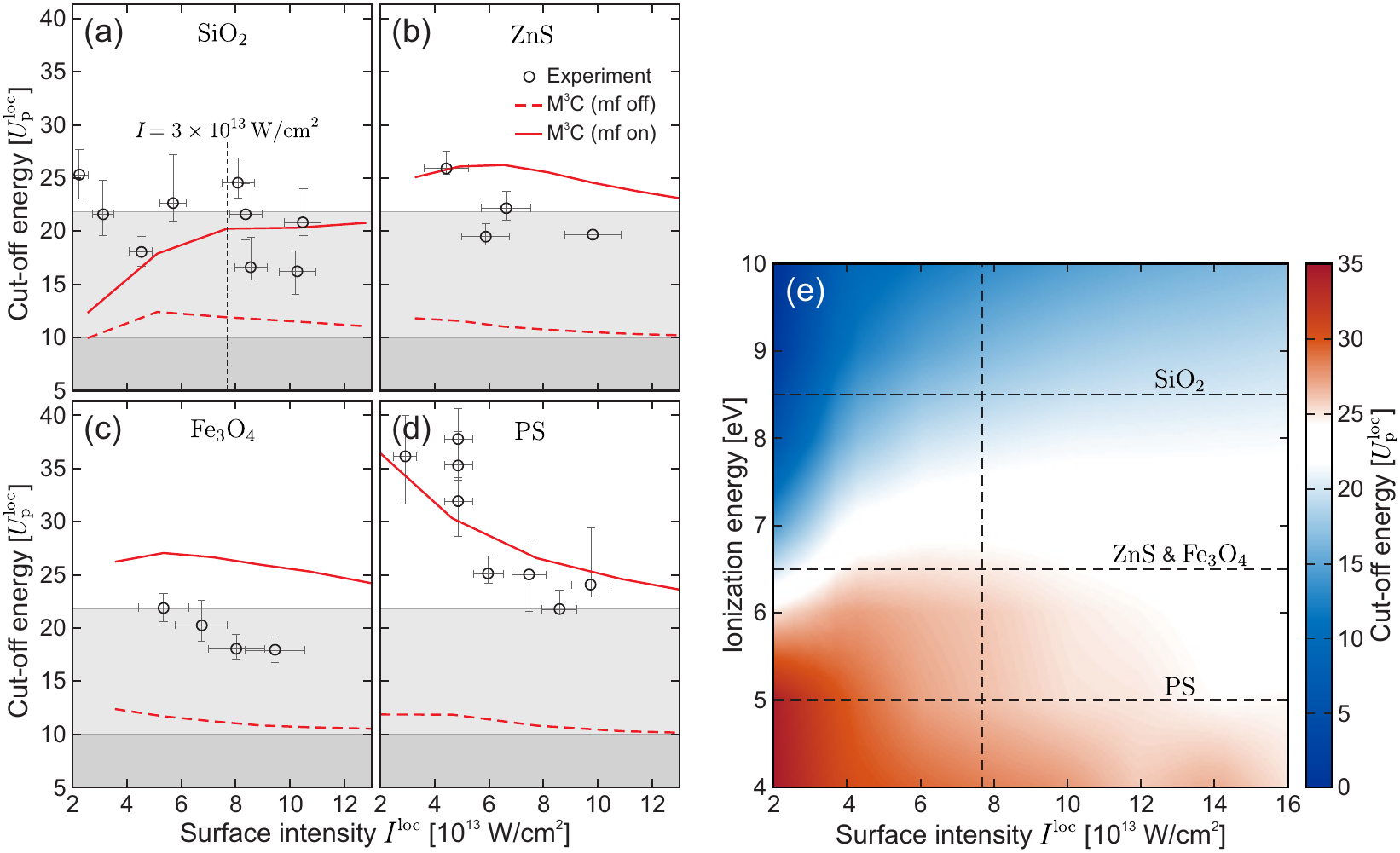}
\caption{Quenching the material dependence in strong-field emission from nanospheres. \panel{a-d} Measured cut-off energies of electrons emitted from small (a) \silica, (b) ZnS, (c) Fe\textsubscript{3}O\textsubscript{4} and (d) PS nanospheres in dependence of the peak intensity at the surface $I\textsubscript{loc}$ (symbols) and as predicted by \MCCC simulations excluding (dashed curves) and including (solid curves) charge interaction. Experimental error bars reflect uncertainties of the laser intensities and in the cut-off evaluation. Gray shaded areas indicate the $\SI{10}{\Uploc}$ cut-off and an energy of $\SI{22}{\Uploc}$. 
\panel{e} Cut-off energies of photoelectrons from $d = \SI{100}{nm}$ spheres under fewcycle pulses (\SI{5}{fs}, \SI{720}{nm}) in dependence of ionization energy and peak intensity of the enhanced linear near-field as predicted by \MCCC. The vertical line marks the surface field intensity corresponding to a laser intensity of $\SI{3e13}{W/cm^2}$ for \silica. Horizontal lines indicate the ionization energies of different dielectric nanoparticles (as indicated).
Data published in~\cite{Rupp_JMO64_2017}, figure adapted from~\cite{Seiffert_PhD_2018}.}
\label{fig:Rupp_JMO64_2017}
\end{figure}

\Figure{fig:Rupp_JMO64_2017}{a-d} shows the extracted cutoff energies in units of the respective local ponderomotive energies associated with the linear response result for the enhanced surface fields in dependence of the surface fields intensity (black symbols). The results of respective \MCCC simulations excluding and including charge interaction are shown as dashed and solid red curves. For all materials, the predicted cut-off energies converge to \SI{10}{\Uploc} (cf. dark gray area) at high intensities when neglecting charge interaction. At low intensities, slightly higher energies are possible resulting from the finite tunneling exit. For \silica the cutoff decreases for the lowest intensities hinting at quenching of backscattering resulting from very large tunnel exits due the high ionization energy. However, for all materials neglecting charge interaction underestimates the measured cutoffs severely. When including charge interaction the agreement becomes much better and is best for \silica and polystyrene at high intensities where ionization becomes dominated by charge interaction instead of the specific properties of the tunneling process. At low intensities, the deviations could be attributed to experimental and theoretical limitations, such as a transition from the tunneling to the multiphoton emission regime, which is not accounted for in the model. The larger deviations for ZnS and Fe\textsubscript{3}O\textsubscript{4} are most likely caused by deviations of the nanoparticle shapes from spherical geometry, which could result in different enhancements of the local fields. Beside these remaining small deviations, the overall trends in the intensity dependence are very similar. Most importantly, for the highest intensity, all results converge to a common cut-off energy of about $20$ to $\SI{25}{\Uploc}$, supporting the charge interaction induced quenching of the material dependence.

A systematic analysis of the cutoff energies predicted by \MCCC in dependence of ionization energy and surface intensity is shown in \Fig{fig:Rupp_JMO64_2017}{e}. Note that when assuming that effects caused by the classical tunneling exit and charge interaction were negligible the map would show a constant value of \SI{10}{\Uploc}. As both are included in the simulations, however, the map reveals the following features. At low intensities the cutoff depends strongly on the ionization energy and approaches zero for the largest ionization energy, which can be attributed to a vanishing ionization rate and quenching of recollisions due to an increasingly high tunneling exit. For higher intensities, the strong dependence on the ionization energy decreases, which can be understood as follows. At a critical intensity, the trapping field counteracts the linear near-field completely, resulting in a vanishing effective field. Further ionization is therefore limited by the charge interaction-induced trapping field. For the considered parameters (i.e.
nanoparticle size and pulse parameters), \MCCC predicts a material-independent cut-off around \SI{22}{\Uploc} for the highest intensity, in good agreement with the experimentally observed common cutoff (cf. \Fig{fig:Rupp_JMO64_2017}{a-d}).


\section{Field propagation in strong-field photoemission from nanospheres}
\label{sec:large_spheres}
A particularly interesting additional parameter for tailoring the electron dynamics arises if electromagnetic field propagation becomes important. Typically (for non-resonant systems) this requires a size of the nanostructure comparable to the excitation wavelength. For appropriate nanooptical targets, field propagation leads to nanofocusing effects that allow to manipulate the overall near-field structure, including its spatial profile, the position of the hot spots, and the phase evolution across the surface. In what follows, we discuss the impact of field propagation effects that arise when increasing the nanoparticle size and allow to control the directionality of the electron emission~\cite{Suessmann_NatCommun6_2015}. We discuss the resulting possibility of all-optical directional control when utilizing two-color fields~\cite{Liu_NJP21_2019}, additional photoemission channels following multiple recollisions of the electrons~\cite{Seiffert_APB122_2016}, as well as reaction nanoscopy which enables to map the spatial distribution of the reaction yield of molecular adsorbates at the nanoparticle surface~\cite{Rupp_NatCommun10_2019}.

\subsection{Field propagation induced emission directionality}
\begin{figure}[b!]
\centering
\includegraphics[width=0.75\textwidth]{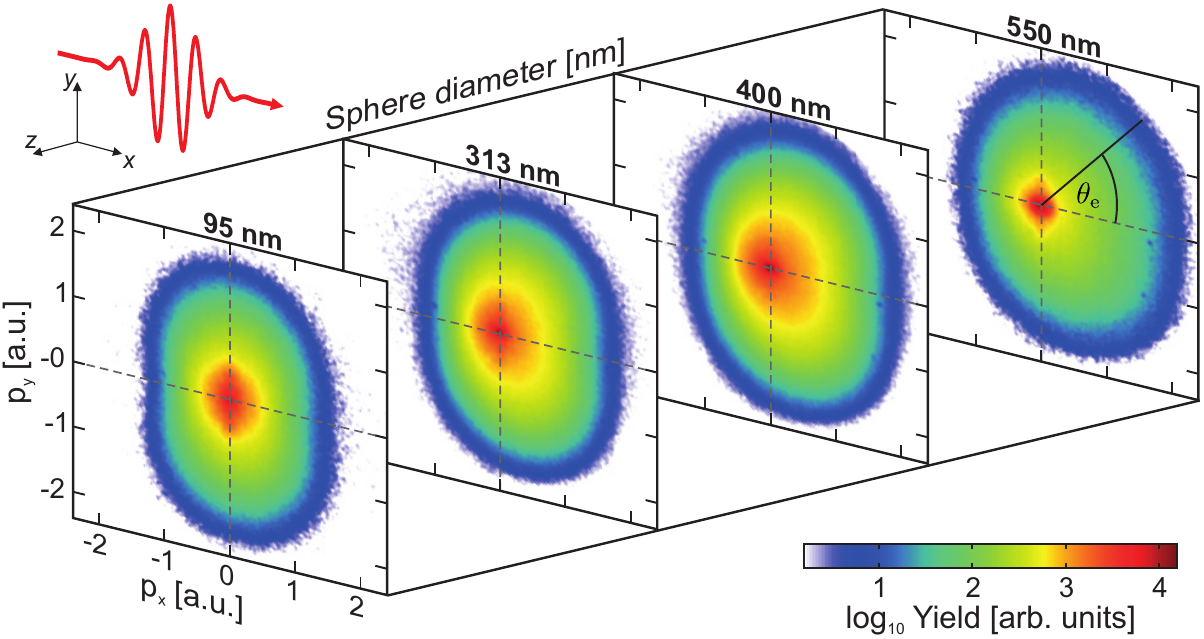}
\caption{CEP-averaged projected momentum distributions measured from \silica nanospheres with 95, 313, 400 and \SI{550}{nm} diameter via velocity map imaging (averaged over millions of laser shots). Note that the intensity of the incident \SI{4}{fs} NIR (\SI{720}{nm} central wavelength) laser-pulses (red curve in the top left corner) has been adjusted to reach maximal momenta around \SI{2}{a.u.} for all sphere sizes. The final emission angle $\theta\textsubscript{e}$ is defined with respect to the pulse propagation direction (x-axis), see \SI{550}{nm} image. Courtesy of F. Süßmann.}
\label{fig:Suessmann_VMI}
\end{figure}
\begin{figure}[b!]
\centering
\includegraphics[width=1.0\textwidth]{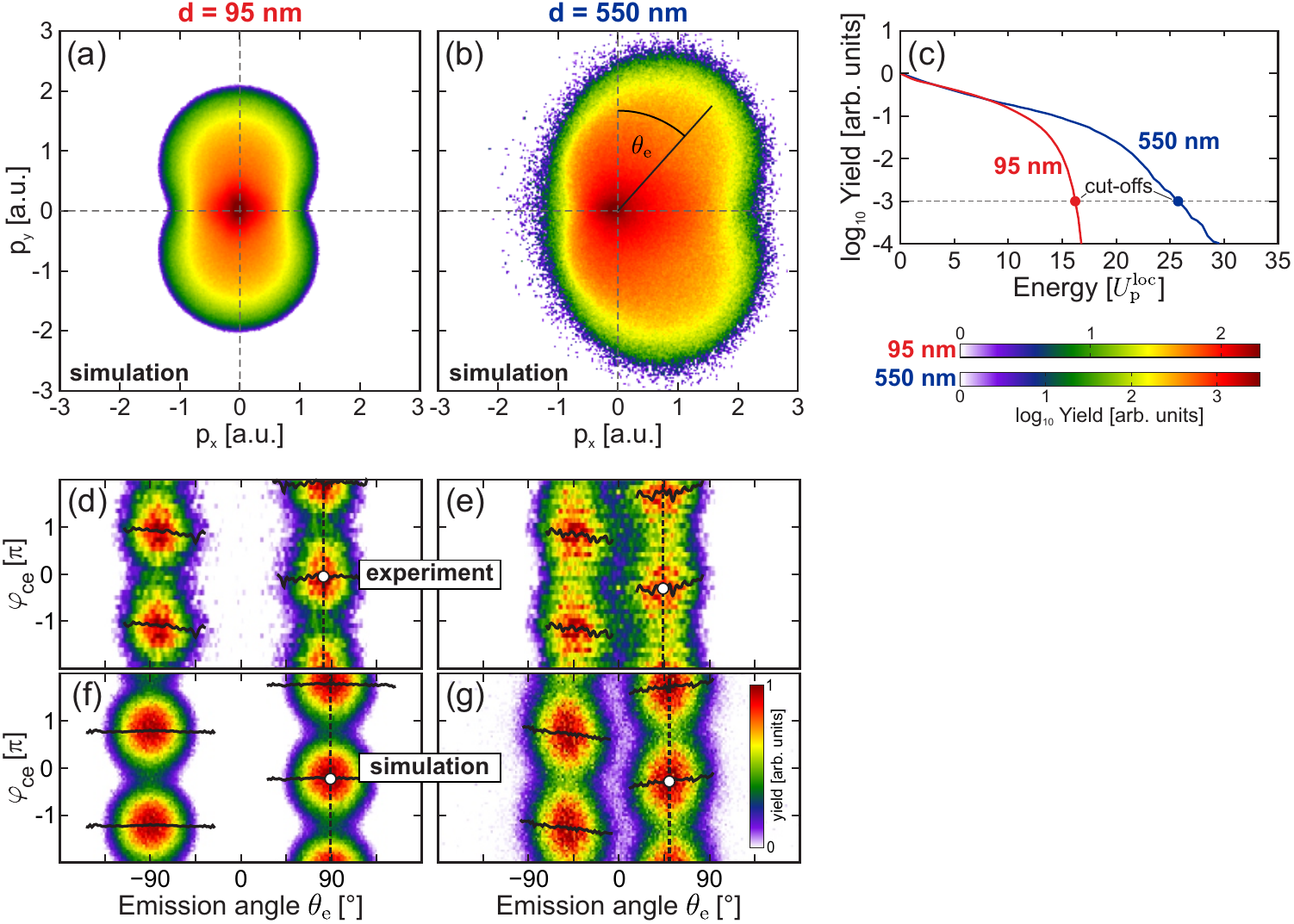}
\caption{\panel{a,b} CEP-averaged projected momentum maps of electrons emitted from $d = \SI{95}{nm}$ (a) and \SI{550}{nm} (b) spheres under \SI{4}{fs}, \SI{720}{nm} few-cycle pulses at \SI{3e13}{W/cm^2} predicted by \MCCC. \panel{c} Corresponding CEP-averaged energy spectra (as indicated). Red and blue symbols indicate respective cut-off energies $E\textsubscript{c}$, defined as the energy where the yield dropped by three orders of magnitude. \panel{d-g} Directionality and phase-dependent switching. (d,e) Yields $Y(\theta_\text{e}, \cep)$ of near cut-off electrons (projected momenta $p>\sqrt{2m_\text{e}E_\text{th}}$ with threshold energy $E_\text{th}=0.5E_\text{c}$) in dependence of final emission angle $\theta_\text{e}$ and CEP \cep, measured from small (d) and large (e) silica nanospheres (parameters as indicated). Critical emission angles $\theta_\text{e}^\text{crit}$ (vertical dashed lines) and phase offsets $\Delta\varphi(\theta_\text{e})$ (solid black curves) are defined via the amplitude and phase of harmonic fits of the data for each vertical slice (for details see original publication). The phase offsets at the critical emission angles define the critical phase $\cep^\text{crit}$ (white symbols). \panel{f,g} \MCCC predictions for the experimental parameters as in (d,e). Adapted from~\cite{Suessmann_NatCommun6_2015, Seiffert_PhD_2018}.
}
\label{fig:Suessmann_switching}
\end{figure}

In order to inspect the impact of nanofocussing on the emission directionality, Süßmann\etal~recorded the electron emission from \silica nanoparticles under intense few-cycle pulses utilizing a VMI spectrometer for variable particle sizes~\cite{Suessmann_NatCommun6_2015}. The measured CEP-averaged momentum projections of the emitted photoelectrons are shown in \Fig{fig:Suessmann_VMI}. In the experiment the laser intensity was adjusted for each particle size to realize comparable cutoffs around \SI{2}{\au}. While the momentum distribution is symmetric with respect to the laser propagation axis (x-axis) for the smallest size, the distributions become more and more deformed with increasing particle size, showing a pronounced shift of the region of maximal momenta towards the laser propagation direction. As a result, the emission angles $\theta\textsubscript{e}$ of the fastest electrons decreases from \ang{90} (i.e. emission in polarization direction) towards smaller values, suggesting close correlation of the emission directionality with the deformation of the near-fields shown in \Fig{fig:Seiffert_Dissertation_2018_Fig5.1}. Respective projected momentum maps obtained from \MCCC simulations are depicted in \Fig{fig:Suessmann_switching}{a,b} for a small and a large nanoparticle. Here, in both cases the laser intensity was \SI{3e13}{W/cm^2} resulting in the emission of faster electrons from the larger nanosphere and a corresponding higher cutoff energy (compare blue to red symbol in \Fig{fig:Suessmann_switching}{c}).

To quantify the combined CEP and directional control of the photoemission, the yields of electrons with final energies near the respective cutoffs in dependence of CEP and emission angle are shown in \Fig{fig:Suessmann_switching}{d,e} for small and large spheres, respectively. The maps strikingly visualize both, the phase-dependent switching between up- and downward emission (i.e. between \ang{-90} and \ang{+90} for the small spheres) as well as the tilt of the dominant emission direction towards the rear-side of the nanosphere for the larger spheres (to about \ang{\pm45}). The excellent agreement of the corresponding \MCCC simulation results (cf. \Fig{fig:Suessmann_switching}{f,g}) allowed a detailed analysis of the physics underlying the strongly enhanced cutoffs and the observed emission directionality.

\begin{figure}[b!]
\centering
\includegraphics[width=1.0\textwidth]{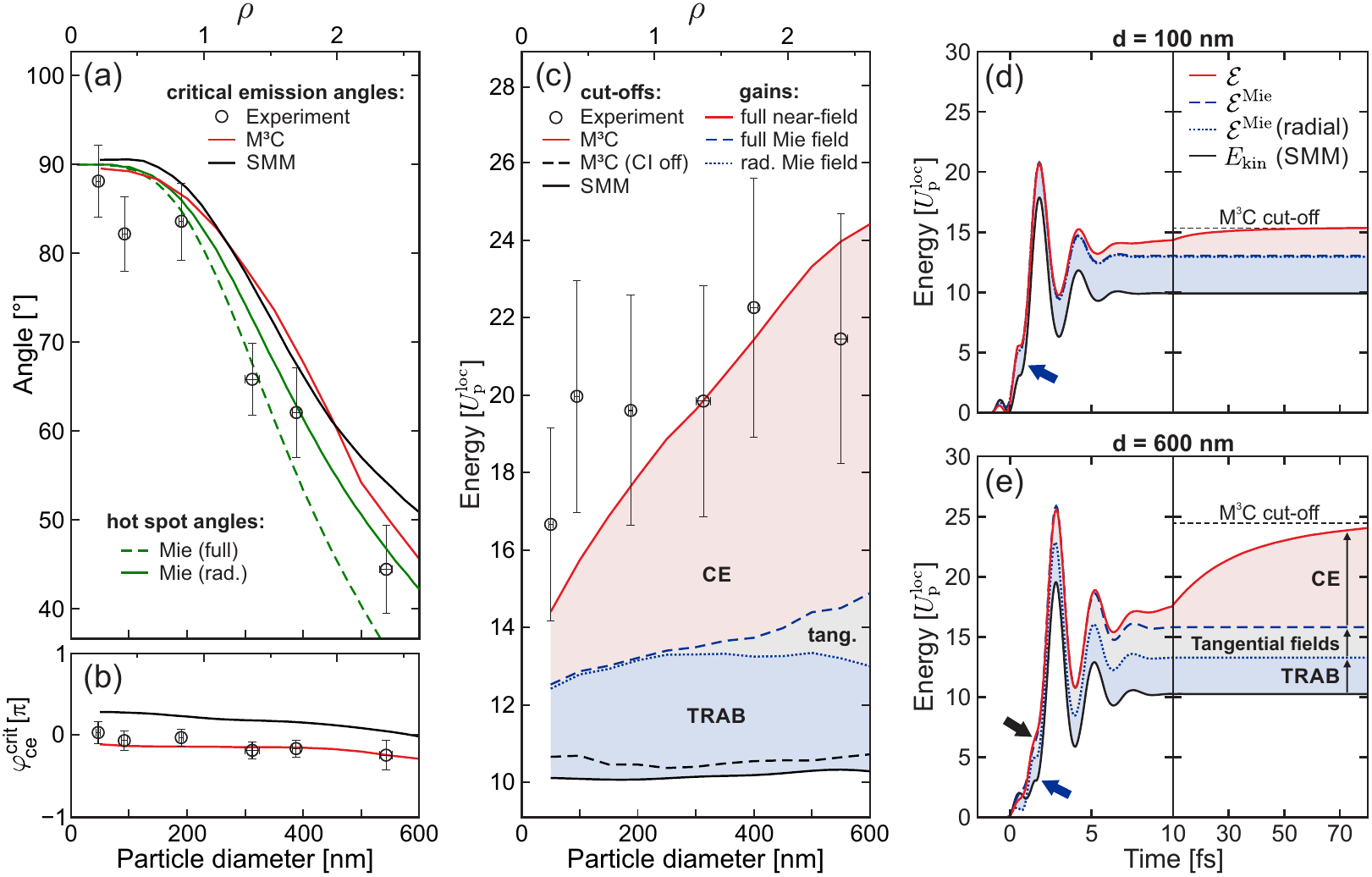}
\caption{
\panel{a-c} Evolution of measured characteristic emission parameters with sphere diameter (symbols) and respective predictions by SMM (solid black curves) and M$^3$C (solid red curves). (a) Critical emission angles and angles of maximal enhancement of the radial (solid green) and full (dashed green) linear near-fields. (b) Critical phases. (c) Cut-off energies and selective energy gains from different field contributions (as indicated). Shaded areas represent the additional energy gains due to TRAB, CE and tangential field effects.
\panel{d,e} Time evolution of the selective kinetic energy gains for two sphere sizes (as indicated). Solid black curves indicate the evolution of the kinetic energies predicted by the SMM. Blue and red curves represent the energy gains from selective contributions to the full near-field (as indicated) for cut-off electrons calculated via \MCCC (averaged for cut-off electrons). Shaded areas indicate the energy gains from TRAB (blue), Coulomb explosion (red) and tangential field effects (gray). Note the different scaling of the time axes before and after the vertical black lines. Adapted from~\cite{Suessmann_NatCommun6_2015, Seiffert_PhD_2018}.}
\label{fig:Suessmann_systematic}
\end{figure}

Thereto, the dependence of three characteristic emission parameters, i.e. the critical emission angle $\theta_\text{e}^\text{crit}$, the critical phase $\cep^\text{crit}$, and the cut-off energy $E_\text{c}$ on the nanoparticle size was investigated, see \Fig{fig:Suessmann_systematic}{a-c}. The analysis of the critical emission angles (\Fig{fig:Suessmann_systematic}{a}) revealed that the predictions of both SMM (black curve) and \MCCC (red curve) agree well with the experimental observation (symbols). Comparing the critical emission angles to the hot-spot angles $\theta_\text{h}$ of the radial Mie-field (solid green curve) and the full Mie-field (dashed green curve) supported that the emission direction is closely correlated with the radial components of the surface field. The measured critical phase varies only weakly with size and is well-reproduced by the \MCCC simulations (compare red curve to symbols in \Fig{fig:Suessmann_systematic}{b}). The SMM prediction shows the correct phase evolution (black curve) but exhibits an almost size-independent offset to \MCCC and experiment. The latter can be attributed to the impacts of the classical tunneling exit and charge interaction. For example, the lacking trapping field can result in slightly different timings for the generation of the optimal backscattering trajectories. However, the only small variation of the phase with the nanosphere size supports the possible application of robust attosecond control of the directional emission. Comparison of the predicted and measured cut-off energies in \Fig{fig:Suessmann_systematic}{c} revealed that the SMM cut-off (solid black curve) and the cutoff extracted from \MCCC simulations with charge interaction turned off (dashed black curve) follow the \SI{10}{\Uploc} law irrespective of sphere size where the small deviation of the \MCCC result is mainly attributed to the finite classical tunneling exit. The considerably higher experimental cut-offs (symbols) can only be reproduced by \MCCC with charge interaction turned on (red curve). In this case, the cut-off energies start around \SI{14}{\Uploc} for the smallest investigated spheres and increase almost linearly to around \SI{24}{\Uploc} for the larges spheres. This increase is not attributed to the linear radial near-field as the energies are scaled with respect to the ponderomotive energy of the maximally enhanced radial near-field. The origin of the additional energy enhancement could be traced back to the impacts of charge interaction as well as tangential field components via a selective energy gain analysis. Thereto for each particle size, the kinetic energy gains
\begin{align*}
    \Delta E_\text{kin} &= \int_{t_\text{b}}^{\infty} \dot{\vec{r}}(t) \cdot \vecfield{E}_i(\vec{r}(t))\,\text{d}t
\end{align*}
were inspected for electrons with final energies close to the respective cutoffs for different contributions $\vecfield{E}_i$ to the total near-field. If the full near-field was considered (i.e. $\vecfield{E}_i = \vecfield{E}_\text{nf}$) the resulting kinetic energy gain reproduced the final energy. Considering only the radial component of the linearly enhanced near-field $\vecfield{E}_\text{Mie}^\text{rad}$ revealed the first contribution to the energy gain (dotted blue curve) that was attributed to modification of the trajectory by the mean-field, while the additional acceleration due to its dynamical evolution is excluded. This allowed to isolate the effect of the trapping field assisted backscattering (blue shaded area) which leads to an additional energy gain of around $\SI{3}{\Uploc}$, in good agreement with the predictions of the simplified TRAB model (cf. \Fig{fig:Seiffert_Dissertation_2018_Fig4.9}{b}). The only minor variations with diameter suggested that the trapping potential depends primarily on the surface charge density and that TRAB is hence  insensitive to the nanosphere size. Considering the full linearly enhanced near-field $\vecfield{E}_\text{Mie}$ (dashed blue curve) results in an additional energy gain resulting from the additional acceleration by tangential components of the local near-fields (gray shaded area). This effect becomes only relevant for large spheres and still remains comparably small, substantiating that the recollision process is predominantly driven by the radial near-field. The remaining contribution to the final energies is attributed to the Coulomb explosion of the departing electron bunches (red shaded area). While the impact of this effect is comparable to TRAB for small spheres, it becomes is the prevailing contribution for the largest spheres. This can be attributed to the increasing amount of electrons within the escaping bunches at larger spheres due to the increasing emission area. This suggests that, different from TRAB, the Coulomb explosion is thus sensitive to the full (non-local) electron distribution. \Figure{fig:Suessmann_systematic}{d,e} show the corresponding time-evolutions of the selective energy gains for small and large spheres. In both cases, the energy gain from TRAB takes place during the recollision phase (indicated by the blue arrows). The additional gain from acceleration due to tangential near-field components for the larger spheres also occures during the recollision (black arrow). In contrast, the Coulomb explosion mediated contribution unfolds on a much longer timescale.

\subsection{Electron emission with long pulses}
\begin{figure}[h!]
\centering
\includegraphics[width=1.0\textwidth]{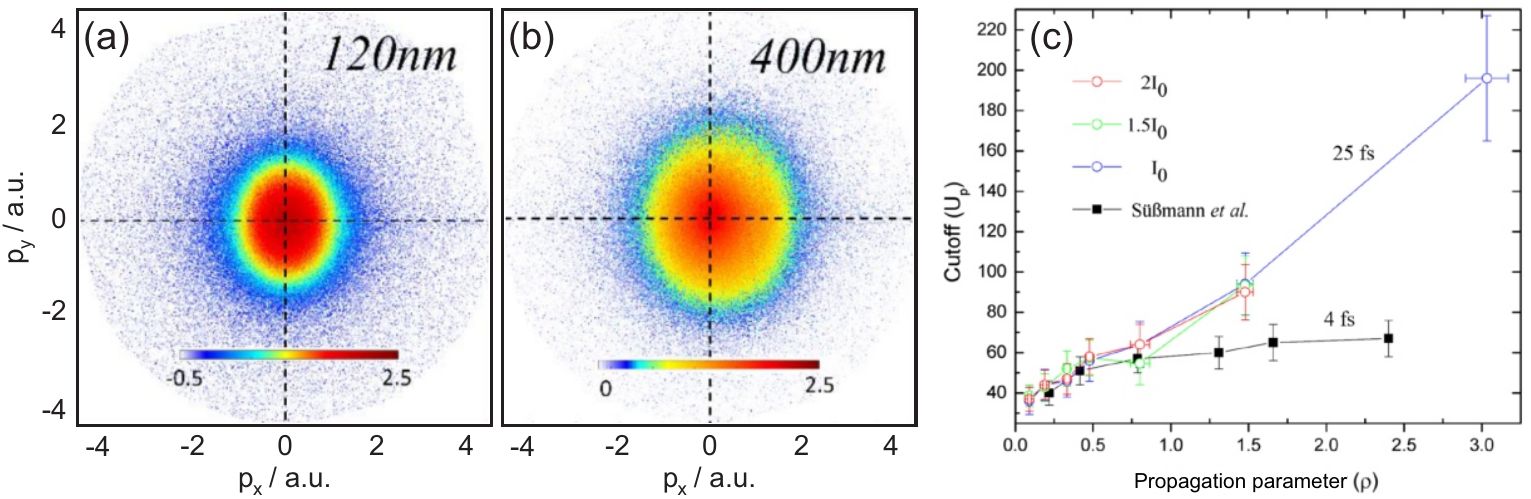}
\caption{Photoemission from \silica nanospheres under long pulses.
\panel{a,b} Averaged projected momentum maps (in atomic units (a.u.) of momentum) obtained from VMI images of photoelectrons emitted from small (a) and large (b) silica nanospheres under \SI{25}{fs} laser pulses pulses (central wavelength \SI{780}{nm}) at \SI{1.32e13}{W/cm^2}. The color bars represent the electron yields on a logarithmic scale. \panel{c} Open circles show size-dependent \Up-rescaled cutoff energies for three different intensities (as indicated, $I_0 = \SI{8.8e12}{W/cm^2}$). Black squares represent the respective data discussed earlier for few-cycle pulses~\cite{Suessmann_NatCommun6_2015}.  Note that the size dependence is expressed via the propagation parameter $\varrho$ to enable comparison with the few-cycle data which was taken at a slightly different wavelength of \SI{720}{nm}. Adapted from~\cite{Powell_OE27_2019}.}
\label{fig:Powell_OE27_2019}
\end{figure}

Besides the above discussed results obtained with few-cycle fields, similar results have also been reported for the electron emission using longer pulses by Powell\etal~\cite{Powell_OE27_2019}. In that case, photoemission from \silica nanospheres was inspected utilizing \SI{25}{fs} (intensity FWHM) pulses at a central wavelength of \SI{780}{nm}. Consistent with the few-cycle results, projected momentum images acquired via velocity map imaging were symmetric with pronounced emission along the laser polarization direction for small spheres and a pronounced shift towards the laser propagation direction for larger spheres, see \Fig{fig:Powell_OE27_2019}{a,b}. For small spheres (propagation parameter $\varrho < 1$) also the extracted \Up-rescaled cutoff energies agree with the few-cycle data for different inspected intensities (compared red, green and blue curves to black curve in \Fig{fig:Powell_OE27_2019}{c}). For larger spheres, however, the cutoffs drastically exceeded the results from the few-cycle experiments. These observations were attributed to the longer fields resulting in (i) a deeper trapping potential enhancing the rescattering process and (ii) more electrons in the escaping electron bunches leading to stronger Coulomb repulsion.  

\subsection{All-optical directional control}
\begin{figure}[h!]
\centering
\includegraphics[width=1.0\textwidth]{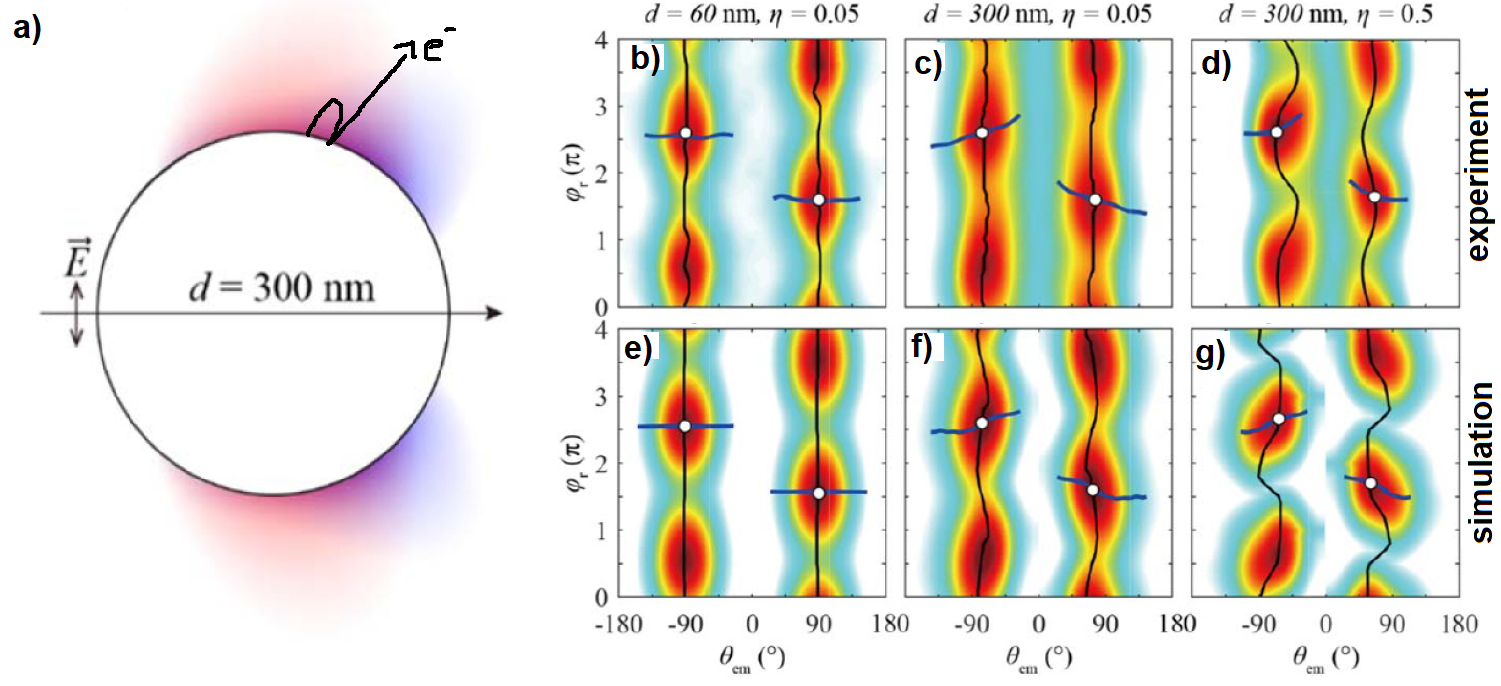}
\caption{All-optical directional control of the photoemission from \silica nanospheres in $\omega$-$2\omega$ laser fields. \panel{a} Schematic representation of the enhanced near-field profiles (radial electric field) for the red ($\SI{780}{nm}$) and blue ($\SI{390}{nm}$) contributions of a two-color laser field at a $d=\SI{300}{nm}$ \silica nanospheres. \panel{b-d} Measured angular and relative phase-resolved electron cutoff energies for different sphere diameters and intensity ratios $\eta = I_{2\omega} / I_\omega$ (as indicated). The IR intensity was $I_\omega = \SI{3e12}{W/cm^2}$. Energies are normalized to the ponderomotive potential of the incident IR field. The solid blue lines are angular dependent phase offsets $\varphi_\text{offs}(\theta_\text{em})$. Black lines show the relative phase dependent optimal emission $\theta_\text{em}^\text{opt}(\varphi_\text{r})$ of the cutoff energies and white dots indicate the critical emission angles $\theta_\text{em}^\text{crit}$. \panel{e-g} Respective maps as predicted by SMM simulations. Adapted from~\cite{Liu_NJP21_2019}.}
\label{fig:Liu_NJP21_2019}
\end{figure}

In the work of Süßmann\etal~the directionality of the photoemission from large nanospheres was controlled by adjusting the near-field deformation via varying the nanosphere size~\cite{Suessmann_NatCommun6_2015}. Since the extent of this modification depends mainly on the propagation parameter $\rho= \pi d/\lambda$ (cf. \Eq{eq:propagation_parameter}) the straightforward complementary strategy would be to tune the wavelength of the driving field. However, both approaches are impractical for potential applications. An alternative route for all optical directional control utilizing two-colors pulses was explored in Liu\etal~\cite{Liu_NJP21_2019}. The central idea is sketched in \Fig{fig:Liu_NJP21_2019}{a}. A phase-locked two-color field $E(t) = E_\omega(t)[\cos(\omega t) + \eta \cos (2\omega t + \varphi_\text{r})]$ consisting of a NIR fundamental and its second harmonic was considered to create spatially separated field hot spots for the two spectral components. Varying the admixture of the second harmonic $\eta$ and the relative phase $\varphi_\text{r}$ between the two contributions enabled to control the sub-cycle electron acceleration process including the associated backscattering energy cutoff via modification of the spatiotemporal waveform of the local near-field. 

Maps of measured \Up-rescaled cutoff energies as function of emission angle $\theta_\text{em}$ and relative phase $\varphi_\text{r}$ are shown in \Fig{fig:Liu_NJP21_2019}{a-c} for differently sized \silica nanospheres and different admixtures $\eta$ (as indicated). Respective results as predicted by SMM are displayed in \Fig{fig:Liu_NJP21_2019}{d-f}. For small spheres  where the hot spots are located around the pole regions for both colors, the fastest electrons are emitted at around $\pm\ang{90}$ irrespective of relative phase, see black curves in \Fig{fig:Liu_NJP21_2019}{a,d}. In this case, the relative phase thus only offers control over the up-down switching, similar to the previously reported switching with CEP-controlled few-cycle pulses~\cite{Suessmann_NatCommun6_2015}. For the large spheres, however, the optimal emission angle of the fastest electrons (cf. black curves) is shifted towards smaller absolute values (i.e. towards the rear-side of the nanospheres) and most importantly varies as function of the relative phase. While this effect is only weak in case of a small admixture (cf. \Fig{fig:Liu_NJP21_2019}{c,f}), for an admixture of $\eta=0.5$ the emission angle could be modified by about \ang{30}, see \Fig{fig:Liu_NJP21_2019}{d,g}.

\begin{figure}[t!]
\centering
\includegraphics[width=0.7\textwidth]{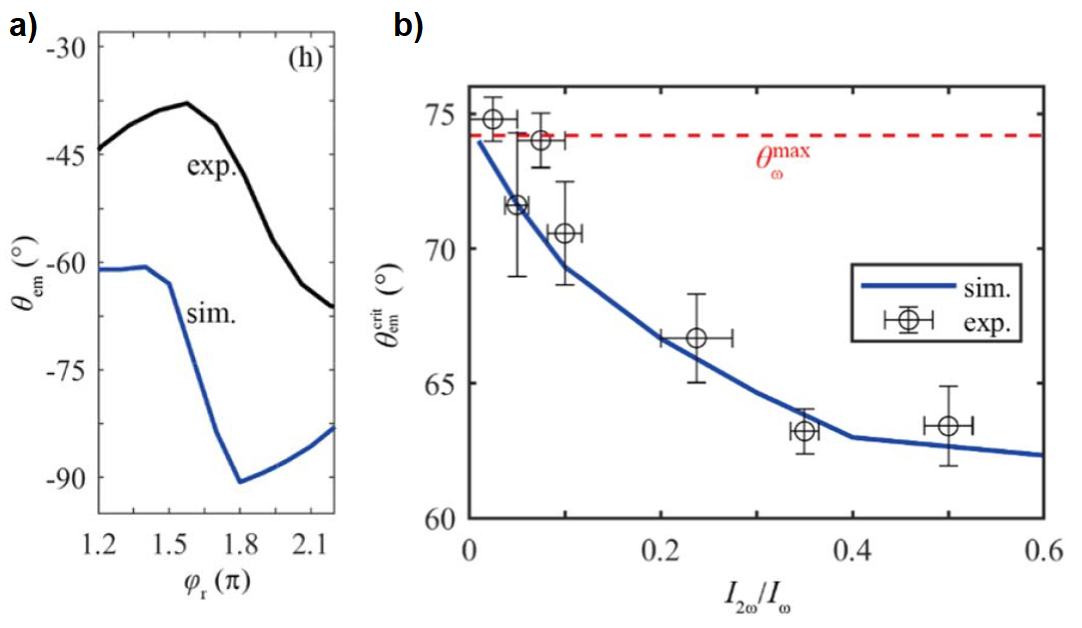}
\caption{\panel{a} Optimal angles for downward emission from (c) and (f), as indicated. \panel{b} Critical emission angles obtained from measurement (circles) and SMM simulations (solid blue line) with \SI{300}{nm} \silica nanospheres. Adapted from~\cite{Liu_NJP21_2019}.}
\label{fig:Liu_NJP21_2019_2}
\end{figure}

The relative phase-dependent directional switching, visualized by the optimal angles for downward emission in \Fig{fig:Liu_NJP21_2019_2}{a}, was well captured by the SMM simulations (compare blue to black curve). The remaining offset was attributed to considered specular scattering in the simplified SMM description. Besides the relative phase, the intensity ratio between the two-color field components provides a second control knob for angular steering. The critical emission angle (cf. white dots in \Fig{fig:Liu_NJP21_2019}{b-g}), which indicates the preferential energetic electron emission direction, is shown by the black symbols in \Fig{fig:Liu_NJP21_2019_2}{b} in dependence of the intensity ratio $\eta$. While it coinsides with the hot spot angle of the fundamental field (red dashed line) in the case of small admixtures, it decreases for higher contributions of the second harmonic component, resulting in a shift of about \ang{10} for an admixture of about \SI{50}{\%}. The results were reproduced quantitatively by the predictions of the SMM (blue curve).

\subsection{Impact of tangential field components and double recollisions}
At small dielectric nanospheres the local near-fields mostly retain the linear polarization of the incident field, resulting in almost exclusively radial orientation of the electric field vector at the hot spot region (cf. \Fig{fig:Seiffert_Dissertation_2018_Fig5.1}{c,d}). At larger spheres the near-fields can develop pronounced ellipticities restuling in substantial tangential field components. While the impact of the latter on the energy of emitted recollision electrons has already been inspected in the initial work of Süßmann\etal~\cite{Suessmann_NatCommun6_2015} (cf. gray areas in \Fig{fig:Suessmann_systematic}{c,e}) a more detailed analysis of the physics behind the additional gain has been reported slightly later by Seiffert\etal~\cite{Seiffert_APB122_2016}. 

\begin{figure}[t!]
\centering
\includegraphics[width=1.0\textwidth]{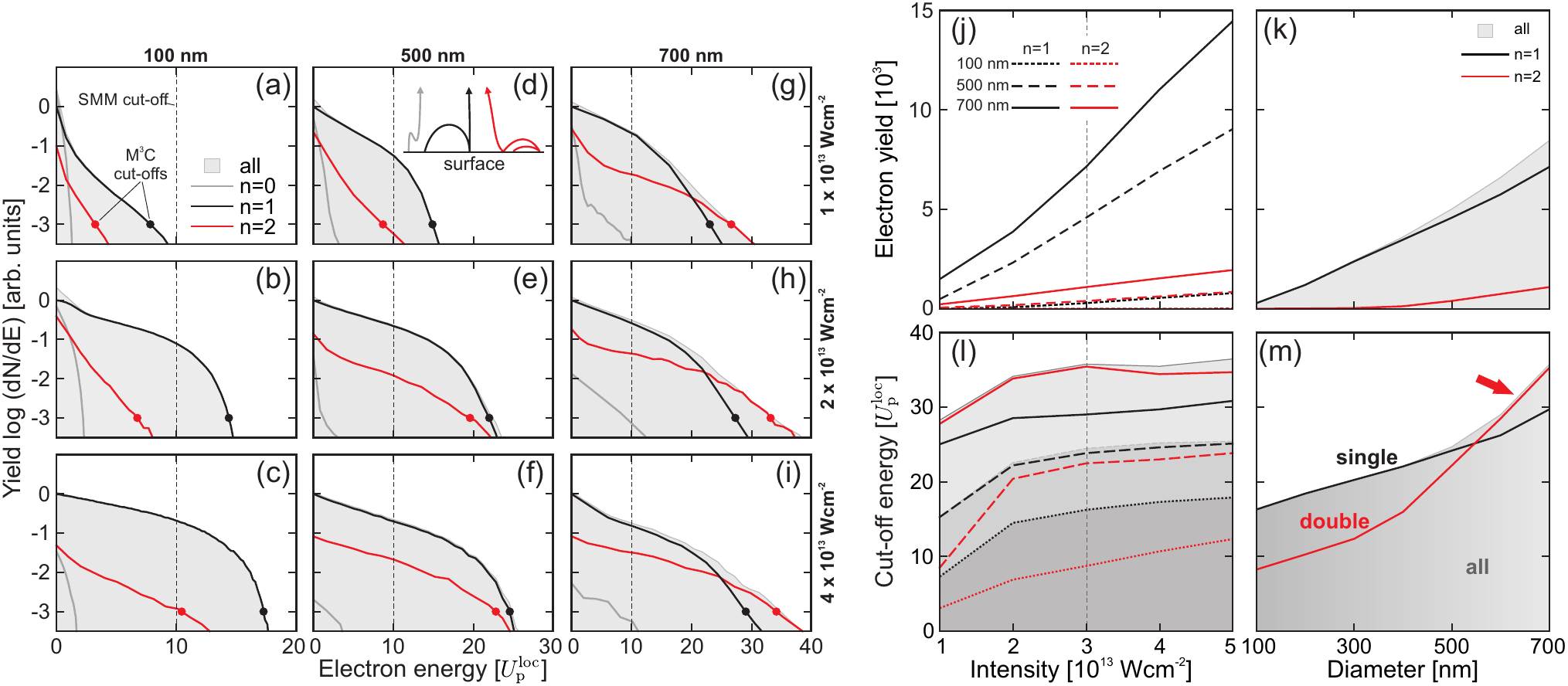}
\caption{
\panel{a-i} Recollision-resolved CEP-averaged \MCCC energy spectra for different sphere diameters and laser intensities (as indicated). Gray shaded areas show full spectra. Solid curves represent selective spectra for direct emission ($n=0$, gray), single recollision ($n=1$, black), and double recollision ($n=2$, red), cf. schematic trajectories in (d). Spectra are normalized to the yield of single recollision electrons at $E=0$. Energies are scaled to the local ponderomotive potential and cut-offs of single and double recollision electrons are indicated as black and red symbols. Dashed vertical lines mark the conventional \SI{10}{\Uploc} cut-off. 
\panel{j,k} Single (black) and double (red) recollision electron yields. (j) Yields in dependence of laser intensity for three nanosphere diameters (as indicated). (k) Yields of all electrons (gray area) and selective yields against nanoparticle diameter for $I = \SI{3e13}{W/cm^2}$ (cf. vertical line in (j)). 
\panel{l,m} Cut-off energies of single and double recollision electrons. (l) Intensity-dependent cutoffs of selective single and double recollision energy spectra and from respective full spectra (dark to light gray areas). (m) Cut-offs against sphere diameter for $I = \SI{3e13}{W/cm^2}$ (cf. vertical line in (l)).
Adapted from~\cite{Seiffert_APB122_2016}.}
\label{fig:Seiffert_APB122_2016_especs_yields_cutoffs}
\end{figure}

\Figure{fig:Seiffert_APB122_2016_especs_yields_cutoffs}{a-i} shows selective electron energy spectra obtained from \MCCC simulations for electrons emitted directly and after one or two surface recollisions from \silica spheres with various sizes under few-cycle pulses with different intensities (as indicated). The results substantiated that direct electrons (gray curves) dominate the spectra for small particles and low intensities as known from earlier studies. With increasing intensity direct emission becomes more and more quenched (i.e. the yield becomes significantly suppressed) due the emerging trapping field. For the largest diameter, direct electrons exceed the conventional \SI{2}{\Uploc} cut-off energy which can be attributed to additional acceleration by tangential components of the local near-field. Most importantly, however, the high energy spectral region consists mainly of electrons with one recollision (black curves) for $d = 100$ and $\SI{500}{nm}$, while electrons with two recollision events (red curves) dominate the cut-off for the largest spheres for all inspected intensities (compare red to black curves in \Fig{fig:Seiffert_APB122_2016_especs_yields_cutoffs}{g-i}).

A systematic analysis of the yields and cutoff energies of backscattered electrons in dependence of laser intensity and sphere diameter is presented in \Fig{fig:Seiffert_APB122_2016_especs_yields_cutoffs}{j-m}. The yields of both, single and double recollision electrons increase essentially linearly with intensity (cf. \Fig{fig:Seiffert_APB122_2016_especs_yields_cutoffs}{j}) substantiating the quenching of tunnel ionization due to charge separation at the surface. The also linear scaling of the yields with sphere size further substantiates the trapping effect, as illustrated by the following intuitive picture. Assuming sequential electron emission, the total charge $Q$ of the sphere increases with each electron which leads to an attractive Coulomb potential proportional to $Q/R$. As a result, a number of $N_\text{crit} \sim R$ electrons may be emitted before the sphere reaches a critical charge $Q_\text{crit}$ = $eN_\text{crit}$ corresponding to a Coulomb potential that overcomes the electrons initial energy quenching further emission. The intensity-dependent cut-off energies of single and double recollision electrons from spheres with different diameters (cf. \Fig{fig:Seiffert_APB122_2016_especs_yields_cutoffs}{l}) first increase for $I \lesssim\SI{2e13}{W/cm^2}$ and show no substantial further enhancement at higher intensities, signifying the quenching of the ionization as discussed in \Section{sec:quenching_material}. Most importantly, \Fig{fig:Seiffert_APB122_2016_especs_yields_cutoffs}{m} shows that with increasing sphere size the cutoff energy of double recollision electrons increases faster and exceeds the cutoff of single recollion electrons for $d \gtrsim \SI{600}{nm}$.

\begin{figure}[t!]
\centering
\includegraphics[width=0.9\textwidth]{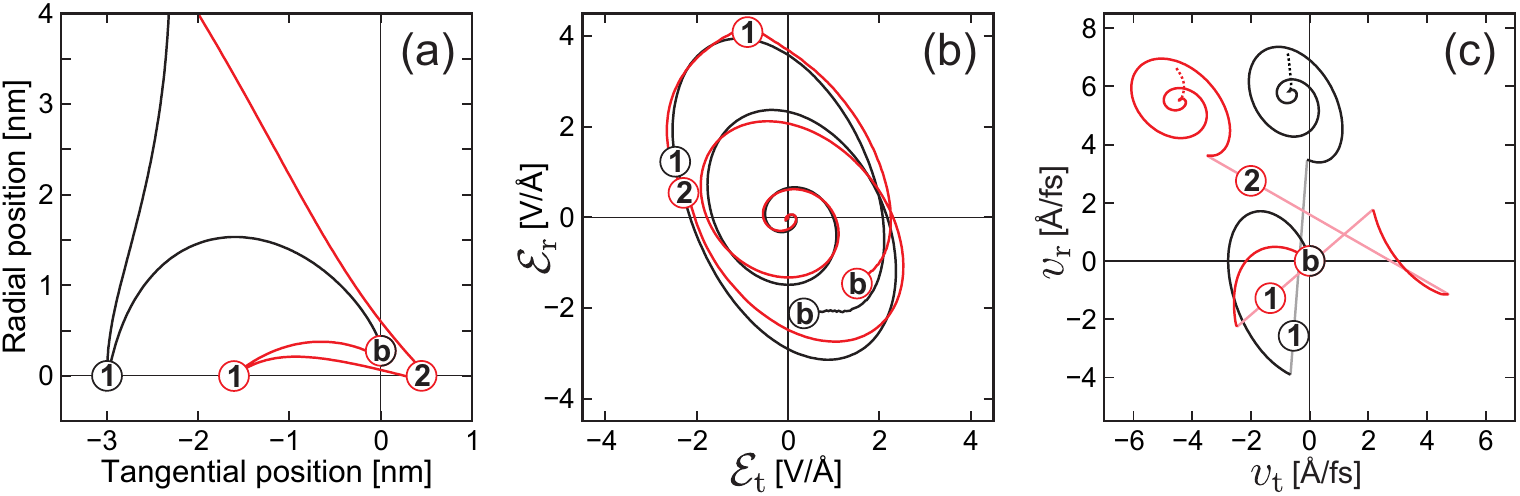}
\caption{Trajectory analysis of single and double rescattering. \panel{a} Evolution of radial and tangential excursion with respect to the birth position of typical trajectories extracted for cut-off electrons emitted from \SI{700}{nm} silica spheres after single (black) and double (red) rescattering. Circles indicate moments of birth 'b', first '1' and second '2' recollision. \panel{b} Evolution of the elliptic local near-fields, sampled along the respective trajectories. \panel{c} Evolution of the velocity components. Jumps at recollisions are indicated as light colored lines. The elliptical features (around $v_r \approx \SI{5}{\AA/fs}$) reflect the trivial radial and tangential quiver motions after the recollision phase. The dotted ends of the curves show the additional (mainly radial) velocity gain mediated by the space-charge driven Coulomb explosion of the escaping bunches. Adapted from~\cite{Seiffert_APB122_2016}.}
\label{fig:double_recollision_trajectories}
\end{figure}

The physical picture for the increasing importance of double recollisions predicted for large nanospheres was unraveled by inspecting typical trajectories of single and double recollision electrons with final energies close to the respective cutoffs for $d = \SI{700}{nm}$ and $I = \SI{4e13}{W/cm^2}$ (cf. \Fig{fig:Seiffert_APB122_2016_especs_yields_cutoffs}{i}) where the effect is most pronounced. The radial and tangential components (w.r.t. the surface) of the single (black curve) and double (red curve) recollision trajectories, the local near-fields sampled along the trajectories and the respective velocities are shown in \Fig{fig:double_recollision_trajectories}. As reported in previous studies~\cite{Suessmann_NatCommun6_2015} the single recollision process is mostly determined by the radial near-field. The respective trajectory is launched at the classical tunneling exit around the minimum of the radial field (cf. labels 'b' in \Fig{fig:double_recollision_trajectories}{a,b}) and recollides after three quarters of the fields cycle close to the zero crossing of the radial field (cf. labels '1' in \Fig{fig:double_recollision_trajectories}{a,b}). The mainly radial movement during the recollision results in a pronounced jump of the radial velocity (light gray line in \Fig{fig:double_recollision_trajectories}{c}). The following elliptic feature in the velocity representation around $v_r \approx \SI{5}{\AA/fs}$ indicates the quiver motion in the elliptic near field, followed by the mainly radial acceleration due to the Coulomb explosion of the escaping bunch (dotted part of the black curve).

In contrast, the double recollision trajectory (red curve in \Fig{fig:double_recollision_trajectories}{a}) proceeds mainly in tangential direction resulting in a 'ping-pong' like motion during the recollision phase. While timing of the second recollison is similar to that of the single recollision trajectory (i.e. at the zero-crossing of the radial field) the first recollision takes place close to the zero crossing of the tangential field, see \Fig{fig:double_recollision_trajectories}{b}. These particular timings result in efficient tangential acceleration before and after the first recollision and efficient radial acceleration during the final escape from the surface (cf. \Fig{fig:double_recollision_trajectories}{c}).

\subsection{Reaction nanoscopy on silica nanospheres}
\begin{figure}[h!]
\centering
\includegraphics[width=1.0\textwidth]{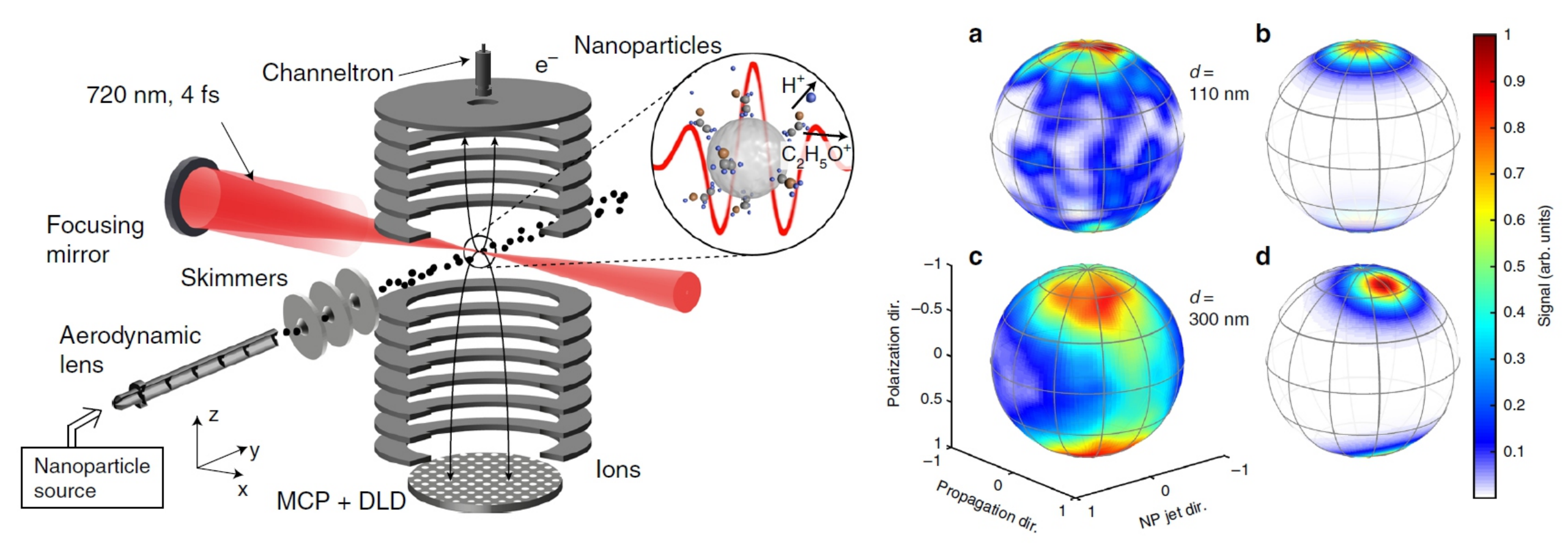}
\caption{Reaction nanoscopy with small and large \silica nanospheres. \panel{a} Schematic setup. Nanospheres and molecular surface adsorbates are ionized by few-cycle laser pulses in the center of the reaction nanoscope. Electrons and ions are recorded in coincidence as fragments arising from molecular photodissociation are accelerated towards the bottom ion detector (microchannel plates (MCP) and delay-line detector (DLD)) whereas electrons are accelerated towards the opposite side and are detected with a channeltron (top). \panel{b-e} Angle-resolved momentum distributions of emitted protons, obtained via radial integration over the 3D momentum distributions. The number of protons per solid angle is encoded in the color scale. Measured (b) and simulated (c) data for $d=\SI{110}{nm}$ spheres and (d,e) for $d=\SI{300}{nm}$ particles, respectively. Adapted from~\cite{Rupp_NatCommun10_2019}.}
\label{fig:Rupp_NatCommun10_2019}
\end{figure}

Fused silica nanospheres have also been utilized for reaction nanoscopy, where the spatially dependent reaction yield of few-cycle induced dissociative ionization of ethanol and water has been inspected via momentum resolved measurement of the photoemission by Rupp\etal~\cite{Rupp_NatCommun10_2019}. In the experiment (cf. \Fig{fig:Rupp_NatCommun10_2019}{a}), the emission of electrons and protons resulting from fragmentation following the dissociation of molecules adsorbed to the nanosphere surface were measured in coincidence after irradiation with \SI{4}{fs} NIR laser pulses ($\lambda=\SI{720}{nm}$, $I\approx\SI{5e13}{W/cm^2}$). Spatially resolving the proton emission allowed to to extract angle-dependent proton momenta, as visualized in \Fig{fig:Rupp_NatCommun10_2019}{b,c} for small and large nanospheres (as indicated). The obtained distributions revealed the spatial variability of the reaction yield on the surface which closely correlates with the amplitude of the local near-field. These obervation were substantiated by comparison with \MCCC simulations (cf. \Fig{fig:Rupp_NatCommun10_2019}{d,e}), which had been extended to include the evaluation of dissociative ionization yields and the integration of proton trajectories starting from the nanosphere surface. The simulations revealed, that the protons were mainly accelerated by the repulsive electrostatic field of the sphere due to its positive net charge following electron emission and takes place after the interaction with the laser pulses. This allowed to reconstruct the nanoscale reaction yield landscape from the measured data, which could enable for spatially resolved characterization of nanoparticle photochemistry.

\section{Subcycle metallization of initially dielectric spheres}
\label{sec:metal_spheres}
\begin{figure}[t!]
\centering
\includegraphics[width=1.0\textwidth]{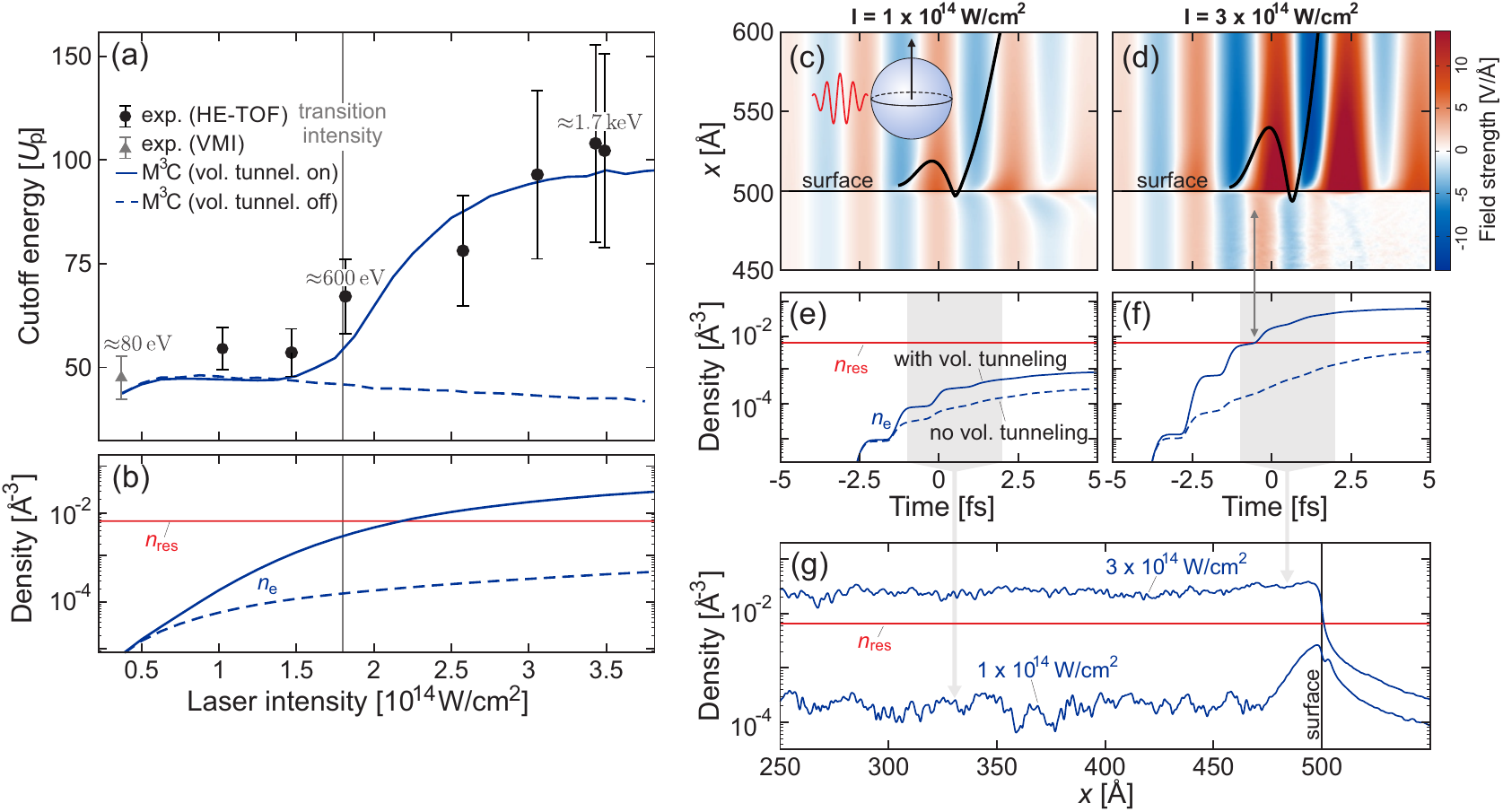}
\caption{Subcycle metallization of \SI{95}{nm} \silica nanospheres under few-cycle NIR laser pulses. \panel{a} CEP-averaged measured electron cutoff energies (black circles ) as a function of laser intensity. Blue curves represent cutoffs predicted from \MCCC simulations for the experimental parameters (including focus averaging) with tunnel ionization enabled only at the surface (dashed) and within the full volume (solid). For the simulations, $\tau = \SI{750}{as}$ was considered to mimic the lifetime of plasmonic excitations (cf. \Eq{eq:el_scattering_plasmonic} and respective discussion in the main text). \panel{b} Number density of free electrons $n_\text{e}$ at the pulse peak from \MCCC simulations with and without volume tunneling (as indicated) as function of intensity. The density for which the frequency of the plasmon matches the laser frequency is indicated as resonant density $n_\text{res}$. \panel{c,d} Evolution of the normal component of the internal ($x<\SI{500}{\AA}$) and external ($x>\SI{500}{\AA}$) near-fields at the surface evaluated along the polarization axis (cf. black arrow in the pictogram in panel (c)) at $\cep=0$ for two different intensities (as indicated). Solid black curves show averaged trajectories of the fastest ten percent of emitted electrons. \panel{e,f} Evolution of the number density of generated electrons from simulations without (dashed) and with (solid) volume tunneling. The red line indicates the electron density at resonance ($n_\text{res}$). \panel{g} Electron density along the x-axis around the upper pole (cf. black arrow in the inset in panel (c)) calculated with volume tunneling for both intensities and averaged during the recollision phase (gray areas in (e) and (f)). Adapted from~\cite{Liu_ACS7_2020}.}
\label{fig:Liu_ACSP_2020_metallization1}
\end{figure}

In this section we will review selected work considering the photemission form spheres with metallic optical response. The onset of subscycle metallization of initially dielectric and semiconducting small nanospheres under intense few-cycle pulses has been reported by Liu\etal~\cite{Liu_ACS7_2020}, where the previous studies have been extended to even higher laser intensities. In particular, a systematic analysis of the cutoff energy of fast photoelectrons measured from \silica particles in dependence of laser intensity revealed a rapid transition from the \SI{50}{\Up} scaling law observed earlier~\cite{Zherebtsov_NatPhys7_2011, Suessmann_NatCommun6_2015} around a transition intensity of $I_\text{trans} = \SI{1.8e14}{W/cm^2}$ followed by saturation around \SI{100}{\Up} (cf. black symbols in \Fig{fig:Liu_ACSP_2020_metallization1}{a}). The connection of this distinctive spectral signature with the onset of subcycle metallization has been enabled by comparison with \MCCC simulations with tunnel ionization enabled in the full volume or only at the surface. While the latter predicted cutoffs around \SI{50}{\Up} for the whole intensity range (cf. dashed blue curve), the transition was only reproduced when allowing volume tunneling (cf. solid blue curve). To 'quantify' the metallization, the number density of free electrons at the time of the pulse maximum was compared to the resonance density associated with the instantaneous plasma frequency $\omega_\text{p} = \frac{n_\text{ion}e^2}{3m_\text{e}\epsilon_0}$ of the sphere (i.e. the Mie plasmon), where $n_\text{ion}=n_\text{e}$ as only single ionization was considered, see \Fig{fig:Liu_ACSP_2020_metallization1}{b}. The comparison revealed that the resonance was crossed around the transition intensity when including volume tunneling, whereas only surface tunneling resulted in insufficient ionization resulting in the system remaining undercritical for all laser intensities. The intuitive picture for the connection between electron density and the observed cutoff energies is that once the volume of the initially dielectric sphere is sufficiently ionized (i.e. $n_\text{e} \gtrsim n_\text{res}$) the liberated electrons lead to the sphere responding to the incident laser field like a metal. As the maximum field enhancement at the poles of a perfectly conduction small sphere is $\gamma_0 = 3$ (cf. \Fig{fig:theory_dipolefields}{e}), the energy of recollision electrons as predicted by the classical SMM is $\gamma_0^2 \cdot \SI{10}{\Up} = \SI{90}{\Up}$.

This picture is supported by inspection of the evolution of averaged trajectories of electrons close to the respective cutoffs and the near-fields at the surface below and above the transition intensity as shown in \Fig{fig:Liu_ACSP_2020_metallization1}{c,d}, respectively. While the in- and outside fields remain in phase, jump by a factor of roughly $\varepsilon_\text{r} = 2.1$ and are only slightly distorted due to charge separation below the transition intensity, above the latter the near-fields are substantially modified. The fields in- and outside run out of phase, the inside field starts to decay around the pulse maximum to become almost completely screened and the outside field gets substantially enhanced, resulting in enhanced acceleration of the electrons during the backscattering process. These signatures clearly indicated a subcycle transition to a metallic state and the temporal coincidence with crossing the resonant density (cf. \Fig{fig:Liu_ACSP_2020_metallization1}{f}) established the link between the metallization and the transition of the electron cutoff energies. The metallization of the whole nanosphere volume was further substantiated by inspection of radial electron density profiles as shown in \Fig{fig:Liu_ACSP_2020_metallization1}{g}. While at low intensities free electrons are localized close to the surface, above the transition intensity the electron density is almost constant and exceeds the resonant density in the full volume.

\begin{figure}[h!]
\centering
\includegraphics[width=0.45\textwidth]{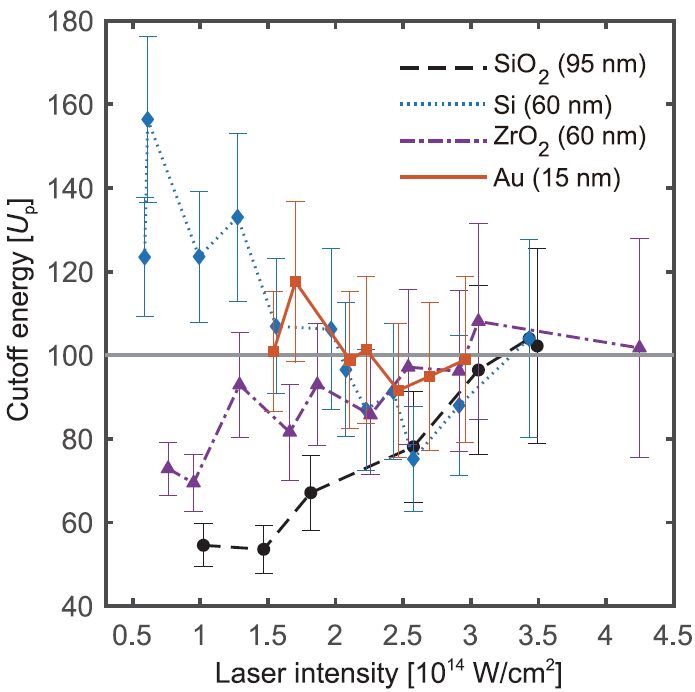}
\caption{Electron cutoff energies measured from different nanoparticles (as indicated) as a function of incident laser intensity. From~\cite{Liu_ACS7_2020}.}
\label{fig:Liu_ACSP_2020_Fig5}
\end{figure}

To inspect the generality of the metallization picture, cutoff energies have been measured from dielectric (\silica, ZrO\textsubscript{2}), semiconducting (Si) and metallic (Au) nanospheres as function of laser intensity, see \Fig{fig:Liu_ACSP_2020_Fig5}. While the cutoff of the already metallic gold nanoparticles remains essentially constant around $\SI{100}{\Up}$ for all intensities, the cutoffs of the three other species differ substantially at low intensities but converge to a common value of around $\SI{100}{\Up}$ with increasing intensity, suggesting that ionization-induced subcycle metallization is a general phenomenon.

\section{Characterization of transport effects in nanospheres via attosecond streaking}
\label{sec:streaking}
Attosecond streaking~\cite{Constant_PRA56_1997, Drescher_Science291_2001, Hentschel_Nature414_2001, Itatani_PRL88_2002, Kienberger_Nature427_2004}, a method in which electrons are liberated from a target via photoionization by an isolated attosecond XUV pulse and subsequently accelerated by a synchronized optical femtosecond pulse, is nowadays well established to characterize both the temporal and spectral characteristics of the driving XUV pulses~\cite{Mairesse_PRA71_2005, Gagnon_APB92_2008} and the light waves of the optical fields~\cite{Goulielmakis_Science305_2004, Kienberger_Nature427_2004}. Moreover, streaking with atomic~\cite{Schultze_Science328_2010, Nagele_JPB44_2011} and molecular~\cite{Hockett_JPB49_2016} targets allowed to study photoemission delays~\cite{Dahlstroem_JPB45_2012, Pazourek_RMP87_2015} which essentially arise from the atomic potentials and contain target-specific information such as Eisenbud-Wigner-Smith-type (EWS) delays~\cite{Eisenbud_PHD_1948, Wigner_PR98_1955, Smith_PR118_1960} and contributions from post-interaction dynamics due to Coulomb-laser coupling (CLC)~\cite{Smirnova_JPB39_2006, Smirnova_JPB40_2007, Zhang_PRA82_2010}. Attosecond streaking was also extended to inspect delays reflecting the electron transport to the surface for clean and adlayer-covered metallic surfaces~\cite{Cavalieri_N449_2007, Neppl_PRL109_2012, Neppl_Nature517_2015} and to semi-conductors~\cite{Siek_Science357_2017}. The study of transport effects using dielectrics surfaces and the streaking method, however, has remained difficult due to the spurious shot-to-shot charge accumulation and resulting space charge fields.

\begin{figure}[t!]
\centering
\includegraphics[width=1.0\textwidth]{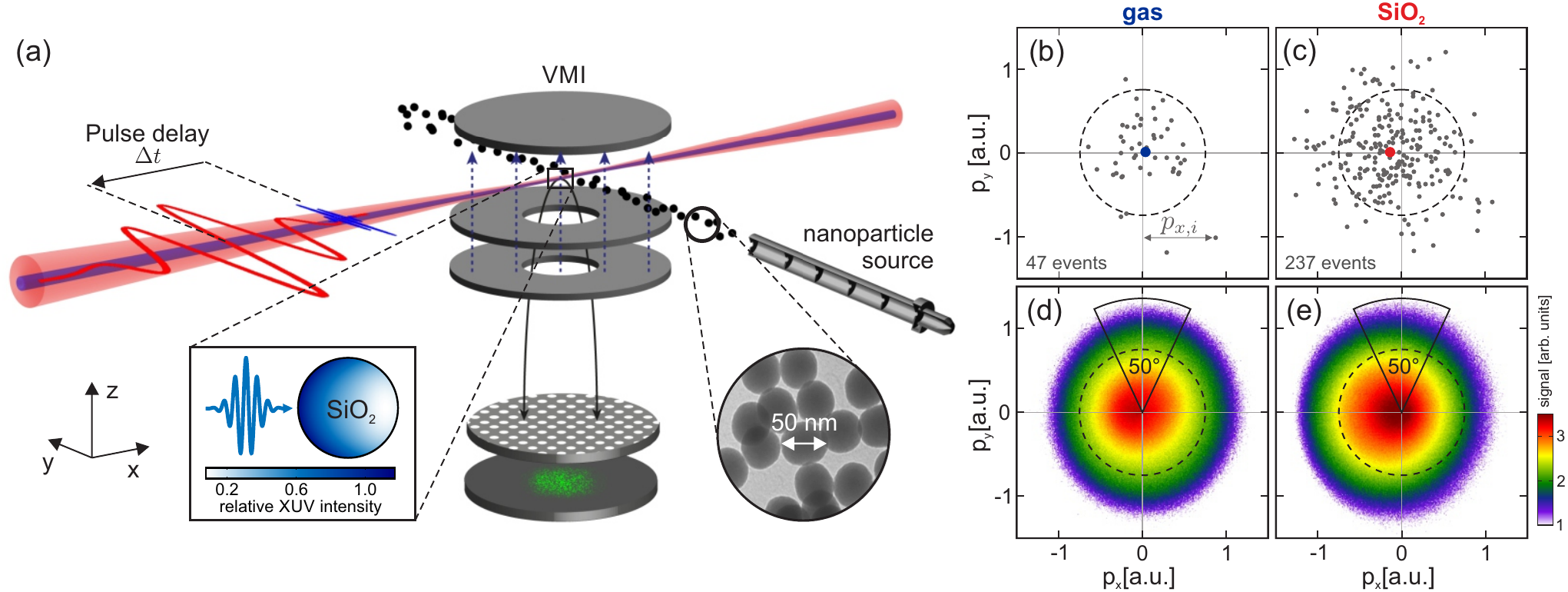}
\caption{Attosecond streaking with dielectric nanospheres. \panel{a} Schematic setup of the experiment. Synchronized attosecond XUV and few-cycle NIR pulses with adjustable delay $\Delta t$ were shot on a beam of isolated \silica nanopspheres. Emitted electrons were recorded by a single-shot VMI spectrometer. \panel{b,c} Typical single-shot VMI momentum projections recorded from a gas-only measurement (b) and when including silica nanospheres (c) for $\Delta t = \SI{0}{as}$. Dashed circles indicate the low momentum region ($< \SI{0.75}{a.u.}$), where the measured signal contains additional contributions from residual gas background. The grey arrow in (b) illustrates the x-component of the projected momentum of the i-th measured event. Blue and red dots indicate respective averaged projected momenta. \panel{d,e} Respective averaged momentum distributions. Adapted from~\cite{Seiffert_NatPhys13_2017}.}
\label{fig:Seiffert_NatPhys13_2017_1}
\end{figure}

Recently, as an interesting alternative, attosecond streaking was employed to inspect transport effects in dielectric nanoparticles, cf.  Seiffert\etal~\cite{Seiffert_NatPhys13_2017}. The schematic setup of the experiment is shown in \Fig{fig:Seiffert_NatPhys13_2017_1}{a}. A beam of isolated $d = \SI{50}{nm}$ \silica nanospheres was irradiated by isolated \SI{250}{as} XUV pulses (central photon energy $\approx \SI{28}{eV}$, peak intensities $\approx \SI{2e12}{W/cm^2}$) and synchronized \SI{5}{fs} NIR pulses (central wavelength $\approx\SI{720}{nm}$, peak intensities $\approx \SI{1e12}{W/cm^2}$) with adjustable delay $\Delta t$. Emitted electrons were recorded via single-shot VMI. The nanosphere density within the incident beam was sufficiently low, such that hits of multiple nanospheres within one laser shot were negligible. Further, only roughly every fifth laser shot hit a nanoparticle, resulting in VMI images including only contributions from the carrier gas and images containing signal from both the gas and a nanosphere. Typical single shot projected momentum images are shown in \Fig{fig:Seiffert_NatPhys13_2017_1}{b,c} for a gas only measurement and when including nanoparticles. Note that the \silica data includes electrons stemming from a nanosphere as well as contributions from the carrier gas. However, due to the higher ionization energy the latter only contributes to the low momentum region (i.e. within the dashed circles). Efficient classification of laser shots hitting only gas or hitting a nanosphere was enabled by inspecting the number of detected photoelectron events and the average electron momentum in the laser propagation direction (cf. blue and red symbols in \Fig{fig:Seiffert_NatPhys13_2017_1}{b,c}, also termed 'shadowing'~\cite{Signorell_CPL658_2016}). For nanoparticle hits, the former is higher and the latter is shifted towards negative values resulting from stronger photoionization at the front side of the sphere where the XUV field is absorbed (cf. left inset in \Fig{fig:Seiffert_NatPhys13_2017_1}{a}). Resulting averaged momentum distributions for gas and nanospheres are shown in \Fig{fig:Seiffert_NatPhys13_2017_1}{d,e}.

\begin{figure}[h!]
\centering
\includegraphics[width=1.0\textwidth]{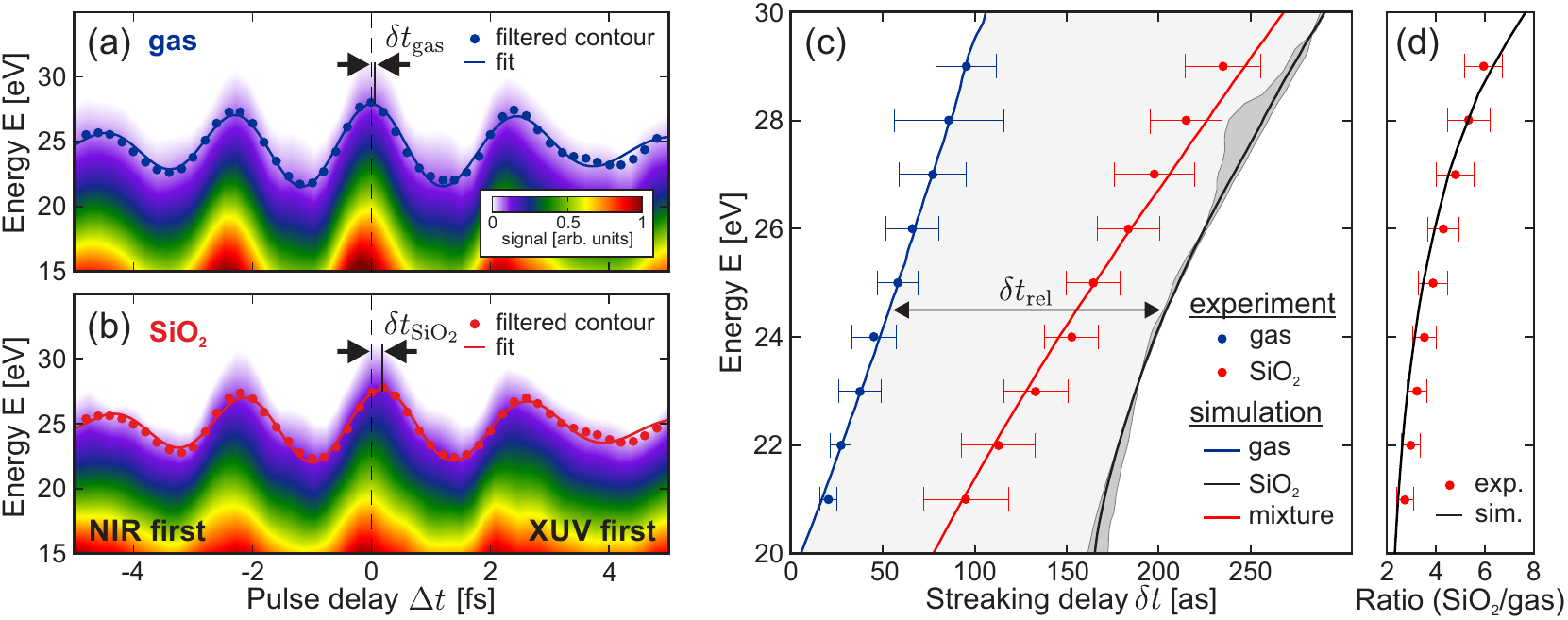}
\caption{Streaking spectrograms and extracted straking delays. \panel{a,b} Attosecond streaking spectrograms obtained from angular integration of projected momentum maps around the laser polarization direction (cf. area indicated in \Fig{fig:Seiffert_NatPhys13_2017_1}{d,e}) for gas (a) and silica nanoparticles (b). To extract the streaking delays energy-dependent frequency-filtered isolines (blue and red dots) were fitted with few-cycle waveforms (blue and red curves) as described in the text. The fits carrier phases define the respective streaking delays $\delta t$. \panel{c} Streaking delays extracted in the high energy range of the measured gas and nanoparticle streaking spectrograms (blue and red dots). Blue and black curves show delays predicted by gas simulations and nanoparticle simulations for the experimental parameters, respectively. The red curve represents the delay extracted from a mixed spectrogram. The dark gray shaded area reflects the maximal variation of the extracted streaking delay when performing the nanosphere simulations with charge-interaction enabled or anisotropic elastic collisions. \panel{d} Ratio of nanoparticle to gas signal in dependence of energy extracted from the experiment (red symbols) and from the simulations (black curve). Adapted from~\cite{Seiffert_NatPhys13_2017}.}
\label{Seiffert_NatPhys13_2017_2}
\end{figure}

Streaking spectrograms for gas and nanospheres were extracted by inspecting the corresponding energies of electrons emitted along the laser polarization direction (obtained via integration over a \ang{50} full opening angle) as function of the pulse delay, see \Fig{Seiffert_NatPhys13_2017_2}{a,b}. Both spectrograms revealed pronounced delay-dependent oscillations as well as a relative delay of the nanoparticle data with respect to the gas. To quantify the individual streaking delays, contour-lines (i.e. delay-dependent energies with equal signal strength) were determined and filtered to eliminate measurement-induced high frequency noise. As an example, two filtered contour lines are indicated by blue and red dots in \Fig{Seiffert_NatPhys13_2017_2}{a,b}. The streaking delay is extracted from each contour line via fitting a few-cycle waveform
$E_\text{fit}(\Delta t) = E + A\cos\left(\omega \left[\Delta t - \delta t\right]\right) e^{-\frac{1}{2}\frac{(\Delta t - t_0)^2)}{\tau^2}}$
with asymptotic energy $E$ and streaking delay $\delta t$ (solid curves). The resulting energy-dependent streaking delays $\delta t(E)$ extracted from several contour lines in the high-energy spectral regions ($20$--$\SI{30}{eV}$) of the gas and nanoparticle spectrograms (shown as blue and red symbols in \Fig{Seiffert_NatPhys13_2017_2}{c}) unveiled two prominent features. First, the streaking delays of both gas and \silica nanospheres increased linearly with energy which could be attributed mainly to the chirp of the XUV pulses. Second, and most importantly, the \silica and gas data exhibited an only slightly energy-dependent relative delay of around \SI{100}{as}.

The origin of the relative delay could be clarified by corresponding semi-classical \MCCC simulations for the experimental parameters. As the XUV chirp could not be extracted accurately from the experiment it was determined by comparing streaking delays extracted from a simulation for the reference gas to the experiment. Best agreement was found for a chirp parameter of $\zeta = \SI{-7e-3}{fs^2}$, compare blue curve to blue symbols in \Fig{Seiffert_NatPhys13_2017_2}{c}. Considering this chirp for the nanosphere simulations resulted in an extracted streaking delay exceeding that from the experiment (compare black curve to red symbols). The discrepancy was attributed to the contributions from carrier gas in the nanoparticle signal which results in the measured nanoparticle streaking delay actually reflecting a mixture of the signals from nanosphere and gas. The energy-dependent ratio of nanoparticle and gas signal could be extracted from the experiment via background subtraction, see red symbols in \Fig{Seiffert_NatPhys13_2017_2}{d}. It was accounted for in the simulations by combining the individual streaking spectrograms for gas and nanoparticles in a mixed spectrogram $S_\text{mix}(E,\Delta t) = S_\text{SiO$_2$}(E,\Delta t) + \eta S_\text{gas}(E,\Delta t)$ by adjusting the parameter $\eta$ for best agreement with the experimental ratio (compare black curve to red symbols in \Fig{Seiffert_NatPhys13_2017_2}{d}). The excellent quantitative agreement of the streaking delay extracted from the mixed spectrogram with the experiment (compare red curve to red symbols in \Fig{Seiffert_NatPhys13_2017_2}{c}) support the assumption that the simulations include the relevant physics and can thus be utilized to identify the physical picture behind the essentially  relative delay $\delta t_\text{rel} = \delta t_\text{SiO$_2$} - \delta t_\text{ref} \approx \SI{150}{as}$ between the (pure) nanoparticle delay $\delta t_\text{SiO$_2$}$ and the delay of the gas reference $\delta t_\text{ref}$ (cf. black arrow and gray shaded area in \Fig{Seiffert_NatPhys13_2017_2}{c}).

\begin{figure}[b!]
\centering
\includegraphics[width=0.95\textwidth]{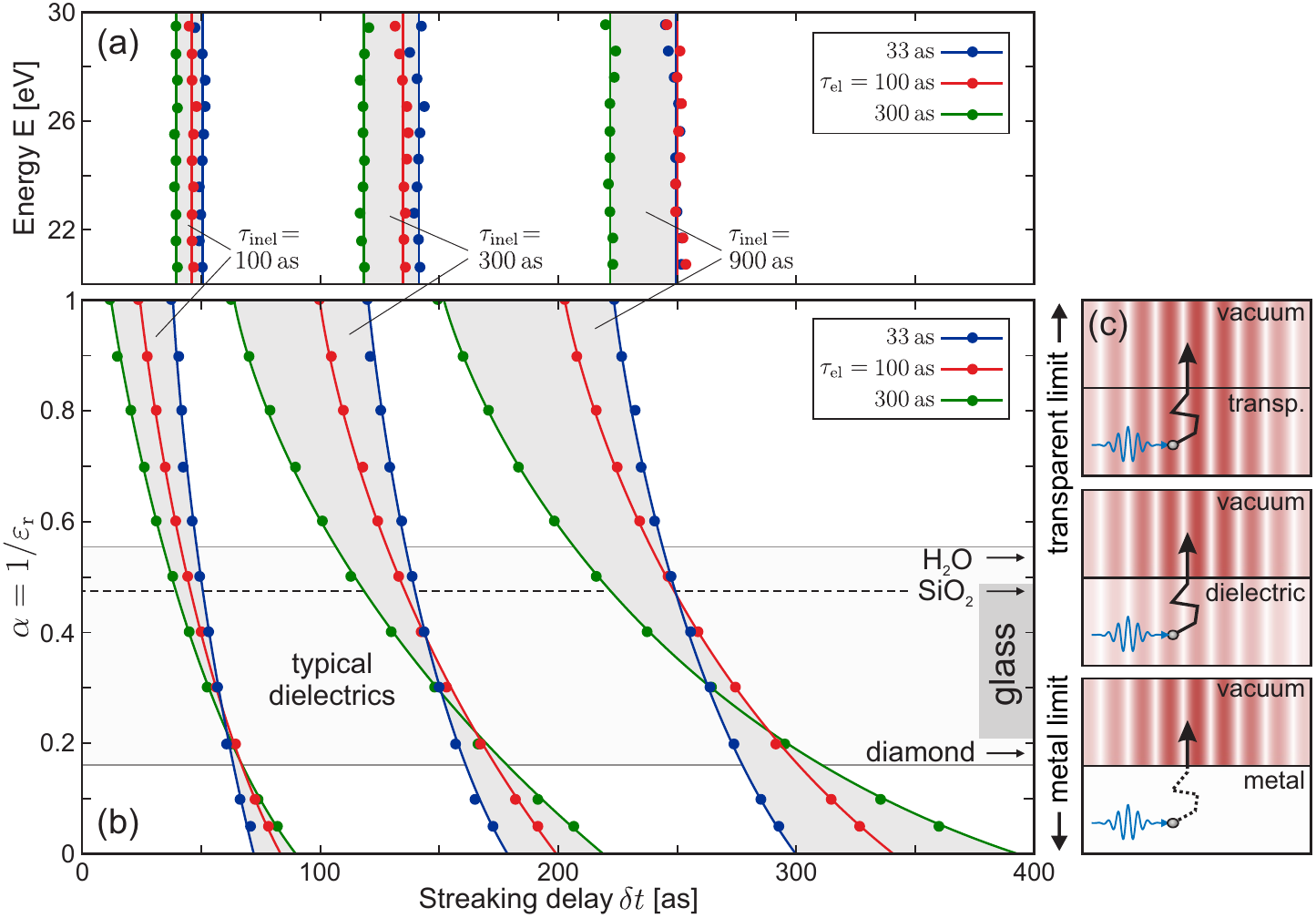}
\caption{Systematic analysis of the collisional streaking delay calculated via \MCCC assuming unchirped XUV pulses and energy-independent scattering times to eliminate energy-dependencies of the streaking delays. \panel{a} Streaking delays in dependence of elastic and inelastic scattering times (as indicated). \panel{b} Streaking delays as function of the material’s attenuation factor $\alpha = 1/\varepsilon_\text{r}$, where $\varepsilon_\text{r}$ is the relative permittivity at the wavelength of the NIR field. Gray shaded areas in both panels visualize the variation of the streaking delay in dependence of the elastic scattering time. The gray rectangle and the black arrows in (b) indicate permittivities of typical dielectric materials. The dashed black line marks the permittivity of \silica. Adapted from~\cite{Seiffert_NatPhys13_2017}. \panel{c} Schematic illustration of attosecond streaking on the surfaces of transparent (top), dielectric (center) and metallic (bottom) targets. After ionization by the XUV pulse (blue) the electron is streaked by the near-field of the optical pulse (red). Adapted from~\cite{Seiffert_PhD_2018}.}
\label{fig:Seiffert_NatPhys13_2017_3}
\end{figure}

A detailed analysis of the impacts of various contributions to the individual delays for the gas and nanospheres revealed that delays imposed by the XUV chirp as well as field propagation and retardation, and near-field inhomogeneities were negligible in the inspected scenario (for a detailed analysis see the supporting material of~\cite{Seiffert_NatPhys13_2017} or~\cite{Seiffert_PhD_2018}). Hence, the relative delay was in good approximation solely governed by the material-specific collisional electron transport dynamics, i.e. $\delta t_\text{rel} = \delta t_\text{coll.}$ To systematically explore the effect of electron transport within the material and its significance for the collisional streaking delay \MCCC simulations were performed with two central simplifications. First, the XUV pulses were considered to be unchirped in order to eliminate the energy-dependent tilt of the streaking delays. Second, electron scattering was modeled via fixed energy-independent scattering times for both elastic and inelastic collisions. These times were considered as $\tau_\text{el} = \SI{100}{as}$ and $\tau_\text{inel} = \SI{300}{as}$ to closely resemble the energy-dependent scattering times in the relevant energy range (cf. \Fig{fig:theory_inelastic_scattering}{d}) and were each varied by factors of $1/3$ and $3$. The nine resulting energy-dependent streaking delays shown in \Fig{fig:Seiffert_NatPhys13_2017_3}{a} allowed three key conclusions. First, neglecting the XUV chirp and assuming fixed scattering times results in essentially energy-independent streaking delays. Second, the delays are strongly sensitive to the inelastic scattering time, where increasing the inelastic scattering time results in larger streaking delays (see three separated groups of curves indicated by the gray areas). Third, the sensitivity to the elastic scattering time is much weaker (compare colored curves within each gray group).

The observed strong sensitivity of the streaking delays to the inelastic scattering time and only weak dependence on the elastic scattering time raised the question if this behavior is general for solid-state attosecond streaking from arbitrary materials. A key difference between streaking at metallic and dielectric systems is the screening of the NIR field within the material and hence its effect on the internal streaking during the electron transport to the surface (cf. \Fig{fig:Seiffert_NatPhys13_2017_3}{c}). The impact of the internal streaking was inspected by repeating the simulations shown in \Fig{fig:Seiffert_NatPhys13_2017_3}{a} and additionally adjusting the strength of the internal field by varying the materials relative permittivity $\varepsilon_\text{r}$ for each run. As the streaking delays were energy-independent, each individual simulation for a set of parameters ($\tau_\text{el}$, $\tau_\text{inel}$ and $\varepsilon_\text{r}$) resulted in one (energy-averaged) value for the streaking delay. The results of this systematic analysis are shown in \Fig{fig:Seiffert_NatPhys13_2017_3}{b} in dependence on the field attenuation factor $\alpha = 1/\varepsilon_\text{r}$. In the two limiting cases $\alpha \rightarrow 0$ and $1$ the delays were found to be sensitive to both the inelastic and the elastic scattering time, as visualized by the broadening of the gray areas. The dependence on the latter, however, had opposite signs in the two limits. In the transparent case ($\alpha \rightarrow 1$, $\varepsilon_\text{r} \rightarrow 1$), the streaking delay decreased for larger elastic scattering times (compare blue to green curves at the top). In contrast, in the metal limit ($\alpha \rightarrow 0$, $\varepsilon_\text{r} \rightarrow \infty$) larger elastic scattering times resulted in larger streaking delays (compare blue to green curves at the bottom). The elastic collision effect reversed (and thus effectively vanishes) for permittivity values of typical dielectric materials (light gray shaded area). This nearly exclusive sensitivity of the streaking delays to the inelastic scattering time for dielectrics therefore enabled the retrieval of quantitative scattering times by matching the simulation results to experimental data. For the presented scenario and electron energies around \SI{25}{eV} the authors extracted an inelastic scattering time of $\tau_\text{inel}\approx \SI{370}{as}$ which corresponds to an inelastic mean free path of approximately \SI{10}{\AA} in good agreement with values calculated from optical data by Tanuma\etal~\cite{Tanuma_SIA17_1991} and obtained from Monte-Carlo simulations by Kuhr\etal~\cite{Kuhr_JESRP105_1999} (cf. \Fig{fig:theory_inelastic_scattering}{b}).


\section{Conclusions and future perspectives}
The selected set of results from the first decade of strong-field physics with nanospheres has clearly demonstrated that these targets provide a suitable and experimentally as well as theoretically accessible model system to tackle various  fundamental questions in attosecond nanophysics. These range from confirming that many of the established pictures and tools from atomic strong-field physics are also applicable to more complex nanotargets (e.g. the backscattering picture of HATI), over the identification of qualitatively new aspects such as the impacts of charge interaction or field propagation to the development of new metrologies for the quantification of transport properties in dielectrics. For example, the at first glance seemingly spurious charge interaction effects also revealed surprising advantages like further enhancements of the energies of fast recollision electrons or the suppression of certain emission channels (e.g. direct photoemission) and even the realization of a material-independent photoemission response at high laser-intensities.

Although nanospheres will probably not be very suitable for future technical  applications themselves, it is expected that many of the fundamental physical insights first obtained utilizing these simple model systems are also of major importance for more applications-relevant systems and application fields like sub-cycle laser processing or Petahertz optoelectronics employing other nanostructures such as nanotips~\cite{Dombi_RMP92_2020}. For example, only recently the impact of charge interaction on the photoemission, first observed with dielectric nanospheres, has also been demonstrated with metal nanotips~\cite{Schoetz_NP_2021}.

The initial proof-of-principles demonstrated in some of the seminal publications reviewed in this work still leave open a long list of questions for future studies. For example, the attosecond streaking metrology, so far demonstrated for silica nanospheres, could be extended to higher photon energies or other materials for which quantification of inelastic mean-free paths is not accessible via other techniques. New milestones could also be reached by combining phase-dependent excitation of nanospheres via few-cycle fields with X-ray imaging techniques to track plasma dynamics in space and time simultaneously~\cite{Peltz_PRL113_2014}. Another approach to gain further insights into strong-field ionization of solids could be to extend the excitation to light fields even more extreme than few-cycle pulses, for example by deploying sub single cycle fields~\cite{Hassan_Nature530_2016}. Hence, the strong-field physics with nanospheres remains a fascinating platform for future scientific activity in the field of attosecond nanophysics.

\section*{Acknowledgments}
We gratefully acknowledge the invaluable contributions of our colleagues and collaborators on related original work that this article is based on.

\section*{Disclosure statement}
No potential conflict of interest was reported by the authors.

\section*{Funding}
This work was supported by the German Research Foundation (DFG) under grant 652714; S.Z. was supported from the DFG via ZH582/1-1; M.F.K. acknowledges support from the Max Planck Society via the Max Planck Fellow program; We acknowledge support from the DFG within SPP 1391 Ultrafast Nanooptics, SPP 1840 QUTIF and SFB 652/3; T.F. acknowledges financial support from the DFG via Heisenberg Grant 398382624; We are grateful for the computing time provided by the North-German super-computing center HLRN (project mvp00011 and mvp00017)

\bibliography{Seiffert_Review}

\end{document}